\documentclass[11pt,a4paper]{article}
\usepackage[utf8]{inputenc}
\usepackage{amsmath,amssymb,amsthm,mathtools}
\usepackage{graphics} 
\usepackage{graphicx}  
\usepackage{epstopdf}
\usepackage[colorlinks=true]{hyperref}
\usepackage{tikz}
\usepackage{makeidx}
\usepackage{subfigure}
\usepackage{caption}
\usepackage{setspace}
\usepackage{a4wide}
\usepackage{epstopdf}
\usepackage{subfigure}

\usepackage{natbib}
\bibpunct[, ]{(}{)}{;}{a}{}{,}
\renewcommand\bibfont{\fontsize{10}{12}\selectfont}
\usepackage{lineno,hyperref}
\usepackage{amsfonts}
\usepackage{amsmath}
\usepackage{amsthm}
\usepackage{cases}
\usepackage{pgf-pie}
\usepackage{comment}

\usepackage{bbm}

\usepackage{tikz,pgf,pgfplots}

\usetikzlibrary{arrows,shapes,snakes,automata,backgrounds,petri,fit,intersections,quotes,positioning}

\usepackage{tikz-3dplot}

\pgfplotsset{compat=1.14}

\theoremstyle{plain}

\theoremstyle{definition}

\theoremstyle{remark}
\newtheorem{remark}{Remark}

\definecolor{cof}{RGB}{219,144,71}
\definecolor{pur}{RGB}{186,146,162}
\definecolor{greeo}{RGB}{91,173,69}
\definecolor{greet}{RGB}{52,111,72}

\newcommand{\R}{\mathbb{R}}

\newcommand{\C}{\mathcal{C}}

\begin{document}


\title{Contrarian effect in opinion forming: insights from Greta Thunberg phenomenon}


\author{Elisa Iacomini\footnote{{ Institut f\"{u}r Geometrie und Praktische Mathematik (IGPM), RWTH Aachen University, Templergraben 55, 52062 Aachen, Germany ( iacomini@igpm.rwth-aachen.de)}} ,  \; Pierluigi Vellucci\footnote{{Department of Economics, Roma Tre University, Via Silvio D'Amico 77, 00145 Rome, Italy (pierluigi.vellucci@uniroma3.it)}}}



\date{ }

\maketitle

\begin{abstract}
In recent months the figure of Greta Thunberg and the theme of climate changings quickly became the focus of the debate. This has lead to a polarization effect in opinion forming about the climate subject. Starting from the analysis of this phenomenon, we develop an opinion dynamics model in which several types of contrarians agents are considered. Each agent is supposed to have an opinion on several topics related to each other, thus the opinions being formed on these topics are also mutually dependent. 
The aim of the paper is to investigate the indirect effects of contrarians agents on the collective opinion about these topics. 
Several numerical tests are presented in order to highlight the main features of the model.
\end{abstract}


\section{Introduction}
Nowadays the problem of global warming plays a crucial role on the public debate and opinions exchanges, affecting also the political and the economical scenarios. 

The symbol of this fight, the Swedish schoolgirl Greta Thunberg, quickly became a global phenomenon.  The debate on environment and climate focused mainly on her figure which has lead to a polarization effect in opinion forming about the climate subject. 

In the latest 20 years a great deal of efforts have been devoted to investigate the opinion dynamics in a group of interacting agents. 
The flow of works has been quite steady, with scientific papers having started to appear in significant numbers after 2000, as documented e.g. in the surveys \cite{Albi2017-Review,Lorenz-Review,Cercel:2014:OPO:2611040.2611088,MasVelNa}. 

Beside consensus, also polarization effects should be taken into account. A small subset of these papers focuses on the role of contrarians agents (see Section \ref{sec:lit}). 
The behavior of contrarian agent is defined by its tendency to adopt the opinion opposite to the prevailing opinion of their contacts, whatever this opinion is. An extremist is instead an agent who supports one choice fervently, even when his/her contacts believe a different idea to be a better choice.

In this work, we concern ourselves with all of the aforementioned focuses in order to introduce a new model for opinions' dynamics, starting from the observation of the Greta Thunberg phenomenon. The aim of the paper is to explain this phenomenon by postulating the existence of some mechanisms inside the society. More clearly, we observe the phenomenon induced by Greta Thunberg appearance, and we try to explain it by adopting some hypotheses (i.e. the existence of contrarians in the society) already established in psychology \citep{hovland1953communication} (see e.g. the so-called ``boomerang effect''). Based on these hypotheses, we perform testable predictions that we compare with the results extrapolated from social media data.

Why contrarians agents? Because from the literature, it seems that this kind of agents appears both in climate change debate and in the adults' attitudes to citizen youth movements. (Let's recall that Greta Thunberg is a young environmental activist.) \cite{Boykoff2013} show that, despite 97\% of climate scientists affirm the reality of human-caused climate change, climate contrarians are more influential than their scant numbers and limited expertise would suggest, and exert an outsized media impact. \cite{robin2019greta} detects instead the tendency for the adults' opinions to discredit Greta Thunberg protest.

One of the main novelties introduced here relies on modeling the links between opinions belonging to the same person, about different topics, and study how they may change after interactions with other people, including a group of contrarians agents, i.e. agents that adopt the polar opposite of the opinions of their contacts. To this aim, we model various types of contrarians: those who think the opposite of everyone, which is the most naive definition of a contrarian agent, and those who are against the global majority, which seems to be the proper definition of contrarian existing in literature. We will discuss related papers devoted to the definition of contrarian agent in Section \ref{sec:lit}.
Moreover, we propose a new type of contrarians: those who are opposed only to a group of people (i.e. contrarians with respect to the opinions of a particular group of people). This case refers to the Greta Thunberg phenomenon, where contrarians contest not only Greta's opinion but also her fan club (the group they oppose).

Lastly, we present several numerical experiments which show that the model is capable to describe correctly the insights coming from our study case.

\subsection{Paper organization}
In Section 2 the Greta Thunberg's  phenomenon is investigated and a preliminary sentiment analysis on Twitter's data is presented. 
In Section \ref{sec:model} we introduce the mathematical model and the interaction functions for modelling different kinds of contrarians agents. Finally, in Section  \ref{sec:numerical} we present the numerical tests and we conclude the paper with some comments and an overview on future directions.

\subsection{Related literature on contrarian and extremist agents}
\label{sec:lit}
\cite{GAMBARO2017465}, consider the presence of contrarians agents in discrete three-state kinetic exchange opinion models. The interaction here takes place in a set of two or three agents; in the case of three-agent interaction, contrarian $i$ tries to take the opposite opinion of the pair $(j, k)$.

\cite{JAVARONE201419} considers binary opinions, $\pm 1$, and models the contrarians' behavior in such a way that they assume the opposite opinion of their neighbors. Binary variables are employed also in \cite{Nyczka2012,Nyczka2013}. Here, the authors distinguish between two types of nonconformity: anti-conformity and independence. While independent individuals evaluate situations independently of the group norm, the anti-conformists are the classic contrarians, which adopt the opinion opposite to the prevailing opinion of their contacts (neighbors). 

\cite{DING20101745} propose some games to model binary opinion formation. They adopt a majority evolving rule: a contrarian agent tends to choose the opposite opinion to the dominate one.


Other binary models have been introduced in \cite{Borghesi06,GALAM2004453}. Here a contrarian is assumed to arrive in a group with a fixed opinion. Then the local update of the group takes place by using a \emph{majority rule} to select the new opinion shared by everyone in the group. However, once the contrarian leaves the group it immediately shifts to the opposite opinion.

Another way to model the contrarian behavior could be the introduction of ``negative relationships'' between the agents, as presented in the work by \cite{altafini2012consensus} where some agents are connected with positive links, reinforcing each others' opinions, while other agents are negatively linked, causing their opinions to repel each other.

As for extremists' behavior, \cite{Amelkin2017} considered the attitude extremity as being a major factor deﬁning the strength of conviction, assuming that extreme opinions are more resistant to change than neutral opinions. This may be the case of a democracy composed by two political parties, where agents' states describe the degrees of support for one of the two parties. In this case, whereas extremists are unlikely to change their political affiliation, neutral voters can be successfully attracted toward one or another pole of the opinion spectrum. In this direction, \cite{GomezSerrano2012} modeled the scenario of a company fusion, dividing the workers into an ``undecided'' group and two ``extremist'' factions. A similar situation can be found also in \cite{Torok2013,Vazquez04,BENNAIM200399}. \cite{Mobilia2011} generalized the three-state constrained voter model \cite{Vazquez_2003} by assuming that the interaction between extremists and centrists is characterized by a bias.

\cite{MartinsPhrevE2008} used the word \emph{extremism} in a social sense. Here each agent assigns a probability $p$ to the statement that one of two choices is the best. More exactly, Martins defines as extremist a society whereby both choices survive in the long run and most of the agents that support either choice are extremists. Still Martins, in \cite{MARTINS2008}, defines as extremists those agents have very strong opinions ($p$ very close to 1 or 0).

\cite{Porfiri2007} linked the presence of extremist parties to the persuasibility of the competing agents. Here, the extreme opinions are called minorities and they represent extremist parties (as $\mu$ decreases to 0 they gradually disappear).

For \cite{salzarulo2006}, \cite{WEISBUCH2005555}, \cite{AMBLARD2004725}, \cite{deffuant2002can}, extremists are agents with an opinion located at the extremes of the initial opinion distribution and a low tolerance or uncertainty (i.e the threshold of a bounded confidence model).

\section{Greta Thunberg's phenomenon}
\label{intro}

In recent months the theme of environmentalism has become the focus of the debate, in new and unpublished forms. The figure of Greta Thunberg --- a Swedish schoolgirl who, at age 15, began protesting to combat climate change --- quickly became a global phenomenon. On Friday, March 15, 2019, more than a million students took part in the first Global Climate Strike for Future. After almost a year, Greta had become a social media celebrity with 9,8mln Instagram followers, 2,7mln Facebook followers and 4mln Twitter followers. As a result, the debate on environment and climate focused mainly on her figure which has lead to a polarization effect in opinion forming about the climate subject (see next subsections).

This phenomenon, especially observable on the online social network (like \emph{Facebook} or \emph{Twitter}), provides important food for thought in the field of opinion dynamics.



\subsection{Empirical analysis of Greta Thunberg phenomenon by means Twitter}
\begin{figure}[hbt!]
    \centering
    \includegraphics[scale=0.9]{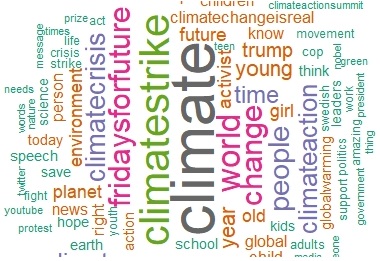}
    \caption{A word cloud of Greta Thunberg phenomenon}
    \label{fig:wcloud}
\end{figure}

The image composed of words depicted in Fig.\ \ref{fig:wcloud} (called ``word cloud'') shows the words used in the world wide web and typically associated with Greta Thunberg. The cloud gives greater prominence to words that appear more frequently in the description of Greta Thunberg phenomenon (the importance of each word is shown with font size). The cloud has been obtained by the process of scraping of Twitter Data. 
Fig. \ref{fig:wcloud} has been obtained from 82,224 tweets in English published between 2018-09-01 and 2019-12-31 and containing the hashtag ``\#GretaThunberg''.

In this paper we will use Twitter data in order to suggest a model in which agents express their opinions according to their connections around the environmentalism theme and the figure of Greta. This kind of data is useful because it has several advantages; one and very obvious advantage is that it is quite easy to analyze due to its boundedness in terms of employed characters (originally and historically, the well-known 140 characters of the Tweets, although this has been made more flexible over time\footnote{See \url{https://developer.twitter.com/en/docs/basics/counting-characters.html}. Accessed: 2019-05-31.}). Another advantage is the high volumes of Twitter data because it is estimated that there are about 500 million tweets per day (that’s 5,787 tweets every second!).

There are also drawbacks when you are using Twitter data. For example, people could use words quite creatively and they make mistakes in their spelling or even create new words so those are factors which make some tweets quite useless. However, we can assume that by scraping a huge number of tweets this is not a problem anymore.


\subsection{Topic and sentiment analysis of Greta Thunberg phenomenon}
\label{sec:sentiment}
In this Section, we perform a preliminary sentiment, topic analysis on Italian tweets concerning Greta Thunberg. The sentiment analysis is used to know if a text, around a certain word, is neutral, positive or negative while the topic modeling is mainly focused on understanding the latent structure in document collections, see \citep{Boyd-Graber:Mimno:Newman-2014}. 



The aim of this Section is to show that the discussions on Twitter regarding Greta are based on various topics and not only on the theme of environmentalism: a user has opinions on different issues, can share them with other people and these opinions are connected and influenced by each other. This insight gives us the idea for the model introduced in Section \ref{sec:model}.

\begin{figure}[hbt!]
    \centering
    \includegraphics[scale=0.42]{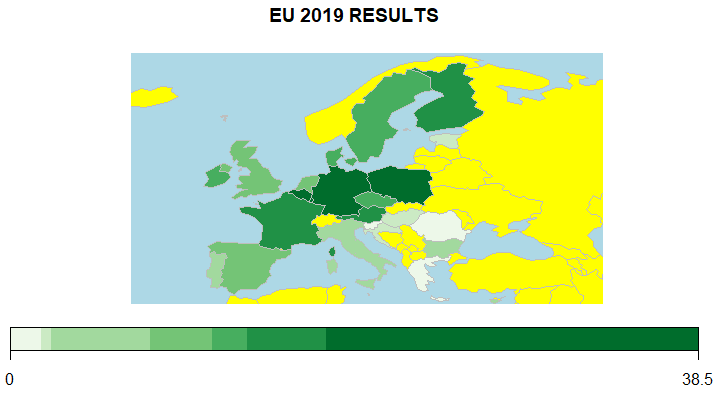}
    \caption{Results of the 2019 European Elections (23 - 26 May) for the Green Party. Source \url{https://europeangreens.eu/2019results}, accessed 2019-06-06. The tone green scale denotes, country by country, the results of Green Party.}
    \label{fig:EUres}
\end{figure}
But why Italy? The answer is depicted in Fig.\ \ref{fig:EUres}. This figure shows the results of the May 2019 European Elections. The election results made clear that the citizens of many European countries voted for change and climate action. Green parties have exceeded expectations in countries such as Germany, France, Ireland, Denmark, Finland and Austria. Anyway, other countries have registered very disappointing outcomes. This is the case of Italy, where the local Green Party achieved only 2.32\% of votes. Since we wish to model the polarization effect induced by Greta Thunberg phenomenon in the presence of contrarian, extremist agents (who are the supporters? Who are the opponents?), the case of Italy offers us an interesting study-case. 

Let us see whether this can be confirmed by the sentiment and topic analysis of Tweets. We collected data from Twitter through our code in Python, focusing on the Tweets containing the hashtags ``\#GretaThunberg'' or ``\#Greta'', or the keyword ``Gretina''\footnote{This is a derogatory term coined by some Italian rightwing newspapers. It merges together the name \emph{Greta} and an Italian word to denote a person of low intelligence.}, including the nickname of users author of the posts. We focused on the Tweets published in Italian between 2018-08-03 and 2019-12-31, for a total amount of 36,529 Tweets posted by 15,110 users. Let us denote this dataset by $\mathcal D_1$. The choice of this temporal range was driven by the fact that, in August 2018, Thunberg became an internationally recognized climate activist after beginning the school climate strikes, with the aim to ask the Swedish government to reduce carbon dioxide emissions as required by the Paris agreement on climate change. A reason to consider 2019 is the presence of European Parliament election in May, an event which has emphasized particularly the environmental issues also in Italy. Finally, we have chosen to stop at 31 December 2019 because at the beginning of 2020 the theme of climate change gave way to concerns due to the covid 19 pandemic.

As to sentiment analysis, it should be noted that there is a lot of research on sentiment analysis and emotion recognition for English language but some methods lack data on different languages, as Italian. For this reason, we employed here the recent approach introduced by \cite{bianchi2021feel}. \cite{bianchi2021feel} created a data set for Italian sentiment and emotion prediction and fine-tuned a BERT model. BERT is one of the most popular neural architectures in Natural Language Processing. In particular, \cite{bianchi2021feel} use UmBERTo, a very efficient Italian BERT model \citep{magnini2006cab,magnini2006annotazione}.

Figs. \ref{fig:emotions} and \ref{fig:sentiments} show, respectively, the results of emotion and sentiment recognition in the tweets of $\mathcal D_1$. As we can see, the result of sentiment analysis according to this approach is a binary, categorical \emph{target} variable --- \emph{positive} (i.e. positive sentiment) vs \emph{negative} (i.e. negative sentiment) for each tweet --- and indeed does not provide a numerical estimation of text polarity sentiment.
\begin{figure}[hbt!]
    \centering
\begin{tikzpicture}
    \pie{64.2/anger, 4/fear, 19.8/joy, 11.9/sadness}
\end{tikzpicture}
    \caption{Pie-chart of emotions distribution}
    \label{fig:emotions}
\end{figure}
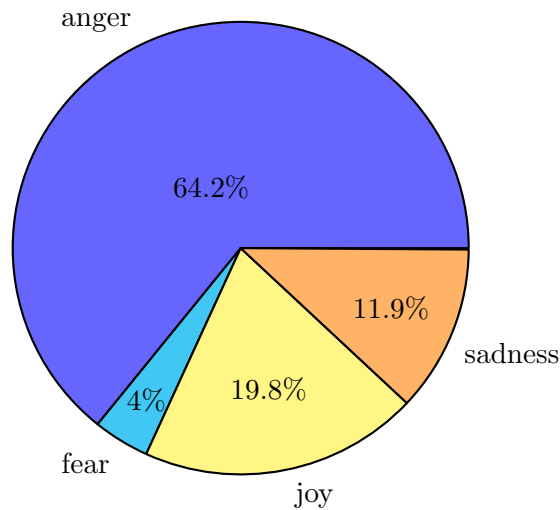
\begin{figure}[hbt!]
    \centering
\begin{tikzpicture}
    \pie{81.2/negative, 18.8/positive}
\end{tikzpicture}
    \caption{Pie-chart of sentiments distribution}
    \label{fig:sentiments}
\end{figure}
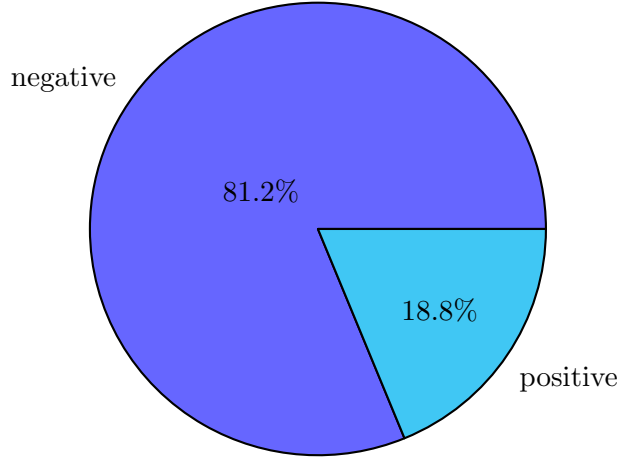

Therefore, we designed a method to calculate text polarity sentiment at the sentence level for the Italian language. For this purpose, we taken inspiration from the approach proposed by \cite{sentimentr} (available instead only for English). Here, unlike the algorithm introduced by \cite{bianchi2021feel}, the result is a number which can be positive (denoting positive sentiment) or negative (denoting negative sentiment). The equation used by the algorithm to assign value to the polarity of each sentence --- see \cite{sentimentr} for further details ---  first utilizes a sentiment dictionary to tag polarized words (carefully designed in Italian by us). Afterward, the algorithm attempts to take into account valence shifters: negators, amplifiers (intensifiers) and de-amplifiers (downtoners). In other words it is an augmented dictionary lookup.

In order to take into account the results obtained through the BERT model, we compared them with the polarity score produced instead by the augmented dictionary lookup. The procedure gone through the following steps:
\begin{itemize}
        \item[a)] Apply the BERT model to dataset $\mathcal D_1$ and obtain a target variable $s_t$ for each tweet $t\in\mathcal D_1$; $s_t\in\{\text{positive},\text{negative}\}$.
        \item[b)] Apply the augmented dictionary lookup to dataset $\mathcal D_1$ and obtain a polarity score $p_t$ for each tweet $t\in\mathcal D_1$; $p_t\in I\subset\mathbb{R}$.
        \item[c)] Compare the results in (a) and (b) and consider those that are consistent with each other. We denote the resulting dataset by $\mathcal D_2=\mathcal{D}_2^{-}\cup \mathcal{D}_2^{+}$, where:
        \begin{align*}
            \mathcal{D}_2^{-}:=\Bigl\{ t\in\mathcal D_1| s_t=\text{negative}\, , \ p_t\leq0\Bigr\} \\
            \mathcal{D}_2^{+}:=\Bigl\{ t\in\mathcal D_1| s_t=\text{positive}\, , \ p_t\geq0\Bigr\} 
        \end{align*}
\end{itemize}
From the 29,683 (resp. 6,845) tweets in $\mathcal D_1$ showing $s_t=\text{negative}$ (resp. $s_t=\text{positive}$), we arrived to $\mathcal{D}_2^{-}$ (resp. $\mathcal{D}_2^{+}$) composed by 20,381 (resp. 4,799) tweets. The resulting set $\mathcal D_2$ is then composed by 25,180 tweets.

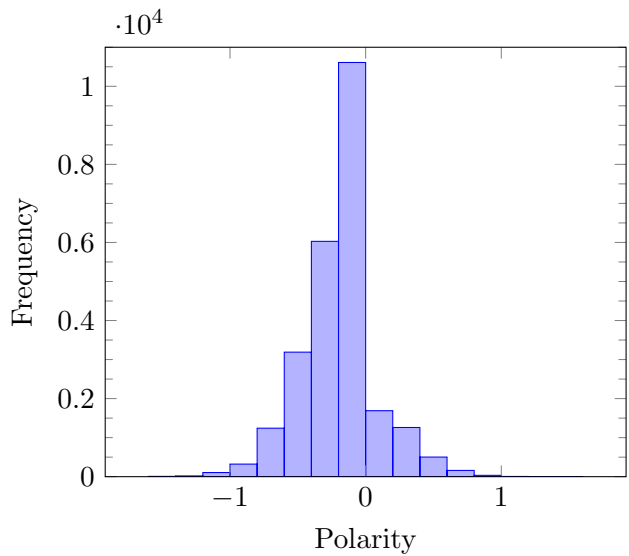
\begin{figure}[hbt!]
    \centering
\begin{tikzpicture}
\begin{axis}[
    ymin=0, ymax=11000,
    minor y tick num = 3,
    area style,xlabel={Polarity},ylabel={Frequency}
    ]
\addplot+[ybar interval,mark=no] plot coordinates { (-1.6, 8) (-1.4,19) (-1.2, 106) (-1.0, 322) (-0.8,1242) (-0.6, 3190) (-0.4,6028) (-0.2,10610) (0,1690) (0.2,1260) (0.4, 502) (0.6,160)  (0.8,34)  (1.0,6)  (1.2,1)  (1.4,1)  (1.6,1)};
\end{axis}
\end{tikzpicture}
    \caption{Histogram containing the results of Greta's sentiment analysis.}
    \label{fig:hist}
\end{figure}

The histogram in Fig. \ref{fig:hist} has been obtained from dataset $\mathcal D_2$. It seems to confirm our previous intuition. The distribution of polarity score is shifted towards negative values. Moreover, we would like to examine the topics emerging from the tweets concerning Greta. Are they centered on environmental issues, or do they also contain opinions on non-environmental issues? From the analyzed tweets, do predetermined political orientations emerge? Are there expressions of offensive contents towards Greta and her supporters? To answer these questions we resorted to the above mentioned topic analysis.

\begin{figure}[hbt!]
    \centering
    \includegraphics[scale=0.4]{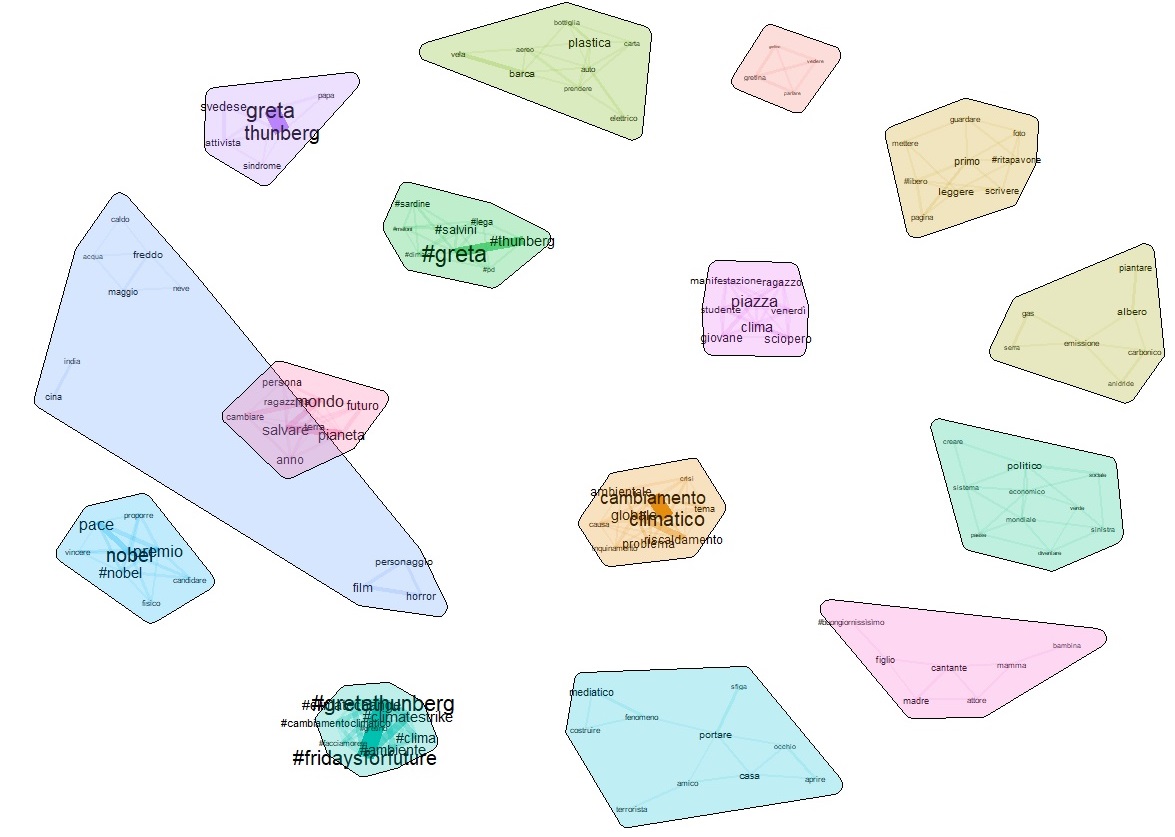}
    \caption{Visualise the topic clusters}
    \label{fig:topics}
\end{figure}
Regarding topic analysis we applied an approach for short texts, like tweets, referred as biterm topic model (BTM). It has been proposed by \cite{Yan2013}, see also \cite{BTMpackage}. Figure \ref{fig:topics} shows the graphical representation of the topics extracted from the dataset $\mathcal D_1$. As we can see: 
\begin{itemize}
    \item Some topics convey a positive sentiment; see e.g. the topic built from words like ``premio'' (prize), ``nobel'', ``pace'' (peace) and so on, which concerns the nomination of Greta Thunberg for Nobel Peace Prize.
    \item Some topics convey a negative sentiment; see e.g. the topic built from words like ``mediatico'' (media), ``fenomeno'' (phenomenon), ``terrorista'' (terrorist), ``sfiga'' (bad luck). According to this topic and the tweets that are part of it, Greta would be just a manufactured media phenomenon (to what end, it is not clear), she would also be bad luck or a terrorist.
    \item One topic in particular (``\#salvini'', ``\#sardine'', ``\#lega'', ``\#pd'', ``\#dimaio'', ``\#meloni'') refers to political party, political movements and their leaders.
\end{itemize}
From the last point of the previous bulleted list, it is clear that the discussions on Twitter regarding Greta are based on various topics and not only on the theme of environmentalism. These topics are related to each other and give us the idea for the model that we will formalize in the next section: an individual usually expresses opinions on different issues and these opinions are connected and influenced by each other.

The provided analysis suggests that there are two groups of people with different/opposite views on the topic of Greta Thunberg, and these groups may be distinguished by whether their members are overly critical of other topics as well (``haters''). This, however, does not mean that Greta Thunberg was the cause of polarization: in our paper she provides a case study  in which it is easy to visualize the polarization, a symptom, but she is not the cause of the polarization itself. More specifically, users of social media could prefer to keep their original opinions, as previously evidenced in \cite{FeiTwitter}, in the sense that, e.g., users that show a hateful or abusive language (\emph{haters}), usually oppose the Greta Thunberg opinions, while people that do not use this kind of language (\emph{non-haters} or \emph{tolerant} users) generally support them. There doesn't appear to be a mixing effect: no haters seem to hear environmental issues, no tolerant users seem to go against these claims. This preliminary analysis may also show the presence of an \emph{echo chamber effect}, where online interactions are conducted in a polarized pattern \cite{prasetya2020model}.

\section{Agent-Based Model}
\label{sec:model}
This section is devoted to the mathematical description of a new model for opinions' dynamics. One of the main novelties introduced here relies on modeling the links  between opinions belonging to the same person, about different topics and study how they may change after interactions with other people, including a group of contrarians. Indeed as it happens in real life, a person expresses opinions on different issues and these opinions are connected and influenced by each other, as introduced also in \cite{LANCHIER20123701,Friedkin321}.

Let us consider a population $\mathcal{N}$ of $\operatorname{card}\left(\mathcal{N}\right)=N\in\mathbb N$ persons and a subset $\C \subset \mathcal{N}$ of contrarians.

From a mathematical point of view, we introduce a vector that represents the thinking of a person: $\textbf{x}_i=(x_{1,i},x_{2,i},\dots,x_{m,i})$, for every $i \in \mathcal{N}$. (For simplicity we will drop the index $i$ when the meaning is clear enough.) We refer to this vector as \textit{mind vector} and it is clear that each person has his own mind vector.

The components of $\textbf{x}$ are the opinions concerning different topics. For example, considering the preliminary analysis performed in Section \ref{intro}, where we have highlighted a possible spillover effect between political bias and opinion on climate change issues, the first component of mind vector, $x_{1,i}$, could be the political bias while the second, $x_{2,i}$, could represent the opinion of agent on climate change issues. So, in order to model this spillover effect, we assign weights $\alpha_{q,k} \in [0,1]$ for $q,k=1,\dots,m$, that quantify how the opinion $x_{q,i}$ influences the opinion $x_{k,i}$, for $i \in \mathcal{N}$. Moreover we assume that $\sum_{k=1}^m \alpha_{q,k}=1$, for every $q=1\dots m$. We refer to these coefficients as \emph{coherence weights}.
Note that the coherence weights are non-negative. This choice is motivated by the meaning of coherence itself: if the sign of opinion $q$ changes, then also opinion $k$ is influenced accordingly.

The components are functions $x_j=x_j(t):[0,T] \to [-1,1]$, where $T>0$ is some fixed final time. At each time $x_j$ belong to $[-1,1]$, that means they can vary with continuity from $-1$ to $1$.

We assume that only the main opinion, i.e. the opinion about the first topic, evolves depending on the interaction with  other agents. This does not mean that the other components are useless, but only that they do not play any role in the first opinion dynamics. Indeed one of the aim of this work is pointing out the indirect influence of the contrarians agents on the opinions concerning other topics (i.e. $x_{k,i}$ for $k>1$). For this reason, in the following, we will assume also  that only the main opinion can influence the others, i.e. only the weights $\alpha_{1,k}$ are considered.

Taking inspiration from \cite{cri2018}, we assume that the evolution in time of $x_{1,i}$, i.e. $\dot x_{1,i}$, is described by the following equation:
\begin{equation} \label{eq:evol_op1}
    \dot x_{1,i}=b_i \left(\frac{1}{card(S)} \sum_{j\in S} I(x_{1,j},x_{1,i})+\sqrt{2\mu}B_i\right) 
\end{equation}
where $b_i$ can be thought of as the inverse of \emph{agent's conviction}, which measures the \textit{strength of confidence} on the opinion $x_1$ for the person $i\in \mathcal{N}$. In other words, $b_i$ is a term expressing the propensity to change opinion $x_{1,i}$: if $b_i=0$, agent $i$ is not willing to change his/her opinion; otherwise, if $b_i=1$, $i$ will show the maximum willingness in exchanging opinions. We assume the underlying idea being that the more convinced the agent (on its initial opinion) the lower the propensity to change it. 
In this case, the seminal model introduced by \cite{friedkin1990social} and ours coincides. 

Moreover $\mathcal S \subset \mathcal{N}$ is the subset of people which the person $i$ interacts with, indeed $j\in \mathcal S$, and $I(x_{1,j},x_{1,i}):[-1,1]^2\to \R$ is the function that models the interaction between the two main opinions. Furthermore $B_i$ is the standard Brownian motion, needed to model the uncertainty and the \textit{self thinking} process.
\begin{remark}
Eq. (\ref{eq:evol_op1}) does not guarantee that $x_{1,i}\in[-1,1]$ $\forall t\in [0,T]$, due to the presence of Brownian motion. In order to overcome this issue, we cut the value of the opinion if it exceeds the boundaries. In other words if $x_{1,i}>1$ we assume $x_{1,i}=1$.
\end{remark}

Of course the change of the main opinion also affects the other components of the mind vector and this influence is measured by the coefficient $\alpha_{1,k}$. Indeed the other components $q=2\dots m$ of the mind vector evolve as the following equation suggests:
\begin{equation}\label{other_op}
    \dot x_{q,i}=\alpha_{1,q}\ x_{1,i}
\end{equation}
for every $i\in \mathcal N$. Our aim is to investigate the time evolution of the opinions' dynamics focusing on the effect due to the presence of a group of contrarians in the population. One could observe that starting from a positive mind vector, after a while some opinions would change their sign, without causing the total switching of signs. In other words it may happen that only some components of the mind vector evolve in time towards the opposite opinion (e.g. from $1$ to $-1$ or vice versa) without affecting the whole mind vector.

Moreover there are different types of contrarians: there are people who think the opposite of everyone, but also who is opposed only to a group of people (i.e. contrarians with respect to the opinions of a particular group of people). Furthermore there exist also people who are against the majority, which can also vary in time: this is the case of nonconformist agents, opposed to the principles and behaviors that predominate within a society.
By changing the interaction function $I(x_{1,j},x_{1,i})$ in Eq.\ \eqref{eq:evol_op1}, we are able to model these different behaviors.

Fix $j,i\in\mathcal N$, in particular $j\in S\subset \mathcal N$. As far as concern contrarians against everyone, we choose $I(x_{1,j},x_{1,i})$ as follows:
\begin{equation}\label{cont_vs_all}
I(x_{1,j},x_{1,i})=\begin{cases}
     -x_{1,j} & \qquad  i\in \C \\
    \frac{x_{1,j}-x_{1,i}}{2} & \qquad i \notin \C.
\end{cases}
\end{equation}

Consider now the case of contrarians opposed to the opinion of a group of people; this case may refer, for example, to the Greta Thunberg phenomenon. As evidenced in Section \ref{intro}, many Twitter users prefer to keep their original opinions: users showing hateful or abusive language usually oppose the Greta Thunberg opinions, while people that do not use this kind of language generally support them. This is also the reason, for example, why, fixed the evolution in time of $x_{1,i}$, the other components of mind vector evolve according to it: in other words, if $x_{1,i}$ represents the political bias and $x_{2,i}$ represents the opinion on environmental policies, with Eq.\ (\ref{other_op}) we postulate that the first of them influence the latter. Moreover, contrarians might contest not only Greta's opinion but also her fan club (this is the reason why we consider the case of contrarians opposed to a group). In fact, as discussed in \cite{HAMILTON2019180}, there seems to be a ``socially constructed silence'' around climate change issues which can contribute to stigmatising people and organisations which attempt to achieve a reduction in energy demand. We called these agents ``fans'' because they believe strongly in their work and share the visions of the climate leaders. So they cannot be influenced by other ones, i.e. $b_i=0$ for every $i$-th fan in Eq.\ \eqref{eq:evol_op1}. The interaction function for this case becomes:
\begin{equation}\label{cont_vs_lead}
I(x_{1,j},x_{1,i})=\begin{cases}
    x_{1,i} & \qquad  i\in \mathcal{P} \\
    -r & \qquad  i\in \C \\
    \frac{x_{1,j}-x_{1,i}}{2} & \qquad i \notin \C \cup \mathcal{P},
\end{cases}
\end{equation}
where $\mathcal P$ is the fan club (the set of fans) and $r$ is a random value in their opinions range.

On the other hand, contrarians could also be against majority, a characteristic attitude of those who do not conform to conventional fashions and lifestyles. In this case we have to be careful since the majority evolves and changes in time. So, at each time step, the most represented opinion, called $M$, has to be computed. In this framework, in order to point out the most represented opinion, the opinions are clustered in sub-intervals as the bars of the histogram.
Thus, the interaction function assumes the following form:
\begin{equation}\label{cont_vs_maj}
I(x_{1,j},x_{1,i})=\begin{cases}
    -M & \qquad  i\in \C \\
    \frac{x_{1,j}-x_{1,i}}{2} & \qquad i \notin \C.
\end{cases}
\end{equation}
In order to be more comprehensive as possible, we will analyse in Section \ref{sec:numerical} these three main contrarians' behaviour with the support of the numerical simulations.

\begin{remark}
Our model fits the literature on multiple opinion dimensions.
\cite{friedkin2011social,friedkin2015problem} introduced a multidimensional extension of Friedkin and Johnsen model, describing the evolution of the agents' opinions on several topics. However, there is no link between different topics, which means that they do not influence each other. In our model instead the opinion on the first topic influences all the others.
On the other hand, \cite{parsegov2016novel} deals with the interdependent issues/topics but all the agents are equal. Our novelty is to introduce different kind of contrarians agents in such a framework.
Also \cite{nedic2012multi} consider homogeneous agents. Furthermore they have the concept of neighbourhood for the interactions which is not suitable in our setting.
\end{remark}

\subsection{Model with feedback effect}
So far, we assumed the opinions on other topics did not affect the evolution of the first component of the mind vector. Now we ask: what is the influence that other topics have on the evolution of the main one? For example, let us assume that the first component of the mind vector is related to the chosen political party. Our aim is to investigate how, interacting with other people, the opinion on, e.g. economics, education, environmental policies and other issues could influence the choice of the party.

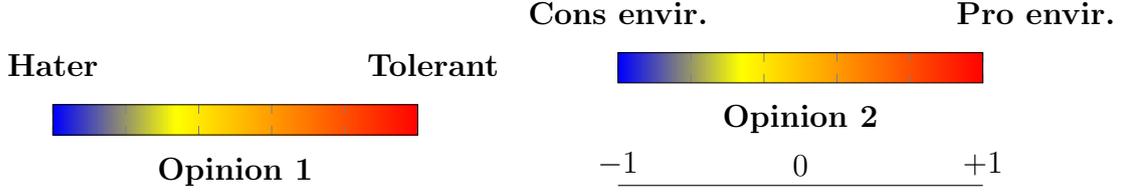
\begin{figure}
    \centering
\begin{tikzpicture}[scale=0.8, font=\large\bfseries]
\begin{scope}
\clip(-0.1,-0.81) rectangle (6.1,0.01);
\pgfplotscolorbardrawstandalone[
    colormap={hot}{
        samples of colormap=(4)
    },
    colorbar horizontal,
    colormap access=map,
]
\end{scope}
\path (5,0) node[above right] {Tolerant} (0,0) node[above] {Hater}
(3,-1) node[below] {Opinion 1};
\end{tikzpicture}
\begin{tikzpicture}[scale=0.8, font=\large\bfseries]
\begin{scope}
\clip(-0.1,-0.81) rectangle (6.1,0.01);
\pgfplotscolorbardrawstandalone[
    colormap={hot}{
        samples of colormap=(4)
    },
    colorbar horizontal,
    colormap access=map,
]
\end{scope}
\path (5.4,0) node[above right] {Pro envir.} (0,0) node[above] {Cons envir.}
(3,-1) node[below] {Opinion 2};
\draw[] (0,-2.5) -- (6,-2.5) node[anchor=west]{};
\draw (0,-2.5) node[anchor=south] {$-1$};
\draw (3,-2.5) node[anchor=south] {$0$};
\draw (6,-2.5) node[anchor=south] {$+1$};
\end{tikzpicture}
    \caption{Very basic political and environmental spectrum. Top: the political spectrum identifies the political bias; on each of the ends of the spectrum is the radical opinion. Bottom: attitude to climate issues.}
    \label{fig:spectrum}
\end{figure}

Hence, in order to model this influence, we modify Eq. \eqref{eq:evol_op1} by assuming the evolution of $x_{1,i}$  as follows:
\begin{align}
\label{eq:feedback}
    \dot{x}_{1,i}&=b_i\Biggl( \frac{1}{card(S)} \sum_{j\in S} I(x_{1,j},x_{1,i}) +\notag\\ &+\sum_{q=2}^{m}\alpha_{q,1}\operatorname{sgn}(x_{q,j})\frac{|x_{q,i}-\operatorname{sgn}(x_{q,i}x_{q,j})x_{q,j}|}{2}
    +\sqrt{2\mu}B_i\Biggr)
\end{align}
where $\alpha_{q,1}$ are the coefficients which take into account the influence of opinion q on the topic 1. In order to explain the meaning of this influence, let us consider e.g. the term 
\begin{equation}
\label{eq:example}
\frac{|x_{2,i}-\operatorname{sgn}(x_{2,i}x_{2,j})x_{2,j}|}{2}\, .
\end{equation}
Eq.\ (\ref{eq:example}) tells us that the evolution of topic 1 for agent $i$ is affected by the difference in the closeness to extreme positions of agents $i$ and $j$. In fact, if opinions $x_{2,i}$ and $x_{2,j}$ have the same sign (i.e. if $i$ and $j$ share the same willingness to be pro or cons environmental issues, see Fig.\ \ref{fig:spectrum}), then Eq.\ (\ref{eq:example}), with $q=2$, is half the absolute value of the difference between them. The same happens when $x_{2,i}$ and $x_{2,j}$ have the opposite sign in Eq. (\ref{eq:example}), because if $x_{2,i}<0$ and $x_{2,j}>0$ then Eq. (\ref{eq:example}) becomes $\frac{|x_{2,i}+x_{2,j}|}{2}$, i.e. $\frac{|-|x_{2,i}|+|x_{2,j}||}{2}$.

\section{Numerical tests}
\label{sec:numerical}
Let us now focus on the discretization of Eq.\ \eqref{eq:evol_op1} and Eq.\ \eqref{other_op}. As far as it concerns the time derivative we will employ the explicit Euler scheme:
\begin{align}
\label{eq:op1discr}
    &x_{1,i}^{n+1}=x_{1,i}^n+ \notag \\
    &\Delta t\ b_i^n\left(\frac{1}{card(S)} \sum_{j\in S} I(x_{1,j}^n,x_{1,i}^n)+\sqrt{2\mu}B_i^n\right) 
\end{align}
\begin{equation}
\label{eq:op2discr}
    x_{q,i}^{n+1}=x_{q,i}^n+\Delta t\ \alpha_{1,q}\ x_{1,i}^n
\end{equation}
where $b_i^n$ is the discrete version of $b(x_{1,i})$. For the approximation of the Brownian motion (or Wiener process) we refer to \cite[Chapter 10]{kloeden2013numerical}. 

In order to overcome the fact that choosing randomly the interacting agents could affect the results, we will consider an average of the results, following the Monte-Carlo approach. 

In the following numerical tests we are interested in pointing out the opinions' evolution when in a population $\mathcal N$ there is a group of contrarians. Moreover we will analyse the different contrarians behaviours presented above, changing the interaction function as described in Eq.\ \eqref{cont_vs_all},\eqref{cont_vs_lead},\eqref{cont_vs_maj}.

In the following, let us fix $N=100$ and assume that the mind vector has three components, i.e. three opinions are considered. 

Without loss of generality, we can assume that the interaction is pairwise because at each step there is just one pair of agents interacting with each other. Then the updating is asynchronous, because not all the agents change their opinion at each time step \cite{MasVelNa}.
For simplicity, we can assume $\Delta t=1$, considering a proper scaling of time.

The interacting agents are chosen randomly within the population, without preferences between contrarians or conformists.

Moreover we assume $b^n_i=1$ and the strength of noise is $0.01$ in the following tests, unless it is specified.

\subsection{Contrarians vs all}\label{subsec:vsall}
In this section we investigate the effect on the opinions' dynamics of a group of contrarians who thinks the opposite to everyone. To do that we assume $I(x_{1,j},x_{1,i})$ as in Eq.\ \eqref{cont_vs_all}.

 First let us assume that, at the initial time, the opinions of half population are equals to $-0.5$ and the others are $0.5$. In particular, for $i\in\mathcal N$:
\begin{equation}
   x_{1,i}=\begin{cases}
   - 0.5 \qquad & i\le50\\
   + 0.5  \qquad & i>50
\end{cases}
\end{equation}
and $x_{2,i}=x_{3,i}=x_{1,i}$.
 
\begin{figure*}[!htb]
\centering
\includegraphics[width=0.32\textwidth]{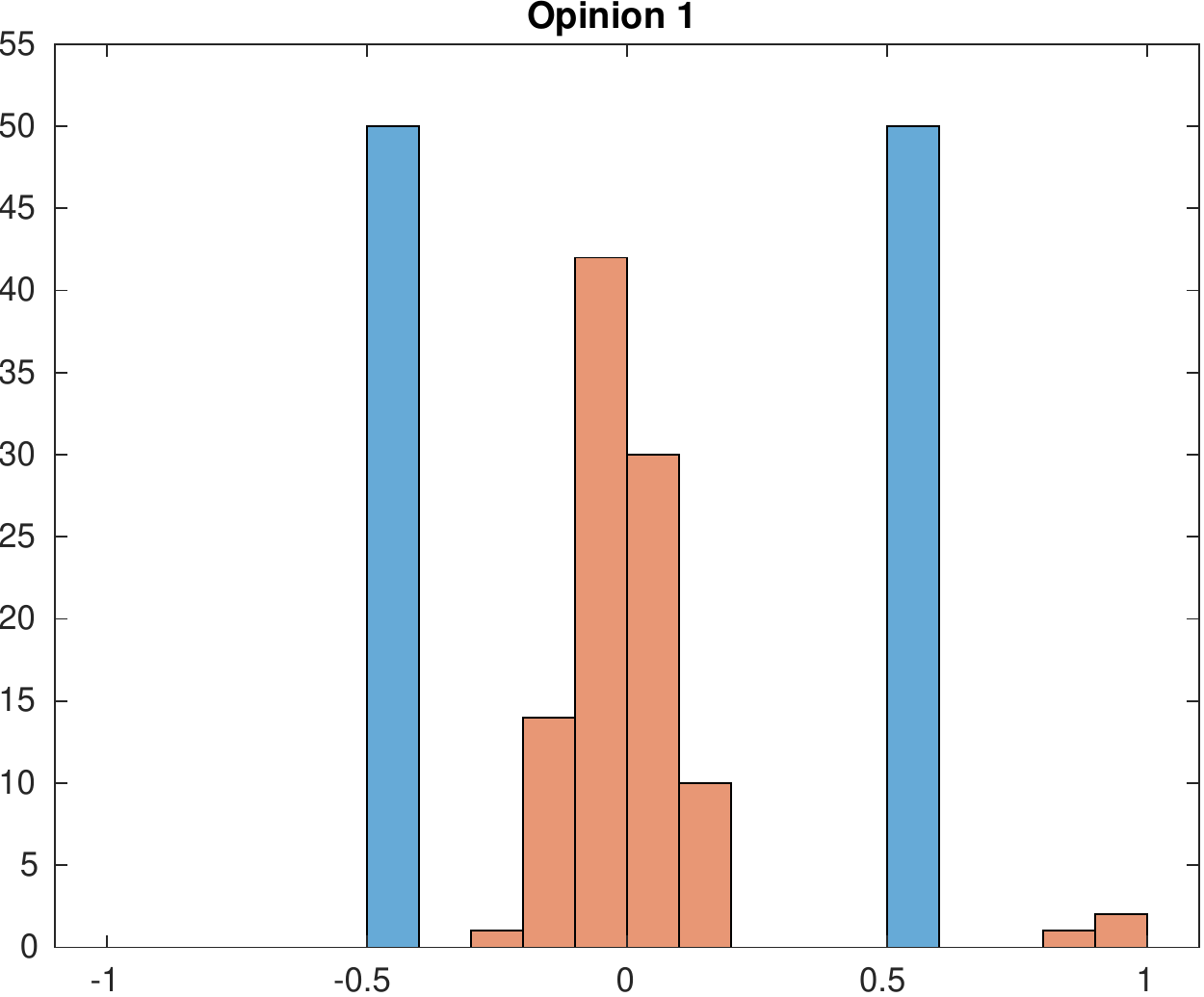}
\includegraphics[width=0.32\textwidth]{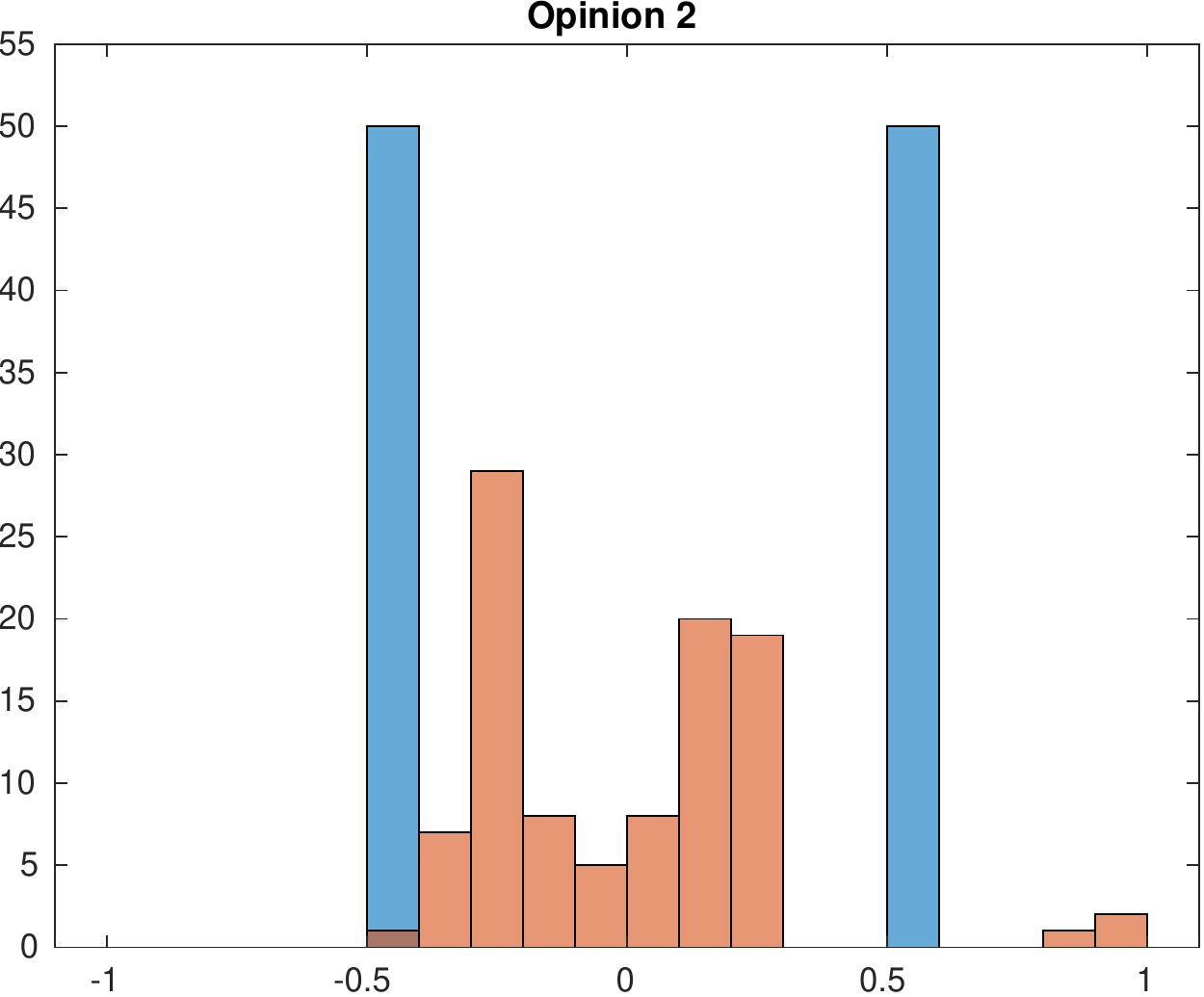}
\includegraphics[width=0.32\textwidth]{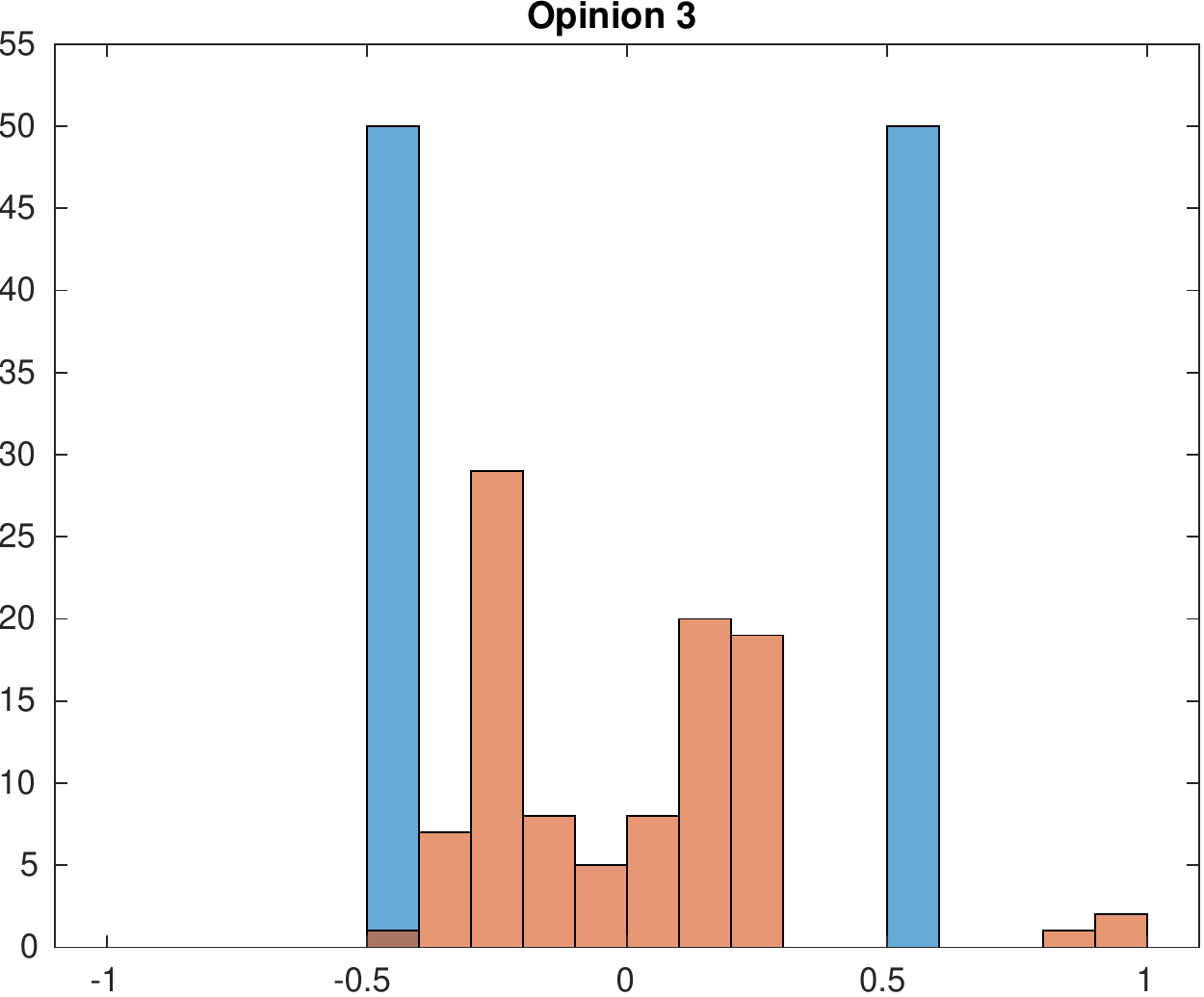}
\caption{Test 1 --- Contrarians vs all. Contrarians percentage $2\%$. The blue bars represent the initial data, while the orange ones the solution at time $T$.}
\label{fig:first_t1_c2}
\end{figure*}

\begin{figure*}[!htb]
\centering
\includegraphics[width=0.32\textwidth]{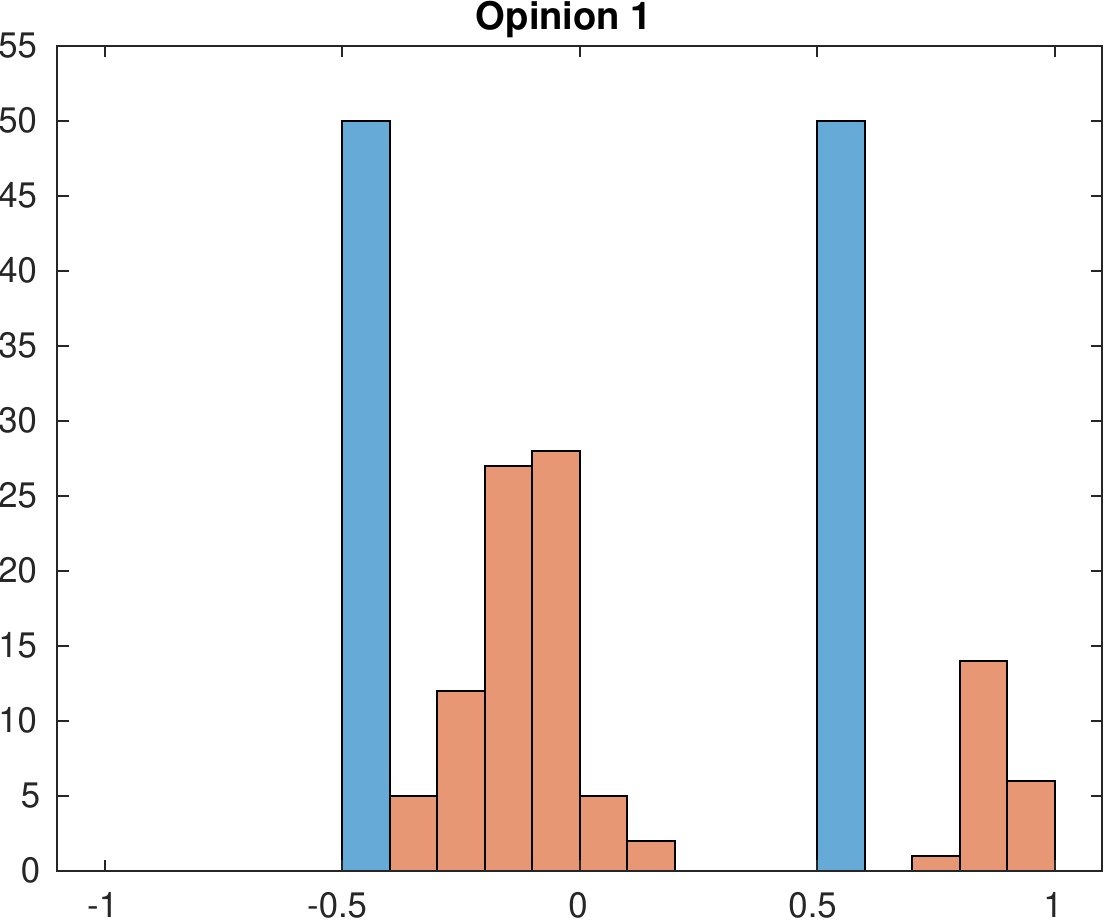}
\includegraphics[width=0.32\textwidth]{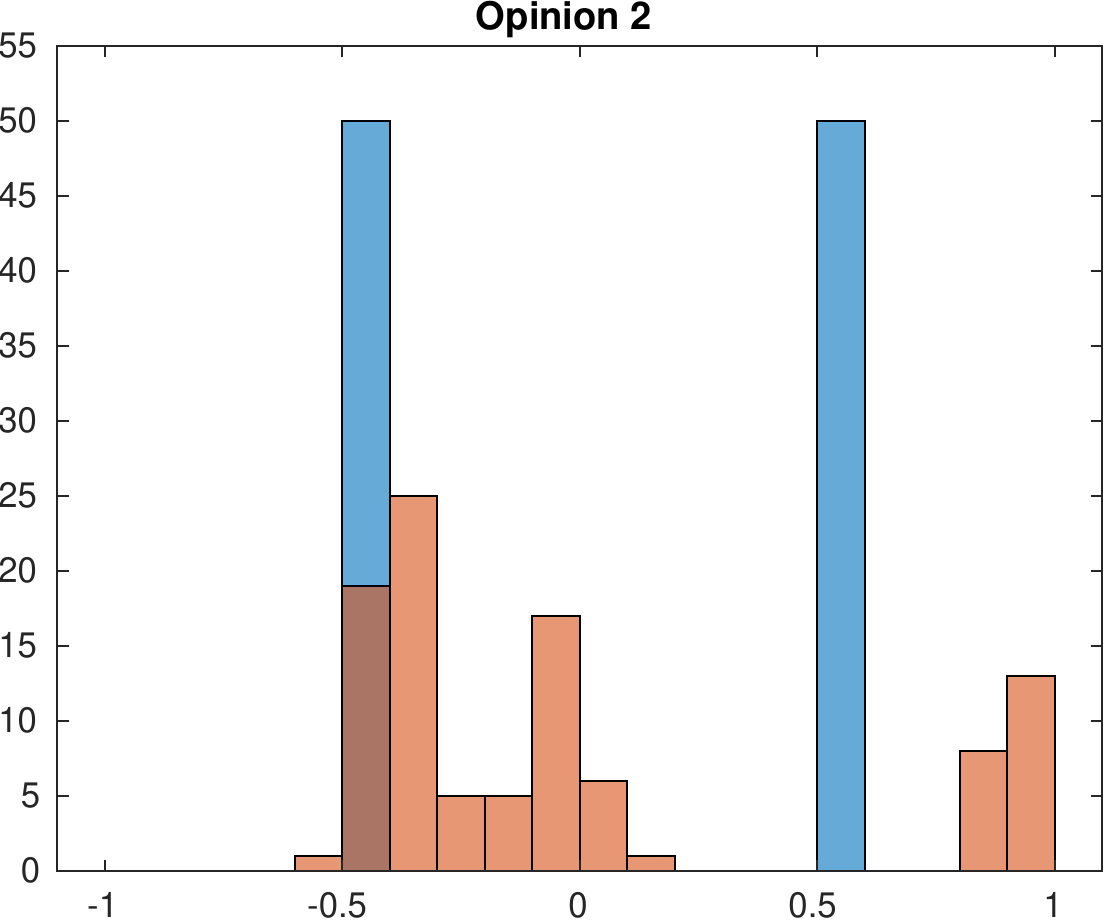}
\includegraphics[width=0.32\textwidth]{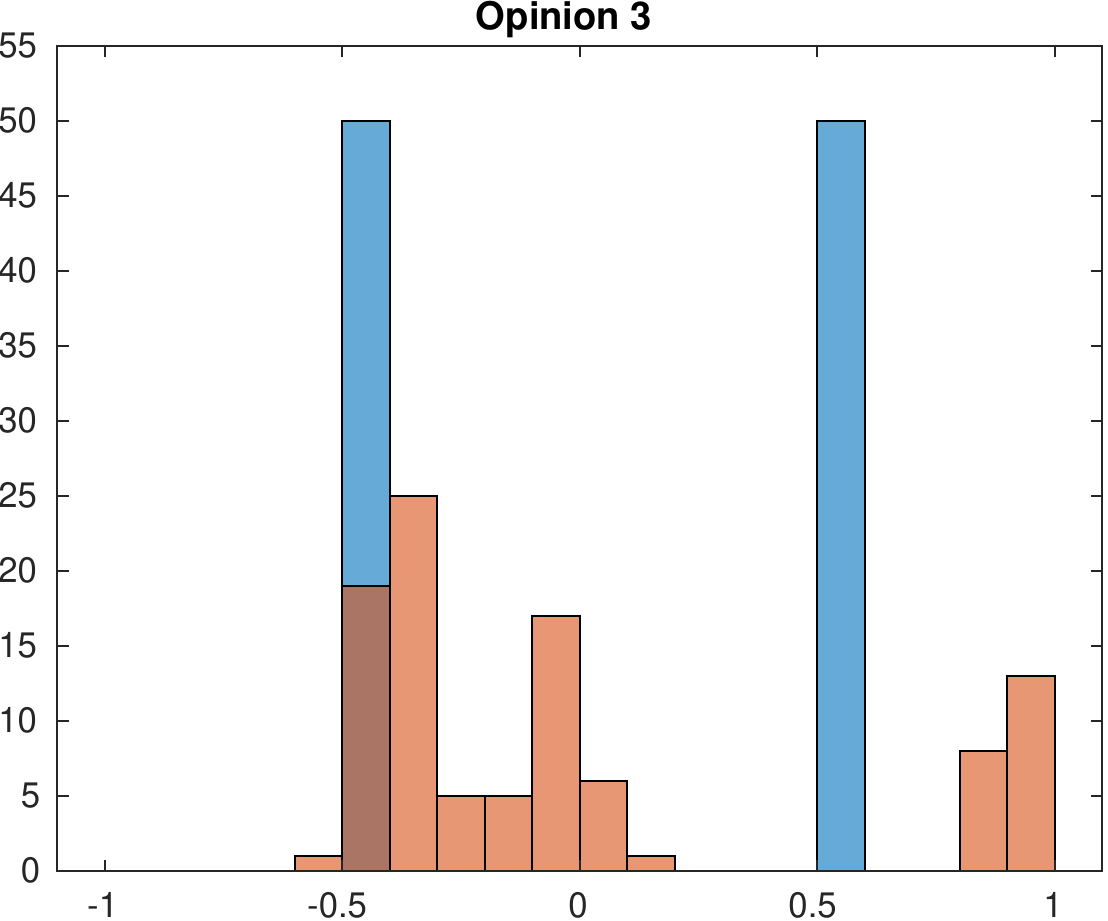}
\caption{Test 1 --- Contrarians vs all. Contrarians percentage $20\%$. The blue bars represent the initial data, while the orange ones the solution at time $T$.}
\label{fig:first_t1_c20}
\end{figure*}

At the beginning, let us also assume that the coefficients $\alpha_{2,1}=\alpha_{3,1}=0.5$. In this way, we expect that the are no differences between the opinion 2 and the opinion 3 since they have the same initial data and coherence weights. This indeed is what we obtain in Fig. \ref{fig:first_t1_c2}-\ref{fig:first_t1_c20} (central-right).

At each time step opinion 1 is updated according to Eq.\ \eqref{eq:op1discr} while opinions 2 and 3 follow Eq.\ \eqref{eq:op2discr}.

Comparing the evolution of the opinions for different percentages of contrarians, we observe a disruptive effect on reaching the consensus as the contrarians population increases.
In other words, focusing on the opinion 1, if the percentage of contrarians is small, i.e.\ $2\%$, the opinion range shows a small fluctuation, see Fig.\ \ref{fig:first_t1_c2}. On the other hand, if $\C$ is $20\%$ of the total population (see Fig.\ \ref{fig:first_t1_c20}), the opinion range is higher. Anyway, we have a significant group of people on the right side of the histogram. This group is caused by the large percentage of contrarians (barely visible also in Fig.\ \ref{fig:first_t1_c2}), whose members are pushed toward extreme positions. 


Moreover, the effects of different contrarians percentages are also evident looking at opinion 2 and opinion 3 in Fig.\ \ref{fig:first_t1_c2}-\ref{fig:first_t1_c20}. In the former, the histogram peaks are higher, while in the latter there are a large variety of opinions values.

Consider now different values for the weights $\alpha_{1,2}$ and $\alpha_{1,3}$ in order to point out the effect of coherence in opinion dynamics problems. 

\begin{figure*}[!htb]
\centering
\includegraphics[width=0.32\textwidth]{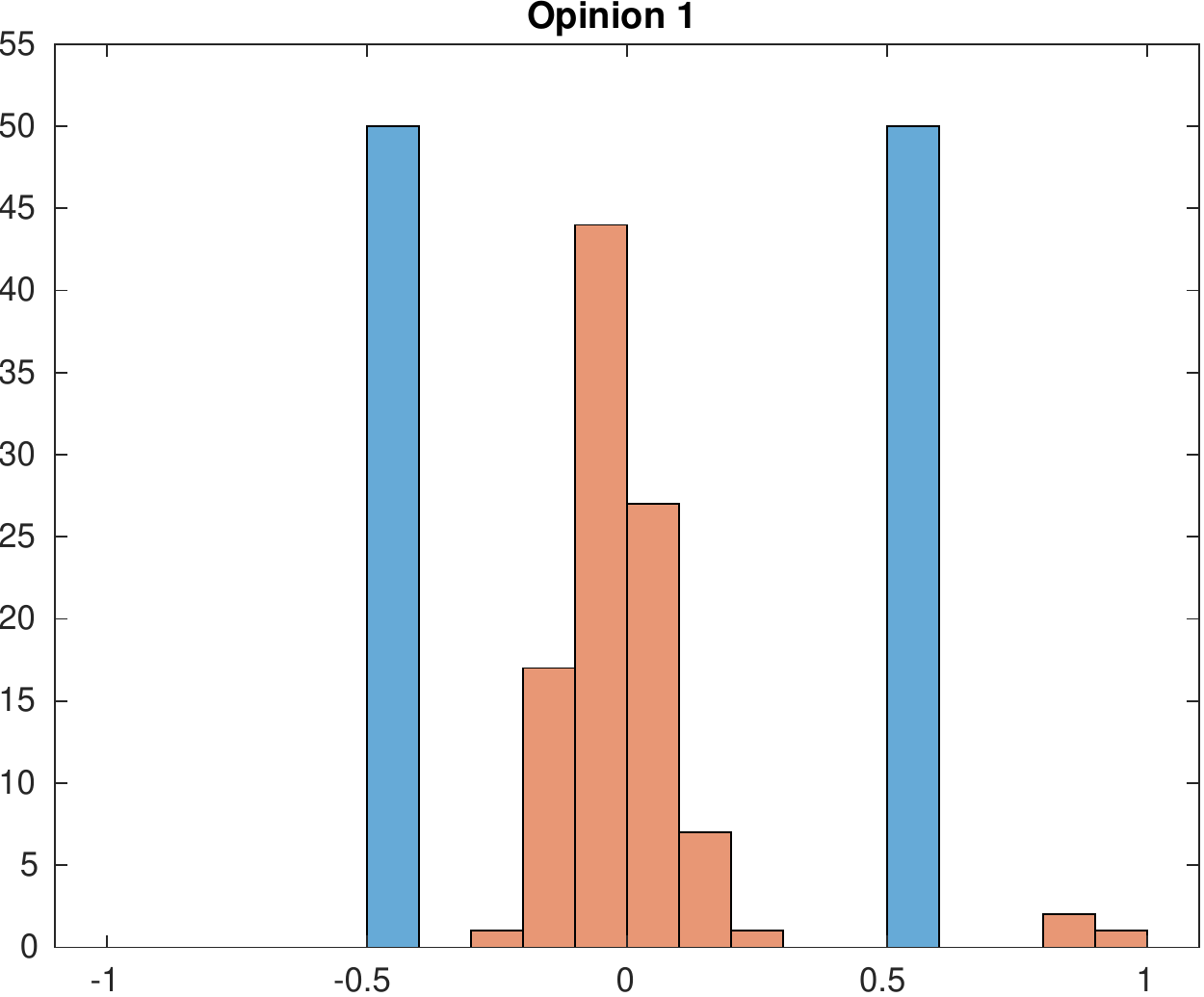}
\includegraphics[width=0.32\textwidth]{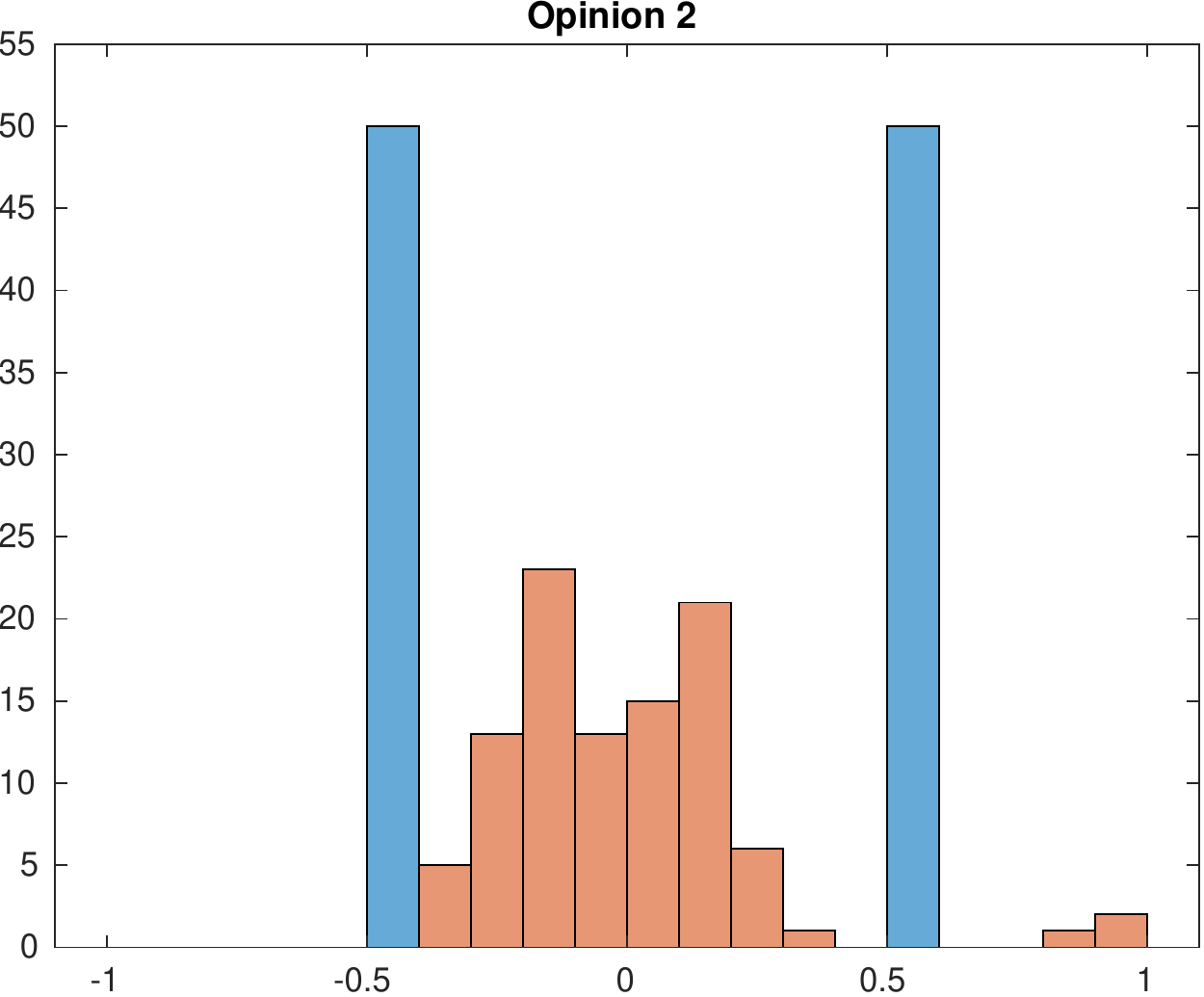}
\includegraphics[width=0.32\textwidth]{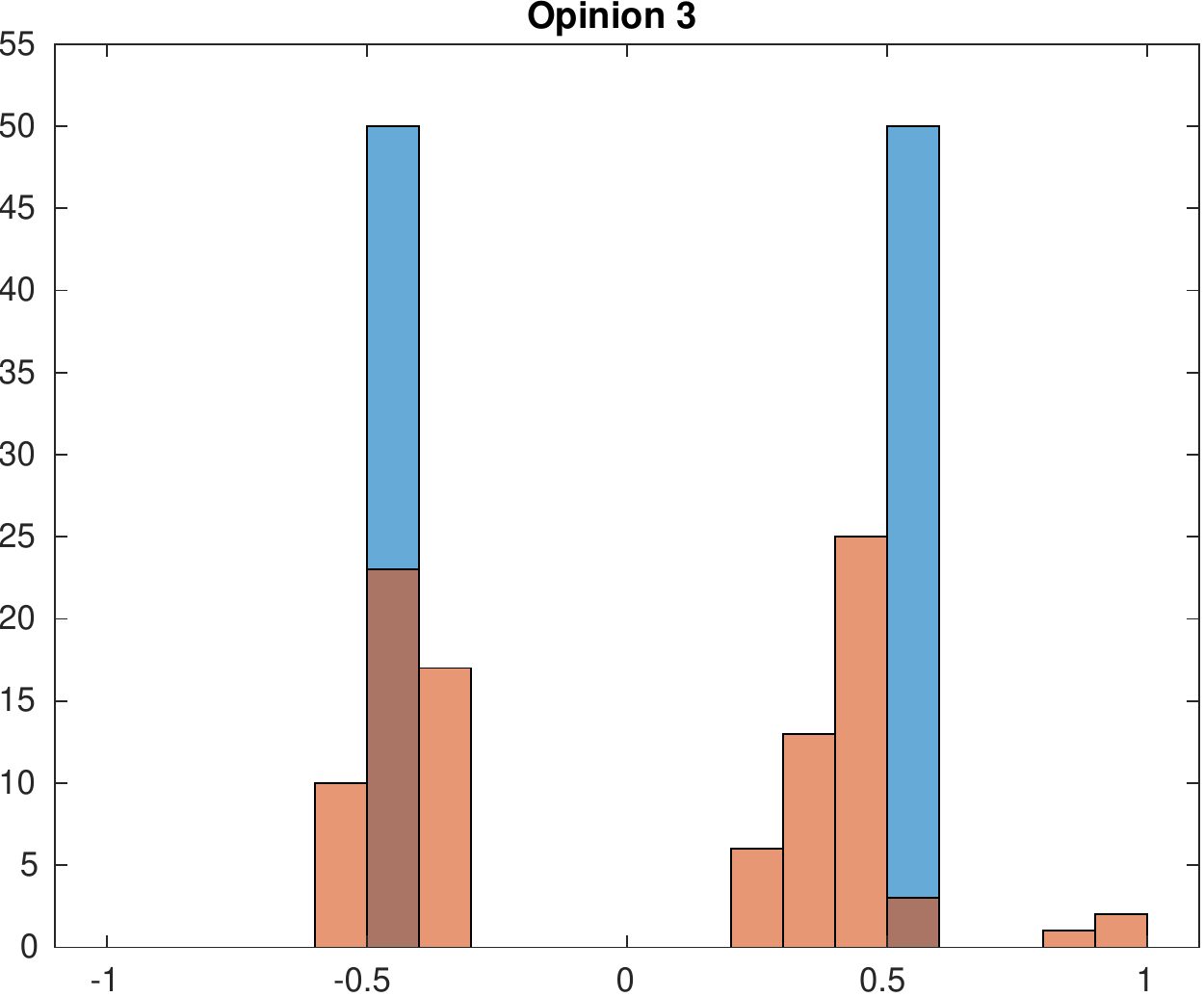}
\caption{Test 1 --- Contrarians vs all. Contrarians percentage $2\%$. $\alpha_{1,2}=0.75$ and $\alpha_{1,3}=0.25$. The blue bars represent the initial data, while the orange ones the solution at time $T$.}
\label{fig:first_t1_c2_alfadiversi}
\end{figure*}

\begin{figure*}[!htb]
\centering
\includegraphics[width=0.32\textwidth]{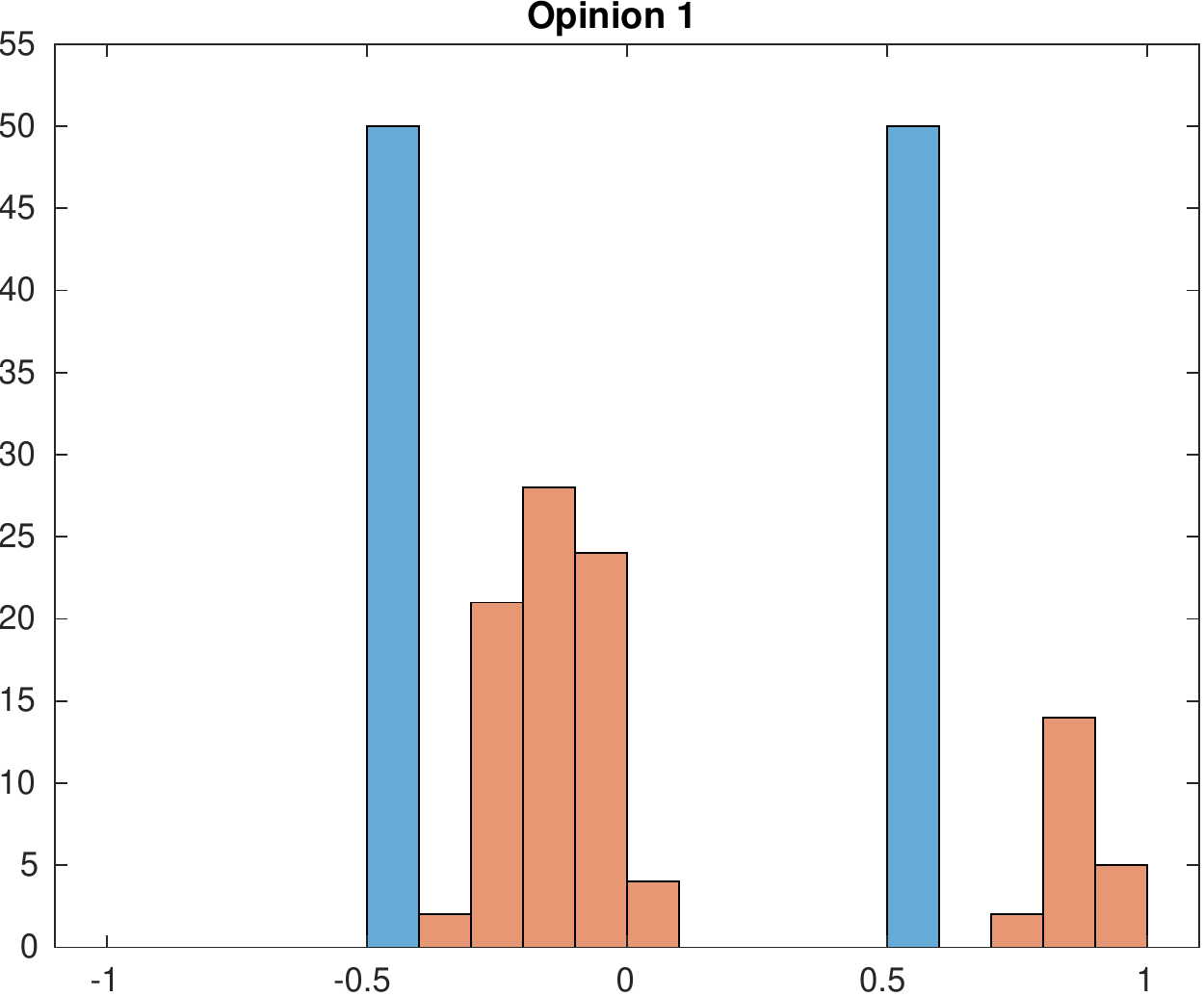}
\includegraphics[width=0.32\textwidth]{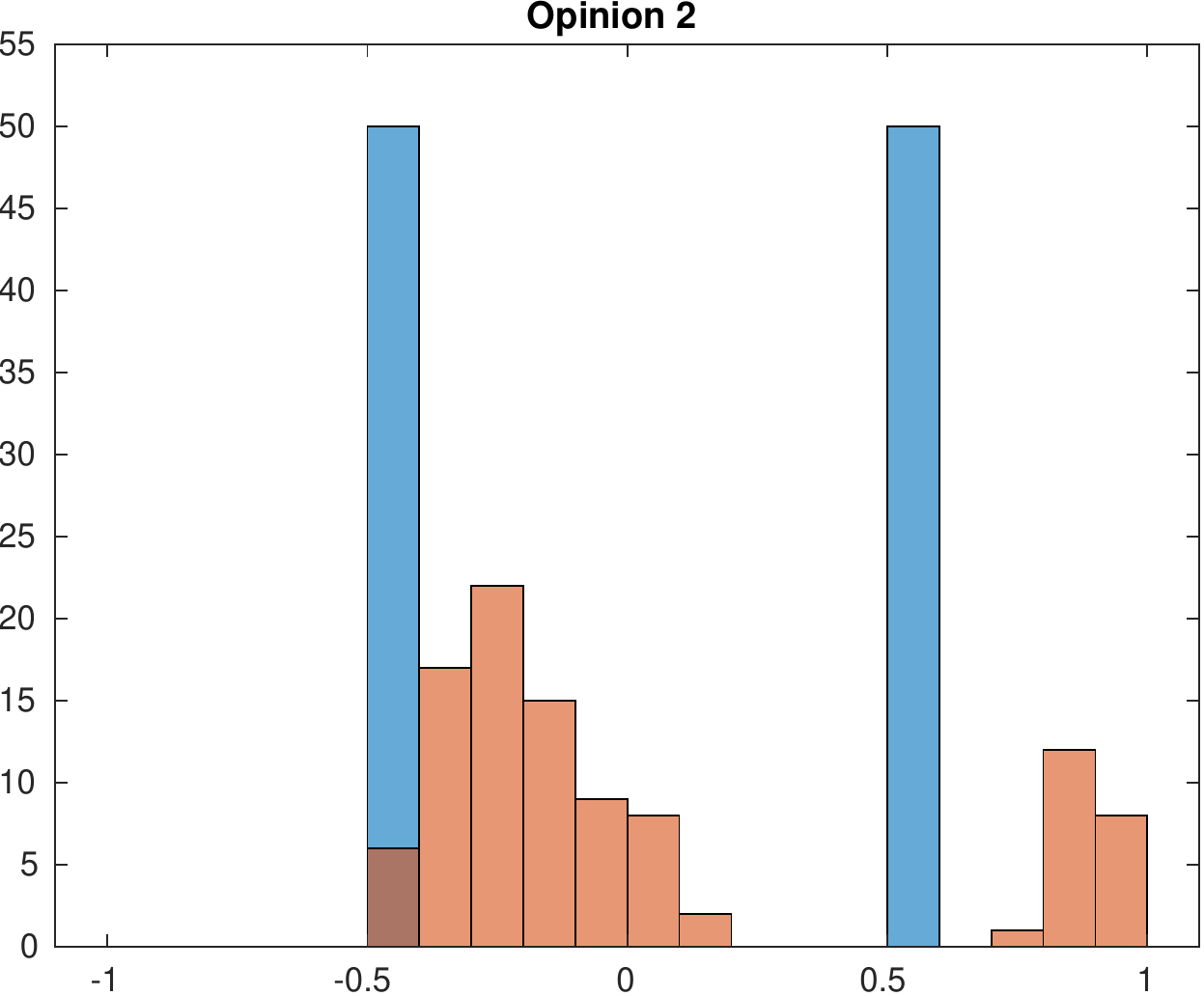}
\includegraphics[width=0.32\textwidth]{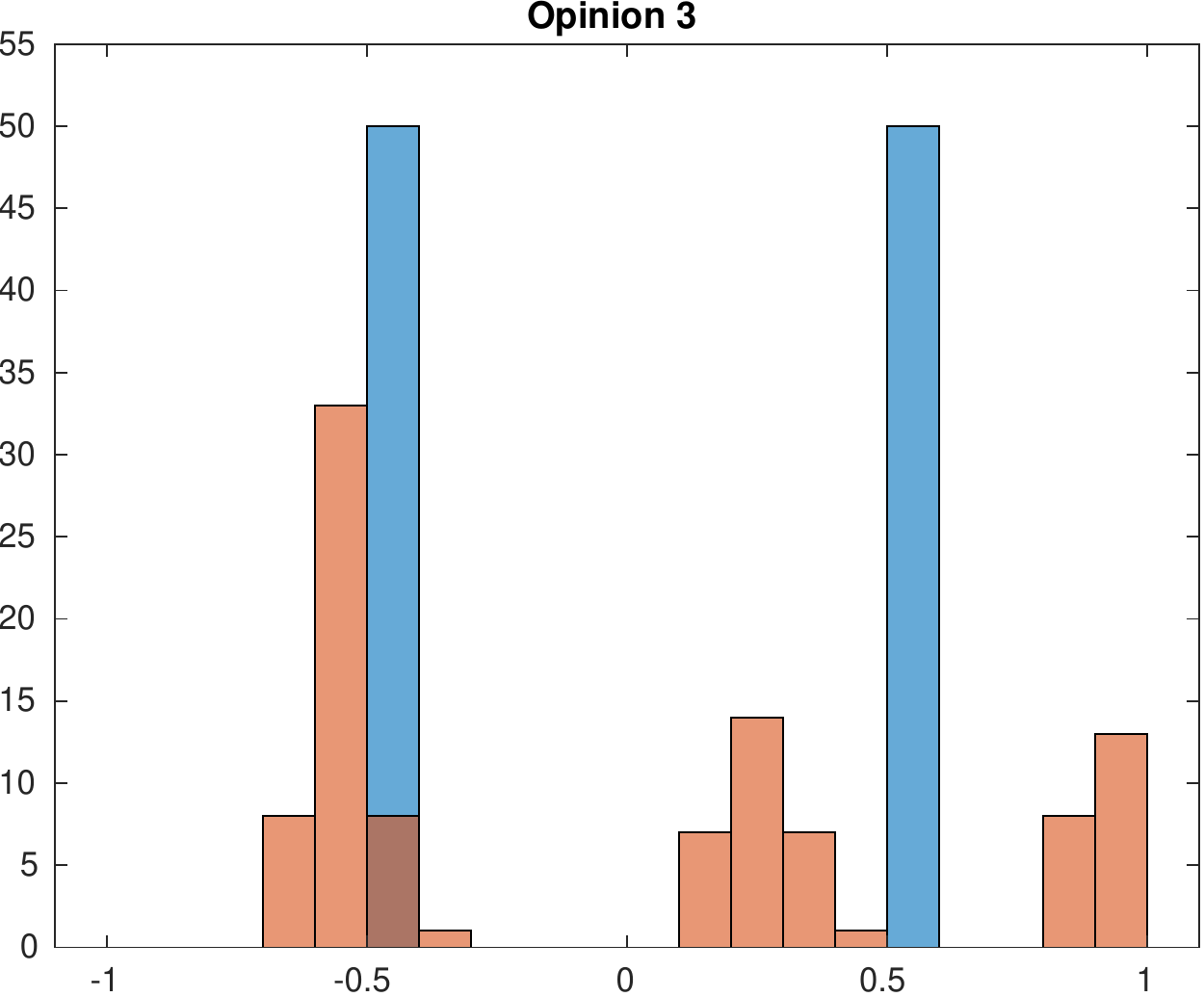}
\caption{Test 1 --- Contrarians vs all. Contrarians percentage $20\%$. $\alpha_{1,2}=0.75$ and $\alpha_{1,3}=0.25$. The blue bars represent the initial data, while the orange ones the solution at time $T$.}
\label{fig:t1_c20_alfadiversi}
\end{figure*}

\begin{figure*}[!htb]
\centering
\includegraphics[width=0.325\textwidth]{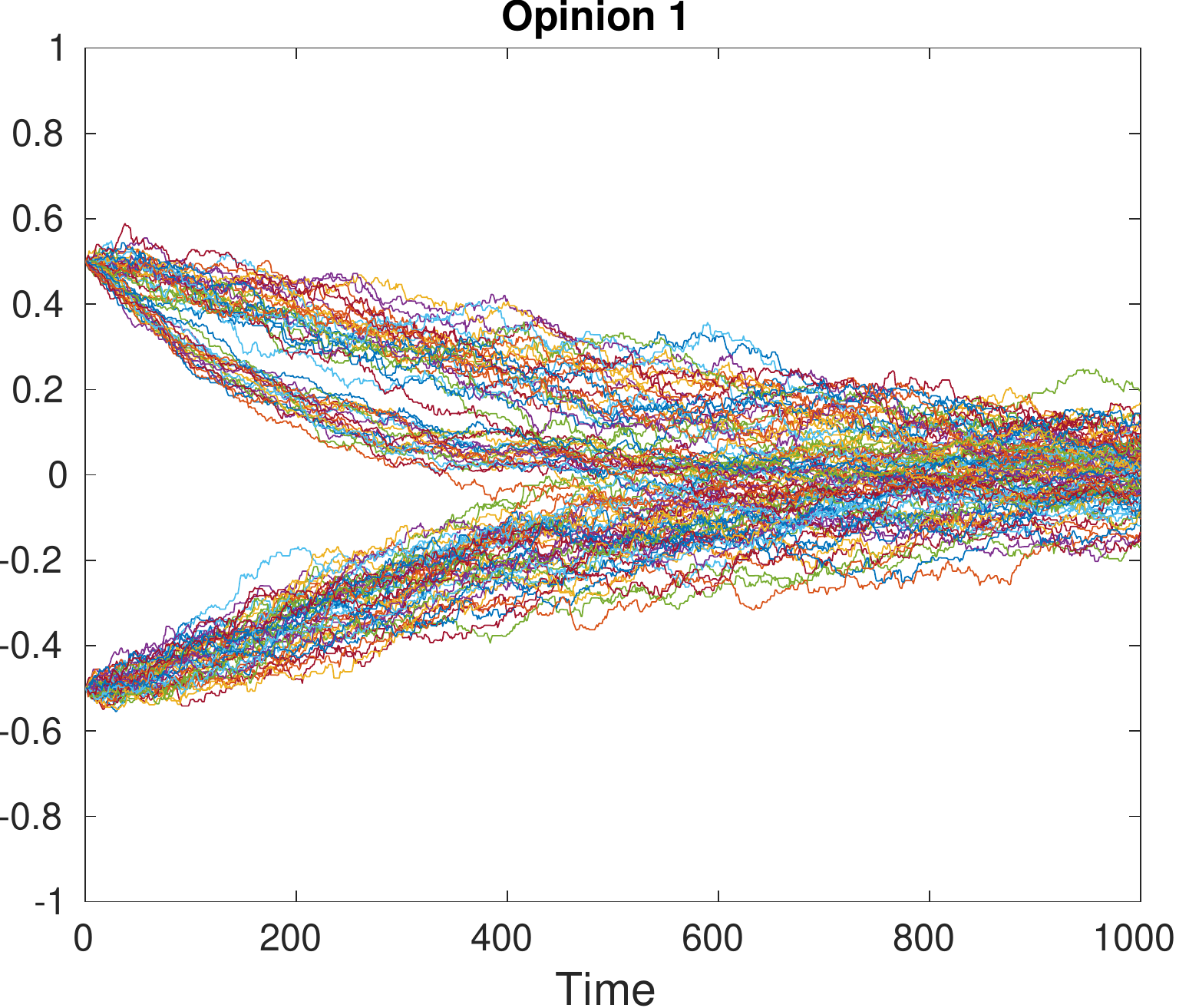}
\includegraphics[width=0.325\textwidth]{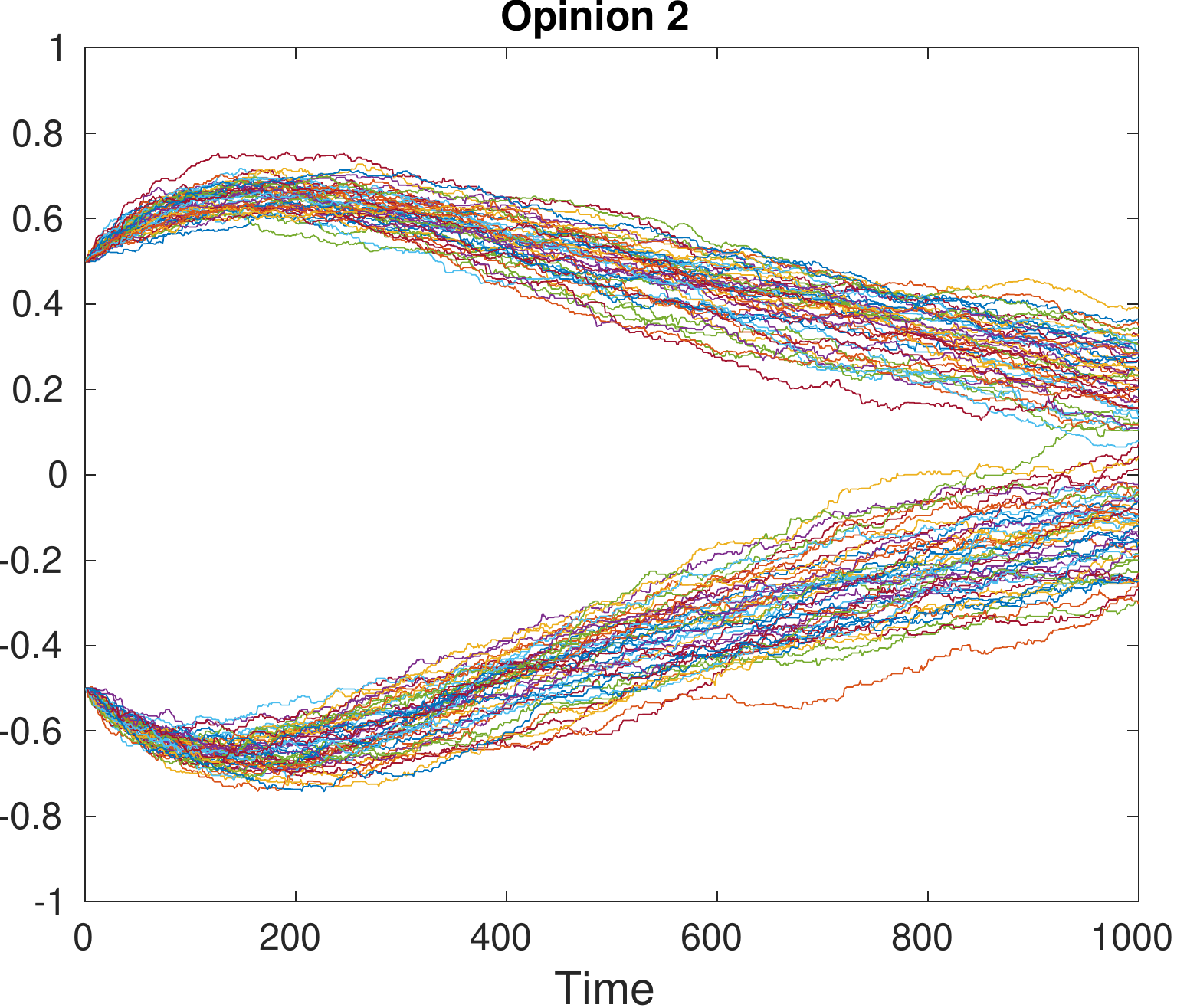}
\includegraphics[width=0.325\textwidth]{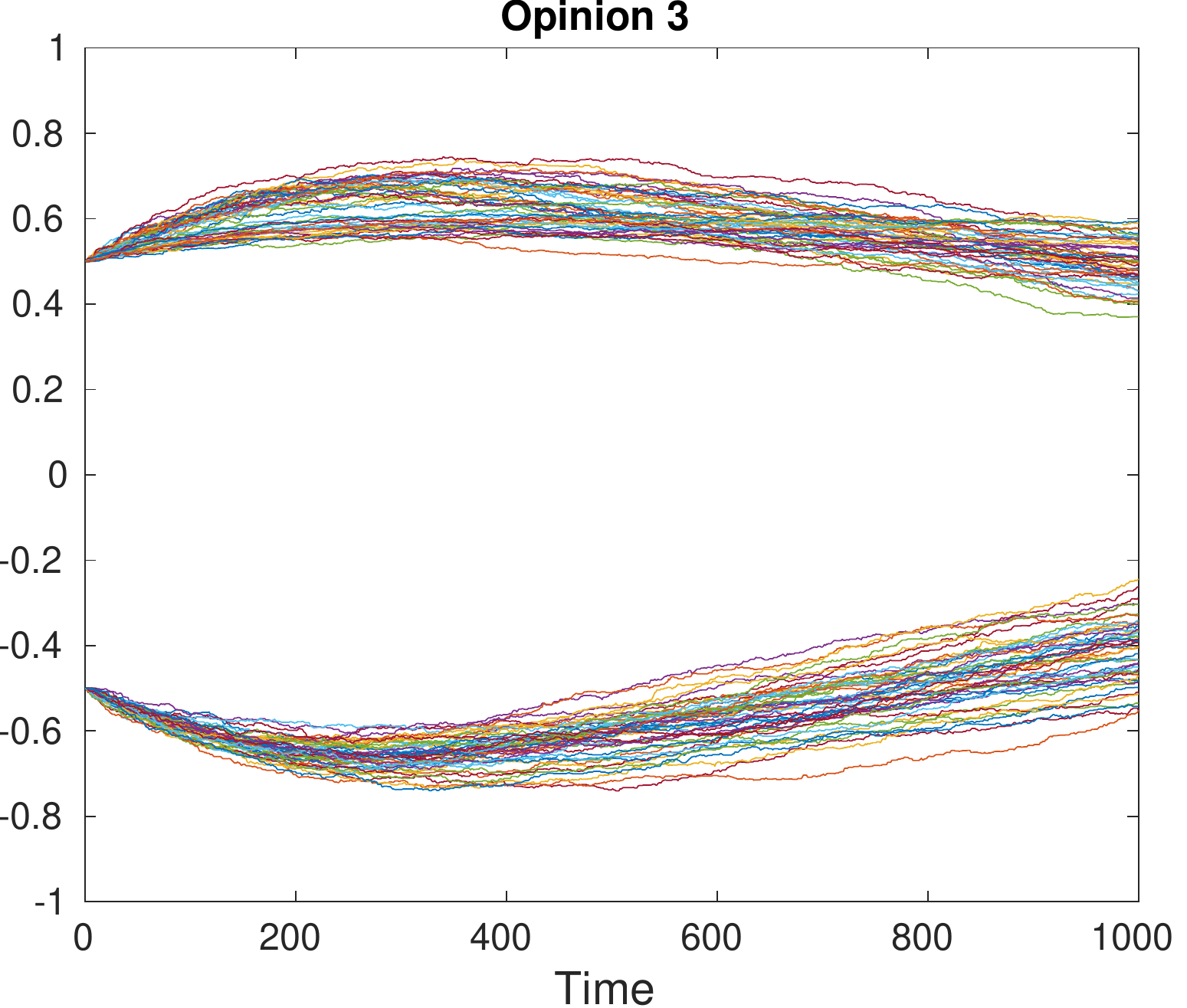}
\caption{Test 1 --- Contrarians vs all. Contrarians percentage $2\%$. $\alpha_{1,2}=0.75$ and $\alpha_{1,3}=0.25$. Evolution of the opinions in time.}
\label{fig:t1_c20_alfadiversi_tempo}
\end{figure*}

Looking at Fig.\ \ref{fig:first_t1_c2_alfadiversi}-\ref{fig:t1_c20_alfadiversi}, we observe that opinion 1 has a strong influence on opinion 2, i.e.\ the corresponding figures exhibit a similar behaviour, indeed $\alpha_{1,2}=0.75$. On the other hand opinion 3 shows a different histogram, and as a matter of fact the weight is lower, $\alpha_{1,3}=0.25$. In Fig.\ \ref{fig:t1_c20_alfadiversi_tempo} the evolution of the opinions in time is shown: the opinion 1 reaches a consensus, which strongly affects the behaviour of opinion 2, whereas the effect on opinion 3 is even less. Indeed the consensus is not reached. This confirms the results of Fig. \ref{fig:first_t1_c2_alfadiversi} 
 
\subsection{Contrarians vs the ``fan club''}\label{sec:vsconag}
Let us now focus on a different type of contrarians agents. Consider the case of contrarians opposed to the opinion of a group of fans and chose the interaction function Eq.\ \eqref{cont_vs_lead} presented above. Taking inspiration from our study-case, the fan club here is devoted to defend environment and support climate action, to adopt virtuous practices in terms of reducing the resources consumption or ask to contrast climate change. Simply speaking, they are assumed to be fans of Greta Thunberg. We will see below that a process of \emph{stigmatization}, where a social disapproval emerges around the positions of fan club, becomes prominent in the population $\mathcal N$.

Then we consider a population $\mathcal{P}$ of fans. These agents are characterized by assuming $b=0$ in Eq. (\ref{eq:evol_op1}) --- they do not change their opinions --- and their opinions on first issue (i.e. opinion 1) are fixed between $0.8$ and $1$. If we report these assumptions on our study-case, we have already said that the first component of mind vector, $x_{1,i}$, could be the political bias while the second, $x_{2,i}$, could represent the opinion of agent $i$ on climate change issues. Therefore $x_{1,i}\in(0.8,1]$ represents positions conveyed by polite or respectful language. We assume instead that opinions $x_{1,i}$ near to the opposite pole, $-1$, correspond to opinions conveyed by hateful or abusive language. Similarly, we can assume that the more (resp. the lower) is the support that agent $i$ gives to climate issues, the nearer to $1$ (resp. to $-1$) are the values of $x_{2,i}$. A very simple picture that describes these assumptions is depicted in Fig.\ \ref{fig:spectrum}. From Fig.\ \ref{fig:spectrum} we can also say that the sign of an opinion describes the willingness to be: hater or non-hater (for opinion 1), pro or cons environmental issues (for opinion 2). Hence the absolute value of $x_{1,i}$ or $x_{2,i}$, denotes instead the closeness to extreme positions of agent $i$; e.g. if $x_{1,i}>0$ (resp. $x_{1,i}<0$), its absolute value $|x_{1,i}|$ measures the closeness to totally tolerant (resp. hater) positions of agent $i$.

Let $\operatorname{card}\left(\mathcal{P}\right)=P\in\mathbb N$ and assume $P=10$. Assume also that the coherence weights are $\alpha_{1,2}=0.75$ and $\alpha_{1,3}=0.25$. As initial data we consider a uniformly distributed random values between $-1$ and $1$, that is the range of all the possible opinions. 

\begin{figure*}[!htb]
\centering
\includegraphics[width=0.32\textwidth]{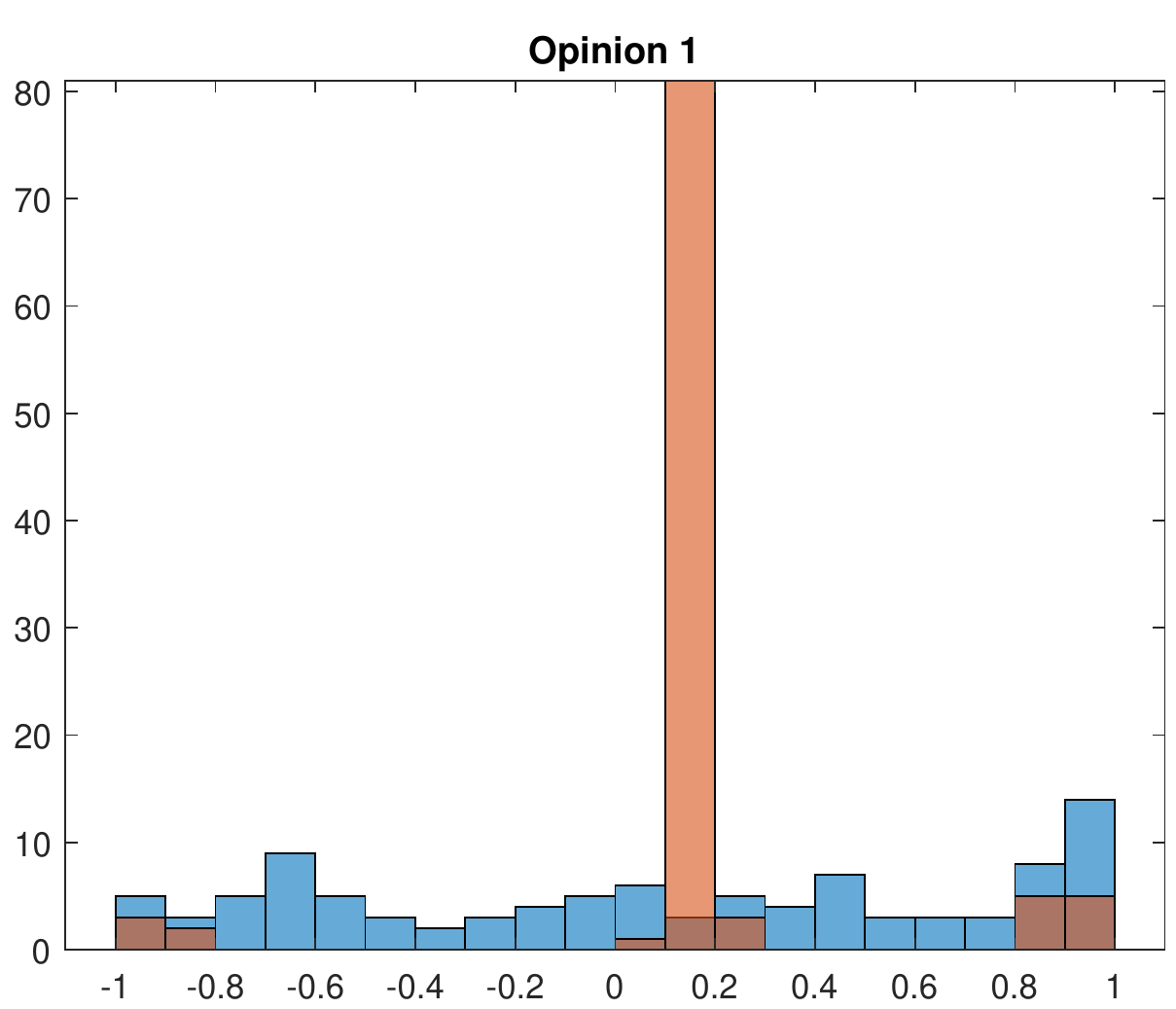}
\includegraphics[width=0.32\textwidth]{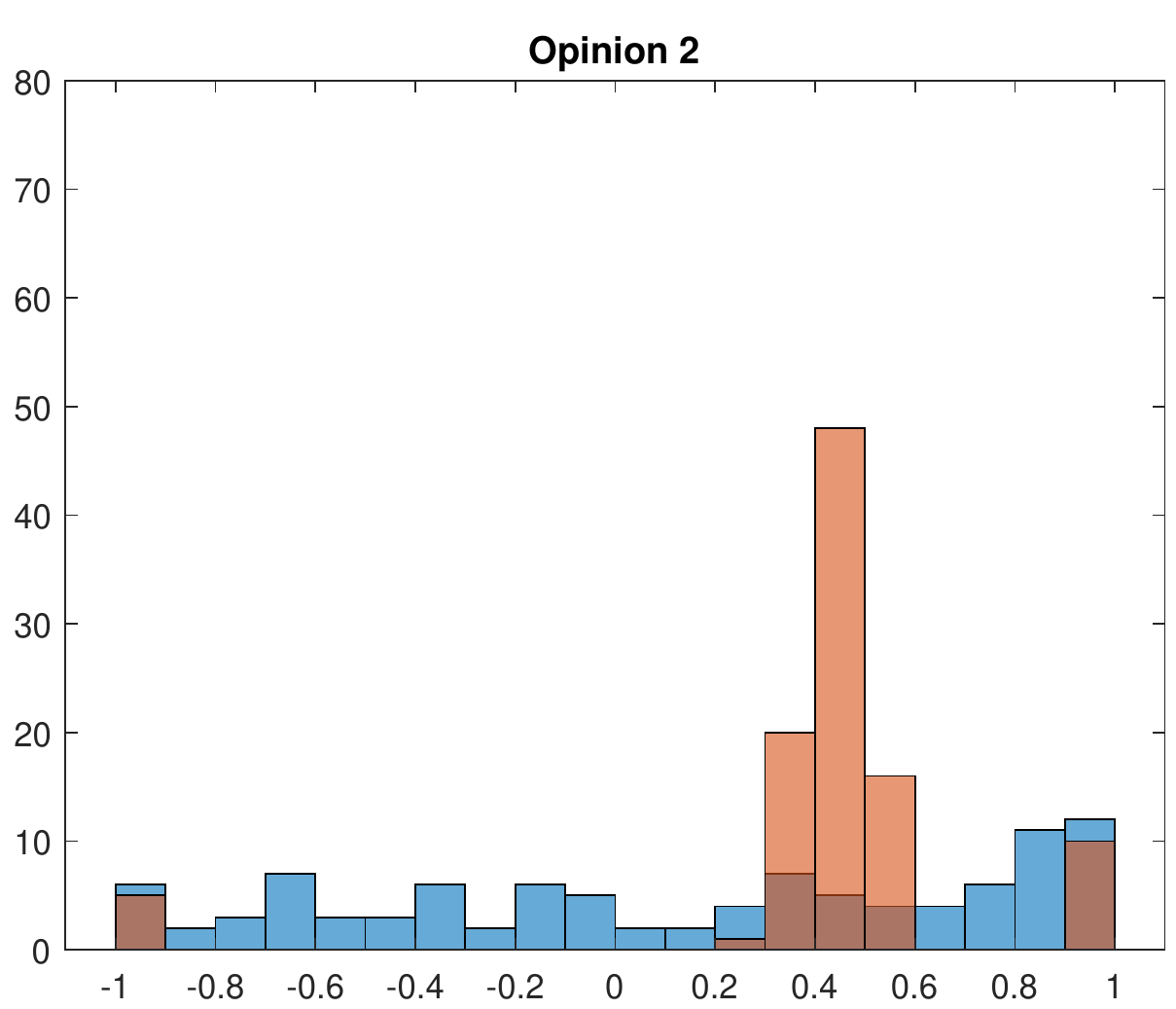}
\includegraphics[width=0.32\textwidth]{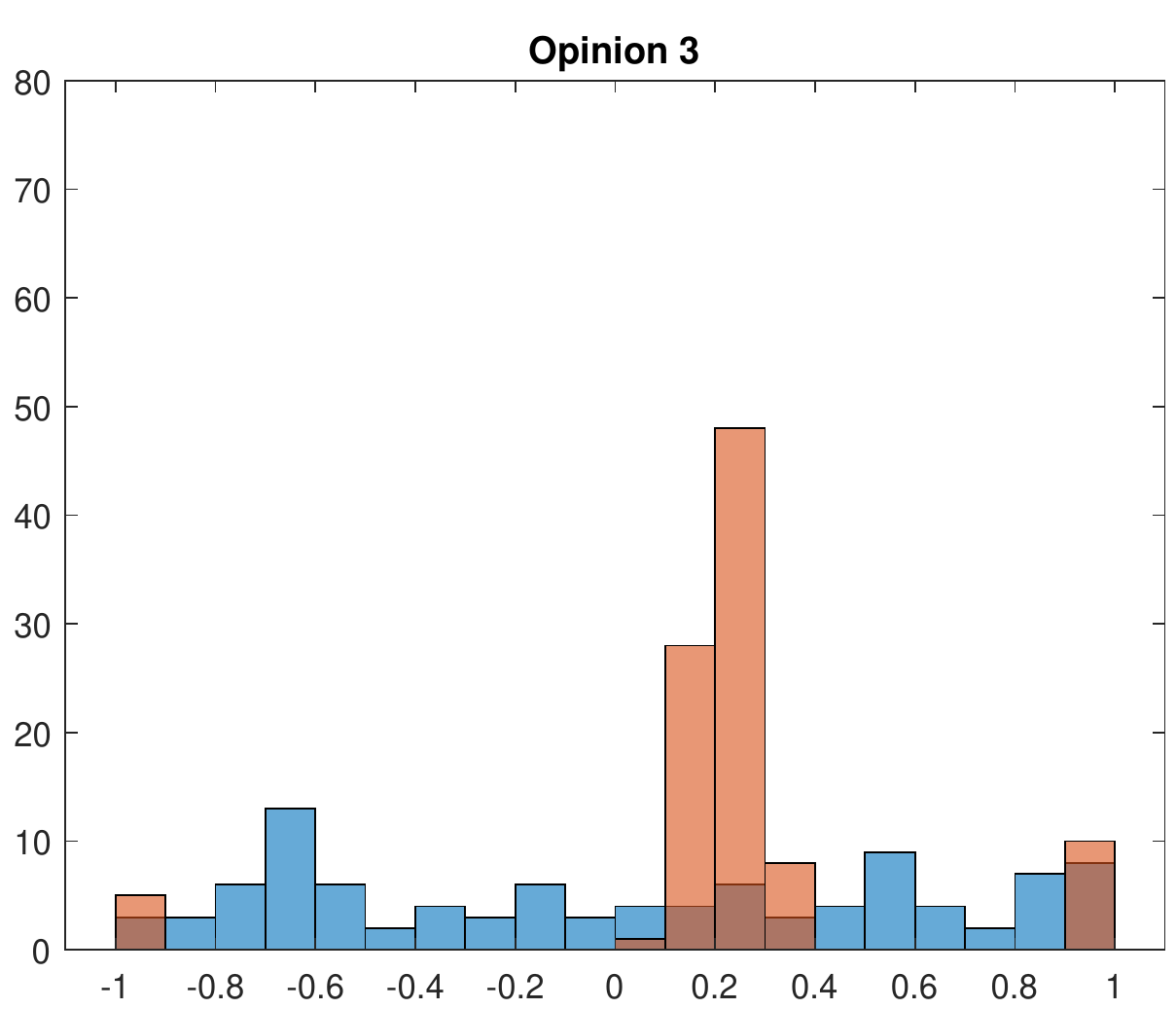}
\caption{Test 2 --- Contrarians vs convinced agents. Contrarians percentage $5\%$. The blue bars represent the initial data, while the orange ones the solution at time $T$.}
\label{fig:T2_c5}
\end{figure*}

\begin{figure*}[!htb]
\centering
\includegraphics[width=0.32\textwidth]{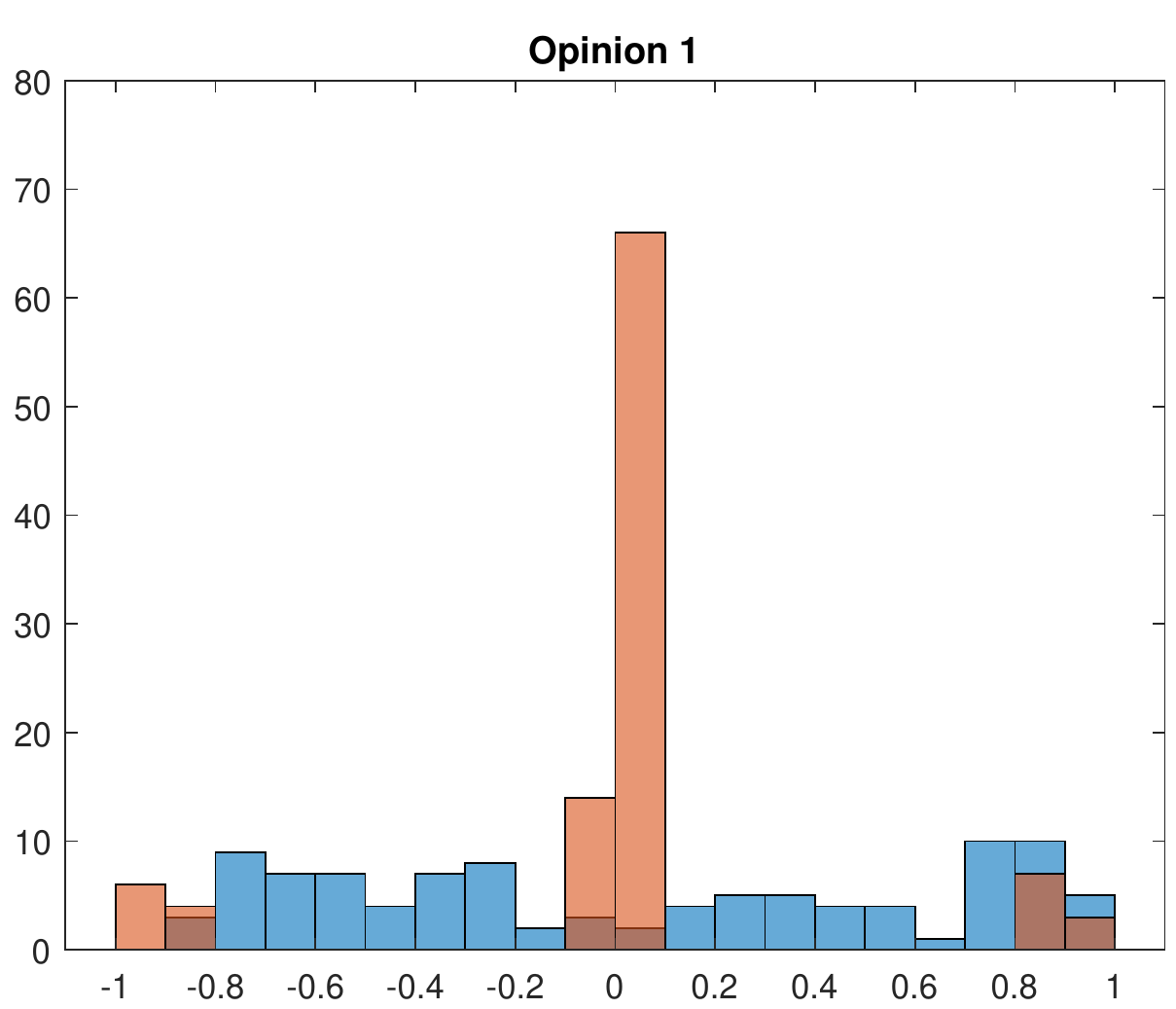}
\includegraphics[width=0.32\textwidth]{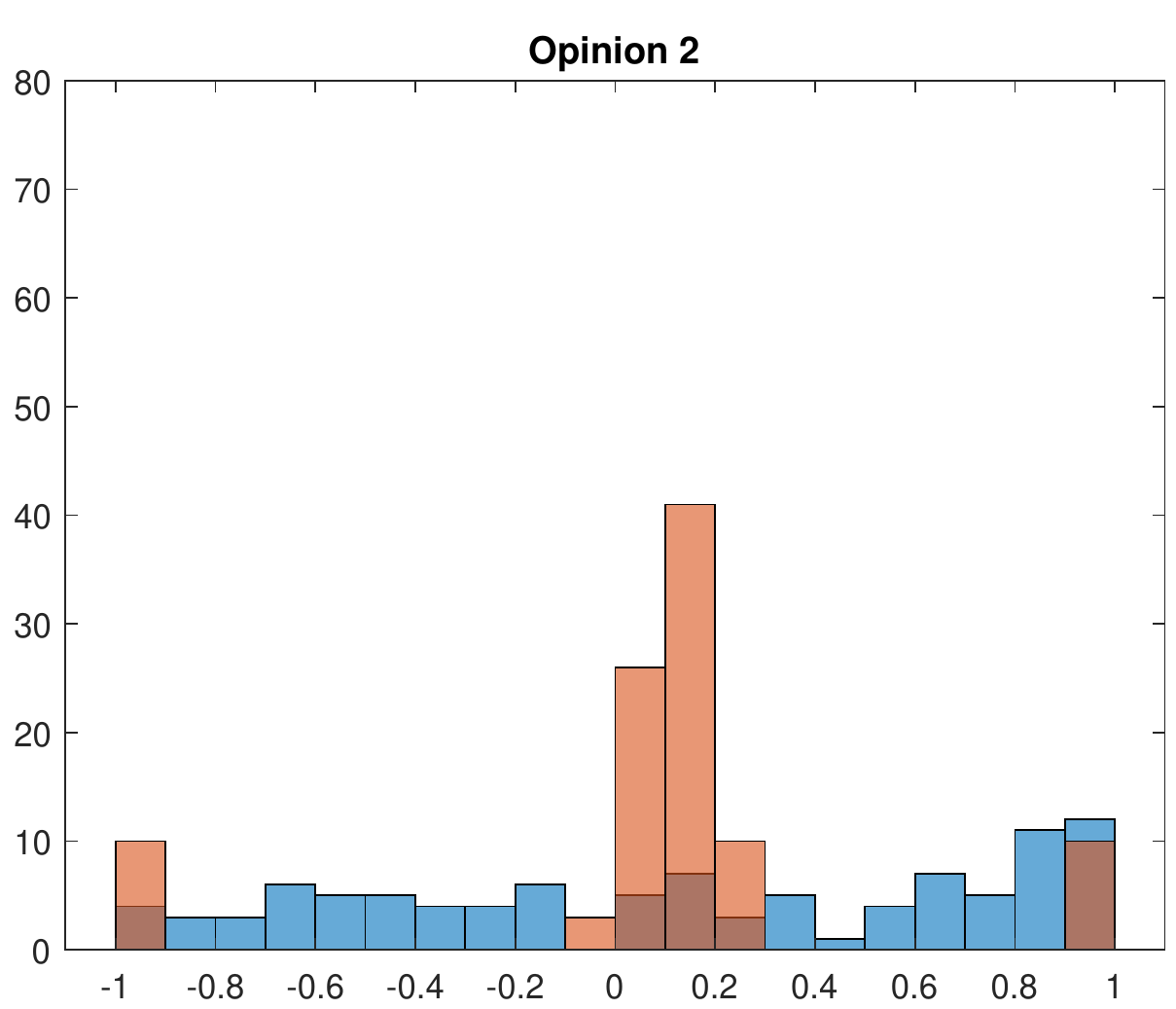}
\includegraphics[width=0.32\textwidth]{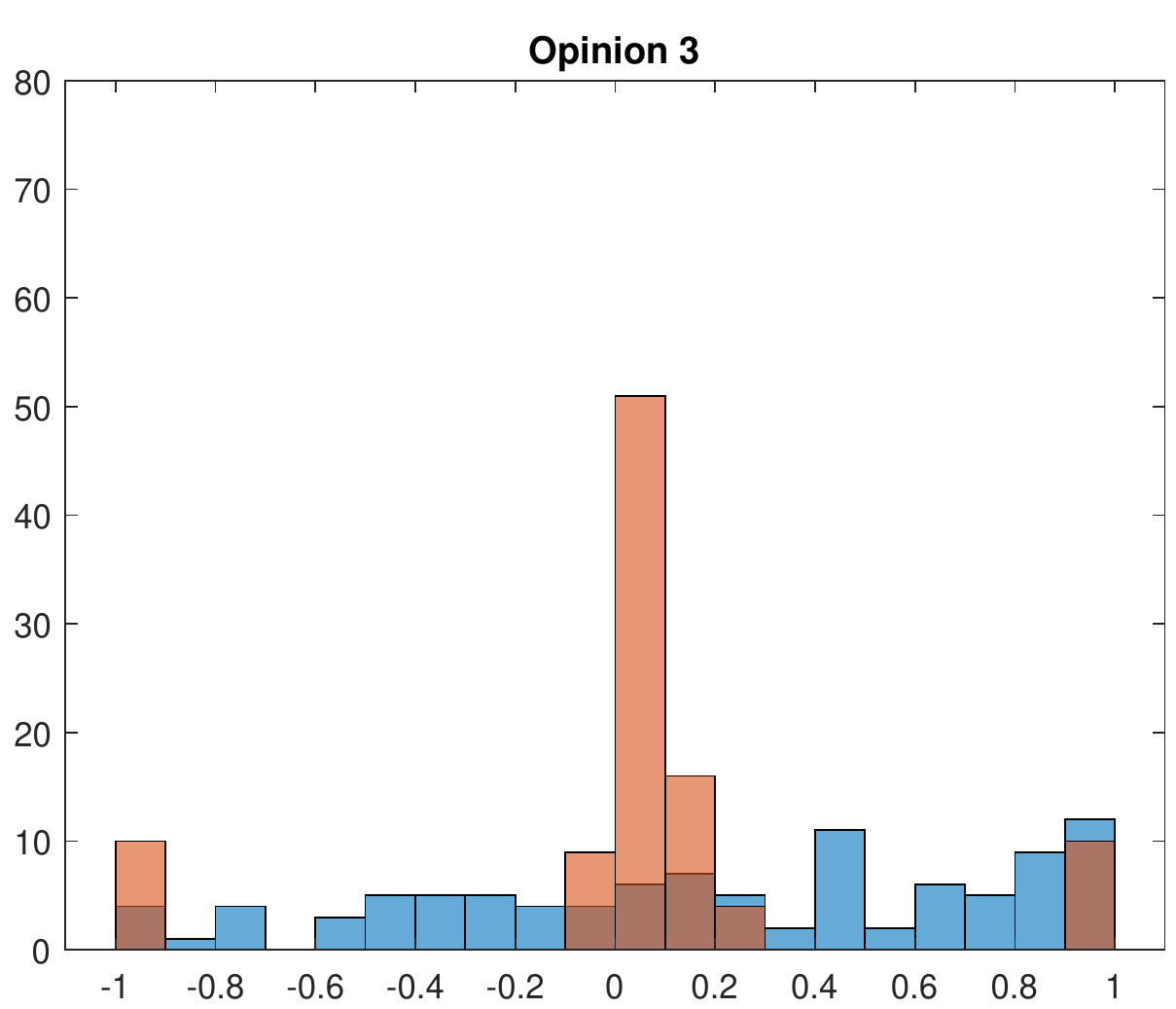}
\caption{Test 2 --- Contrarians vs convinced agents. Contrarians percentage $10\%$. The blue bars represent the initial data, while the orange ones the solution at time $T$.}
\label{fig:T2_c10}
\end{figure*}

We can observe that for small concentrations of contrarians, opinion 1 assumes values close to 0, see Fig.\ \ref{fig:T2_c5}-\ref{fig:T2_c10}. Then, concerning the first component of mind vector, the population reaches a consensus state where almost the whole system adopts one opinion value. This result is less pronounced for the other two components of mind vector, since the agents focus their opinions around a wider range of values.

\begin{figure*}[!htb]
\centering
\includegraphics[width=0.32\textwidth]{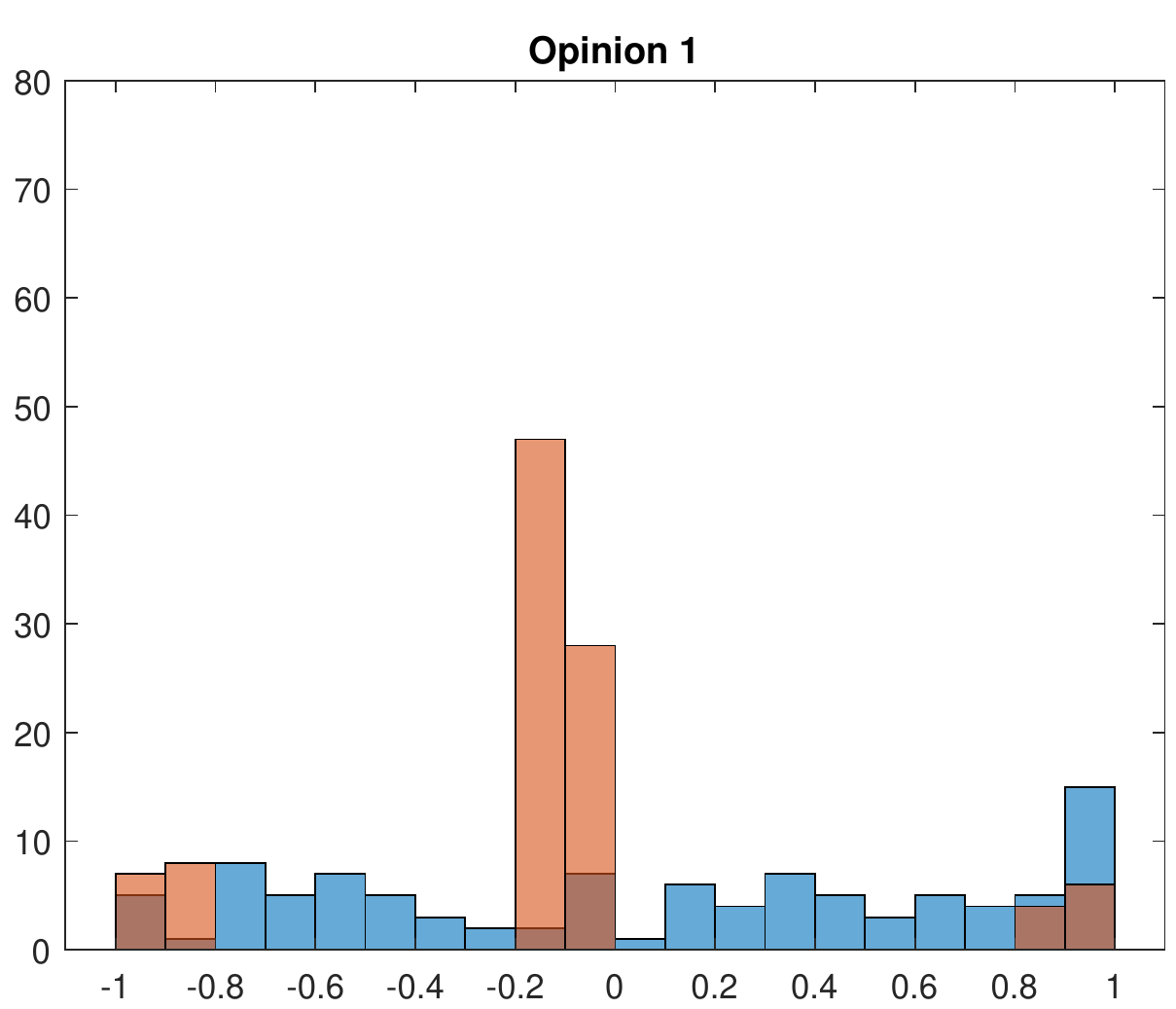}
\includegraphics[width=0.32\textwidth]{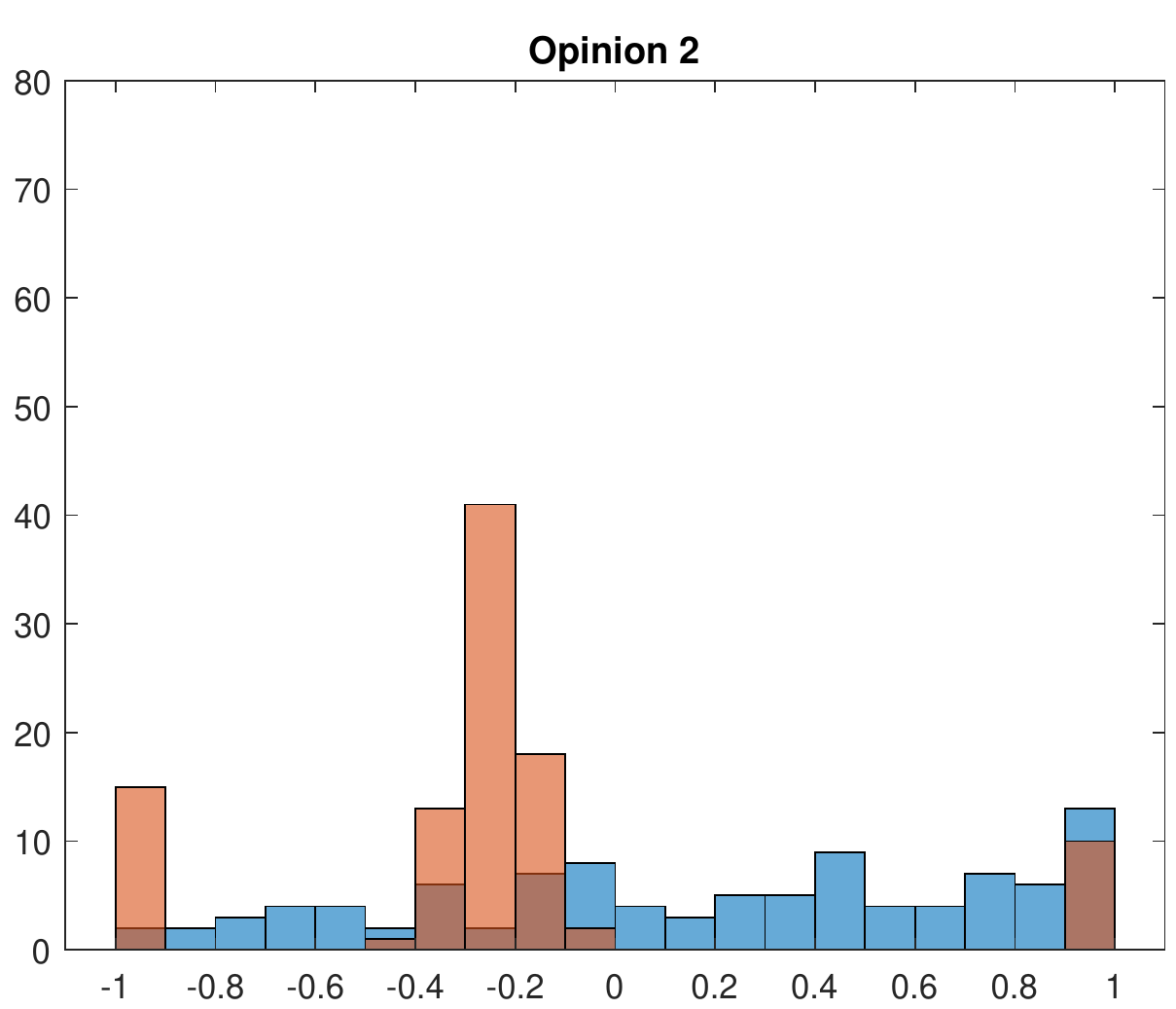}
\includegraphics[width=0.32\textwidth]{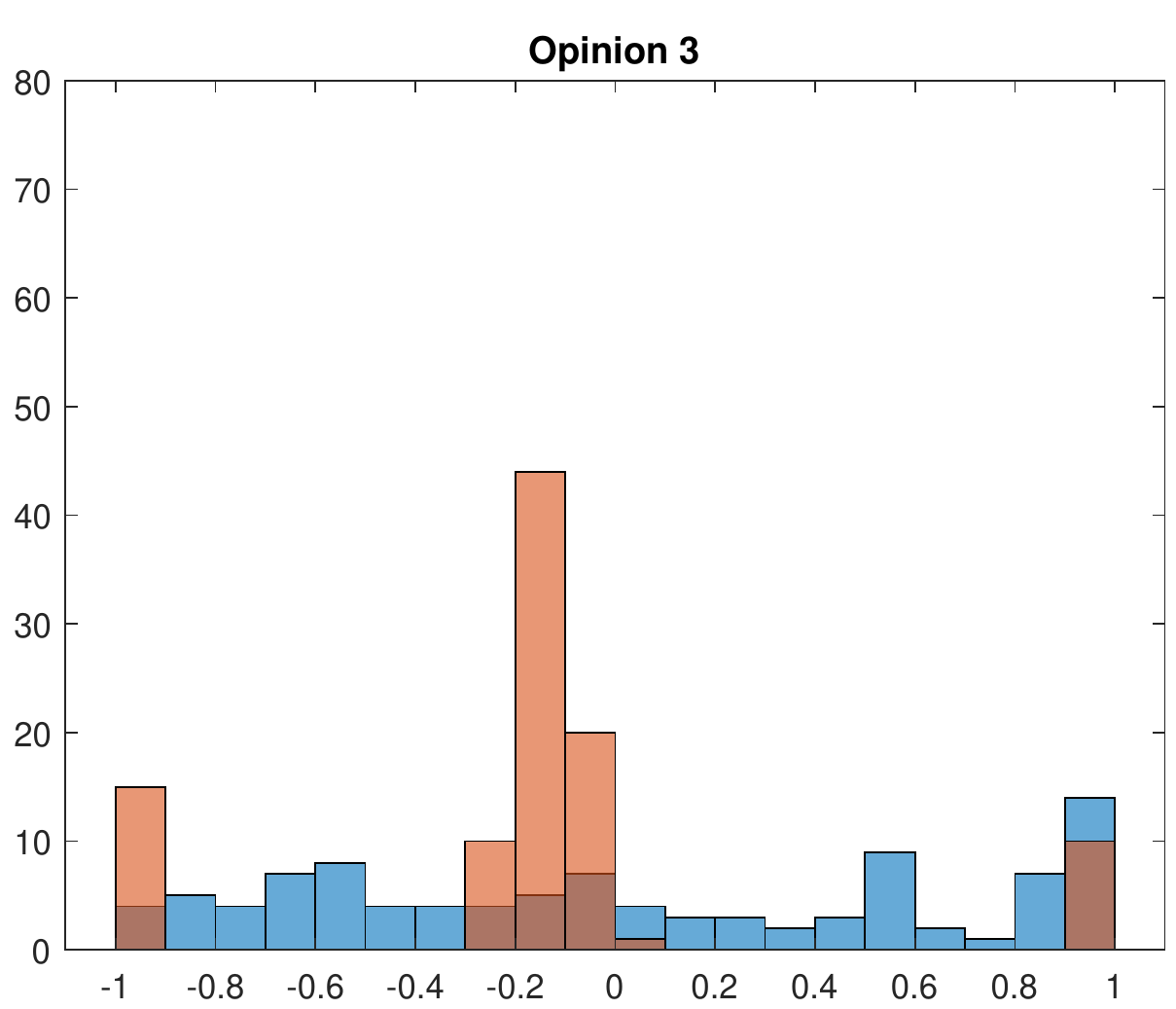}
\caption{Test 2 --- Contrarians vs convinced agents. Contrarians percentage $15\%$. The blue bars represent the initial data, while the orange ones the solution at time $T$.}
\label{fig:T2_c15}
\end{figure*}

\begin{figure*}[!htb]
\centering
\includegraphics[width=0.32\textwidth]{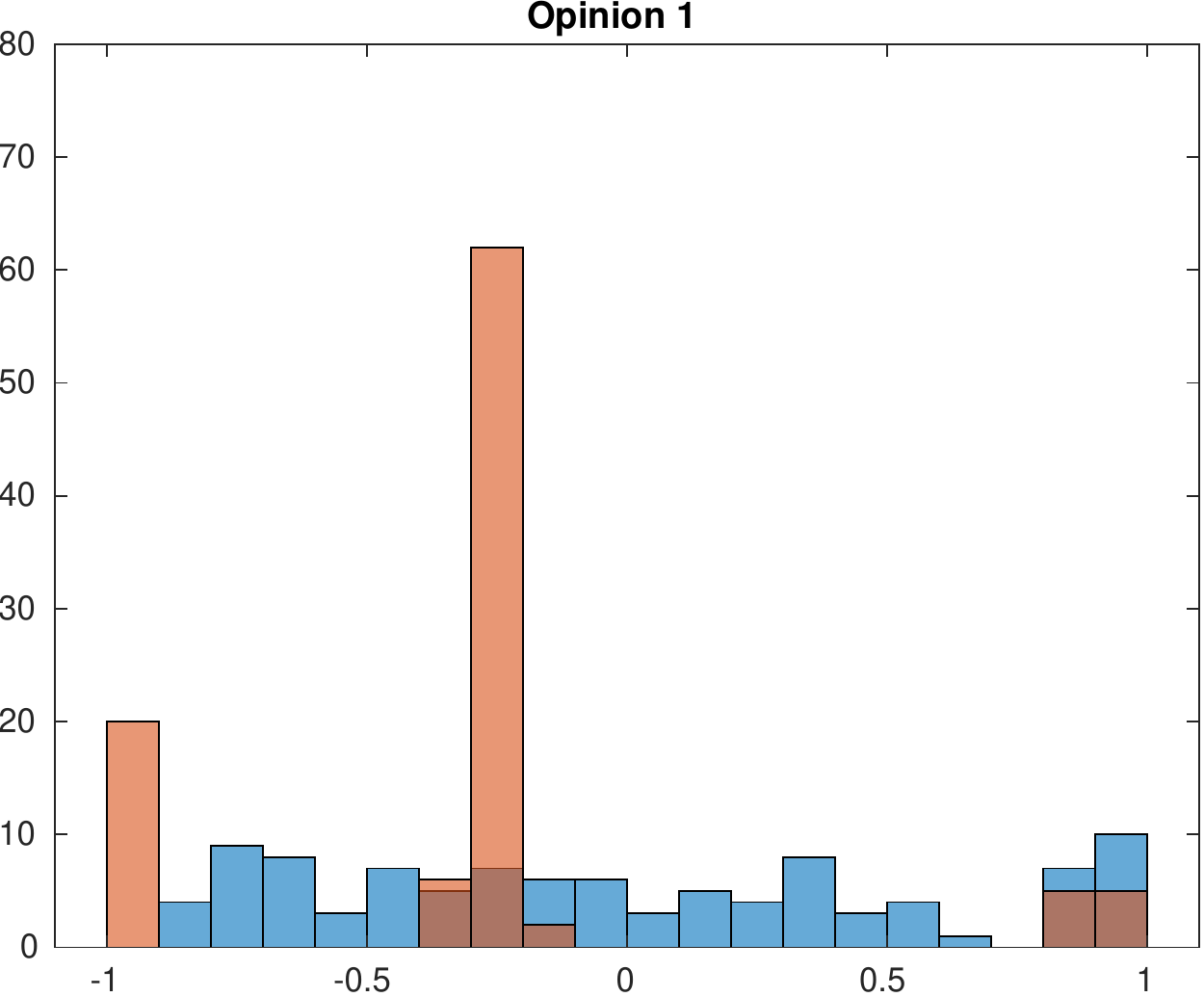}
\includegraphics[width=0.32\textwidth]{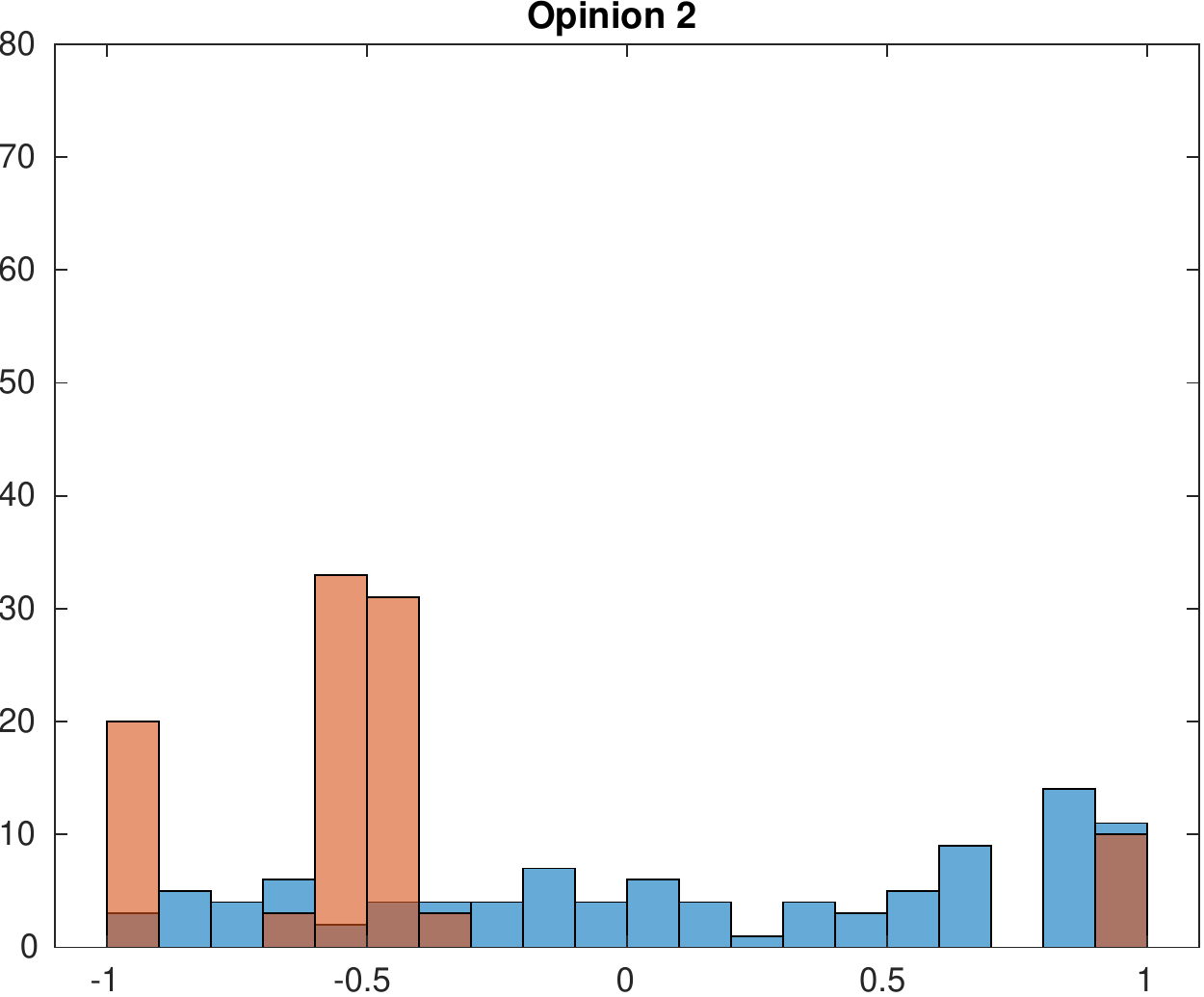}
\includegraphics[width=0.32\textwidth]{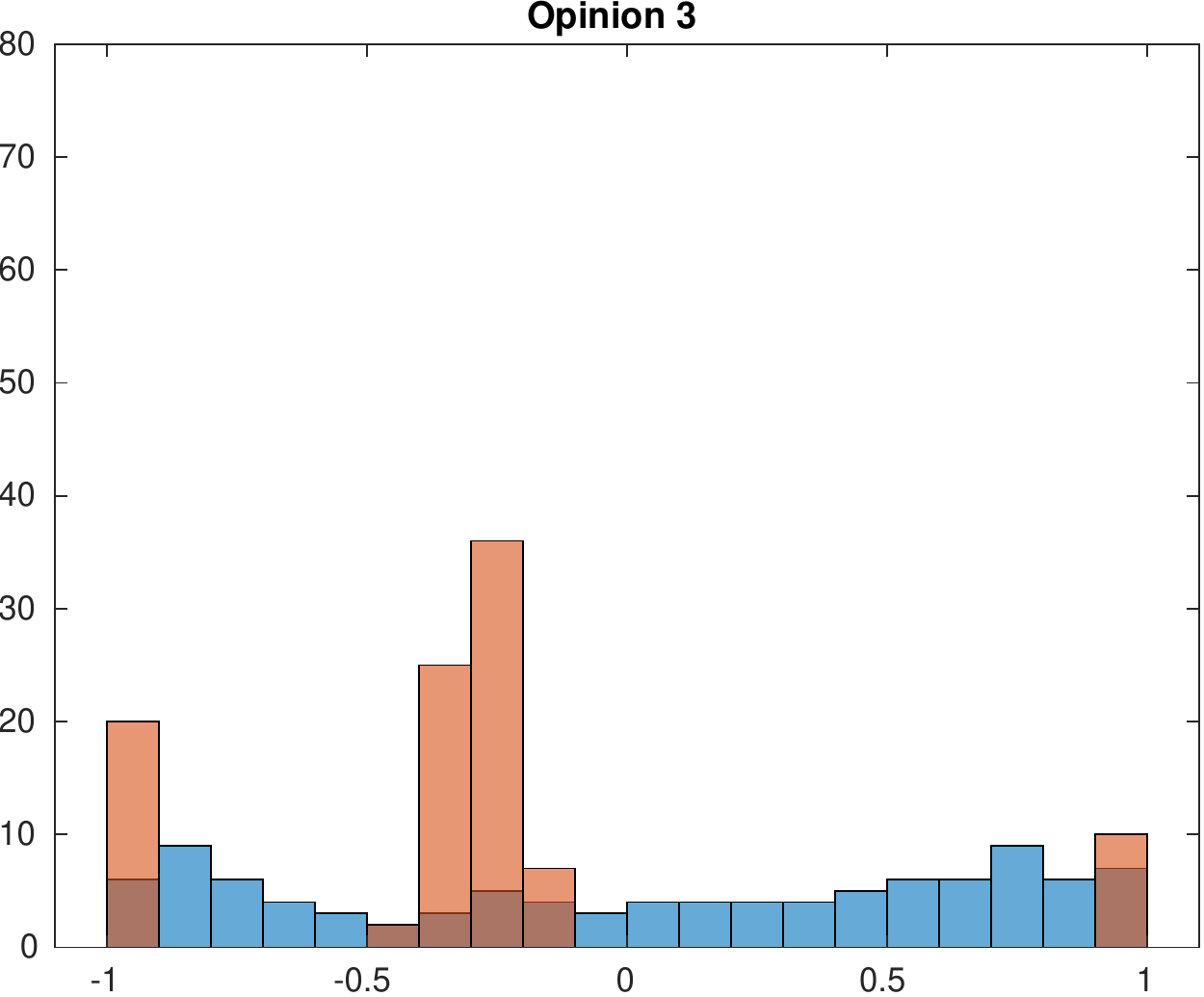}
\caption{Test 2 --- Contrarians vs convinced agents. Contrarians percentage $20\%$. The blue bars represent the initial data, while the orange ones the solution at time $T$.}
\label{fig:T2_c20}
\end{figure*}

As the presence of contrarians increases, opinion 1 assumes negative values, see Fig.\ \ref{fig:T2_c15}-\ref{fig:T2_c20}. This means that the conformists agents are influenced by contrarians and the main opinion of the population is more similar to the contrarians one, rather than the opinion of fans. This behavior is more evident looking at the evolution of opinion 2. For example, comparing Fig.\ \ref{fig:T2_c5}(central) and Fig. \ref{fig:T2_c20}(central), opinion 2 shows a quite dissimilar behavior. Also opinion 3 moves from positive values to the negative ones, even if in a more relaxed way since the coherence weight is low, see Fig. \ref{fig:T2_c5}(right)-\ref{fig:T2_c15}(right).

This case may explain the findings of web scraping and polarization analysis evidenced in Section \ref{intro} around the Greta Thunberg phenomenon. Many Twitter users seem to ``upload'' their opinions on environmental policies (e.g. opinions 2 and 3) according to a pre-determined political bias (e.g. the opinion 1) and this mechanism, translated in formulas by Eq.\ (\ref{other_op}), show, through Fig.\ \ref{fig:T2_c5}-\ref{fig:T2_c20}, the social function of contrarians in the opinion formation of a group of agents. In fact, central plots in Fig.\ \ref{fig:T2_c5}-\ref{fig:T2_c20} show that, by increasing the number of contrarians, it is possible to drive the opinions concerning climate issues toward more critical stances, i.e. it is possible to drive opinions 2 toward negative values. In other words, fixed the time evolution of $x_{1}$ and its influence on the other components of mind vector, the role of contrarians in this case is to increase the stigmatization of people which attempt to achieve a reduction in resources demand, mathematically represented by the shift of opinions 2 toward $-1$. This happens despite the presence of a group of Greta's fans acting as leaders. The polarization effect exerted by large numbers of contrarians advocates particular attention, especially considering that social media services like Facebook and Twitter have recently proven themselves susceptible to various forms of political manipulation \cite{tucker2017liberation,Gorwa2018}. To mention just the most important cases, it has been documented that, during the Brexit referendum and the 2016 U.S. Presidential election, social bots were used to push hyperpartisan news and to affect political conversation \cite{Bastos}. Moreover, at first, we thought that the introduction of new digital technologies such as the internet and social media could make democratic political change easier because of the creation of a larger number of interconnections among people. However, as demonstrated by outcomes of the 2011 Arab revolts or the illiberal turn witnessed in several European countries, this has not been always the case \cite{Dacrema2020}.

\subsection{Contrarians vs Majority}\label{sec:vsmaj}
Contrarians could also be against majority, a characteristic attitude of those who do not conform to conventional fashions and lifestyles. 
The main difference with the previous test is that the opinion of the majority evolves and can change in time. Indeed at each time step we compute the opinion of majority while the interaction function is given by Eq.\ \eqref{cont_vs_maj}.

Let us assume the initial data as follows:
\begin{equation}
   x_{1,i}=\begin{cases}
   - 0.5 \qquad & i<50\\
   + 0.5  \qquad & i\ge 50
\end{cases}
\end{equation}
and $x_{2,i}=x_{3,i}=x_{1,i}$. The coherence weights are $\alpha_{1,2}=0.75$ and $\alpha_{1,3}=0.25$.

At the initial time the opinion of the majority is $0.5$, assuming this as the value of the opinion of half population and one agent more. Thus the contrarians opinion will be the opposite.

\begin{figure*}[!htb]
\centering
\includegraphics[width=0.32\textwidth]{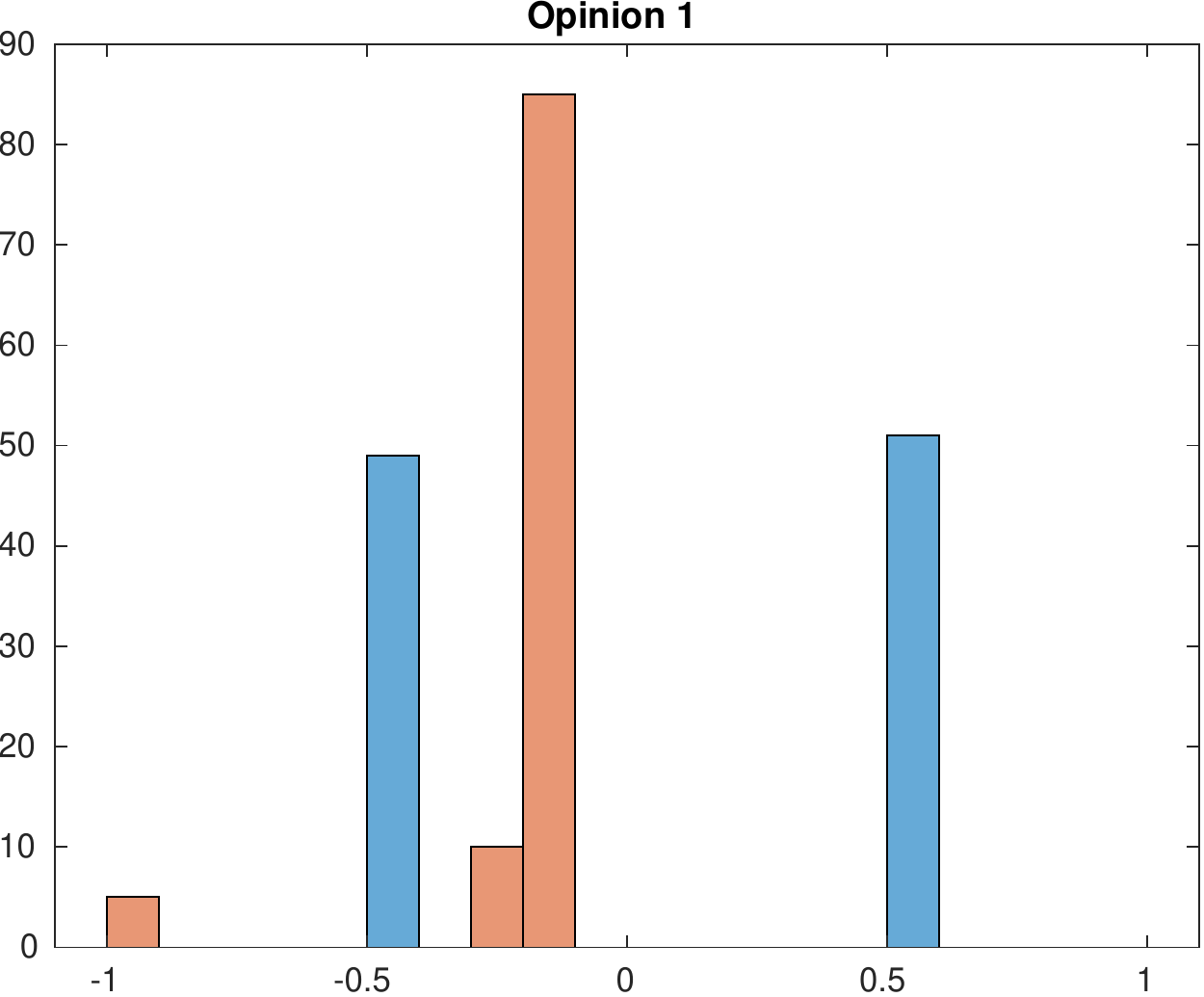}
\includegraphics[width=0.32\textwidth]{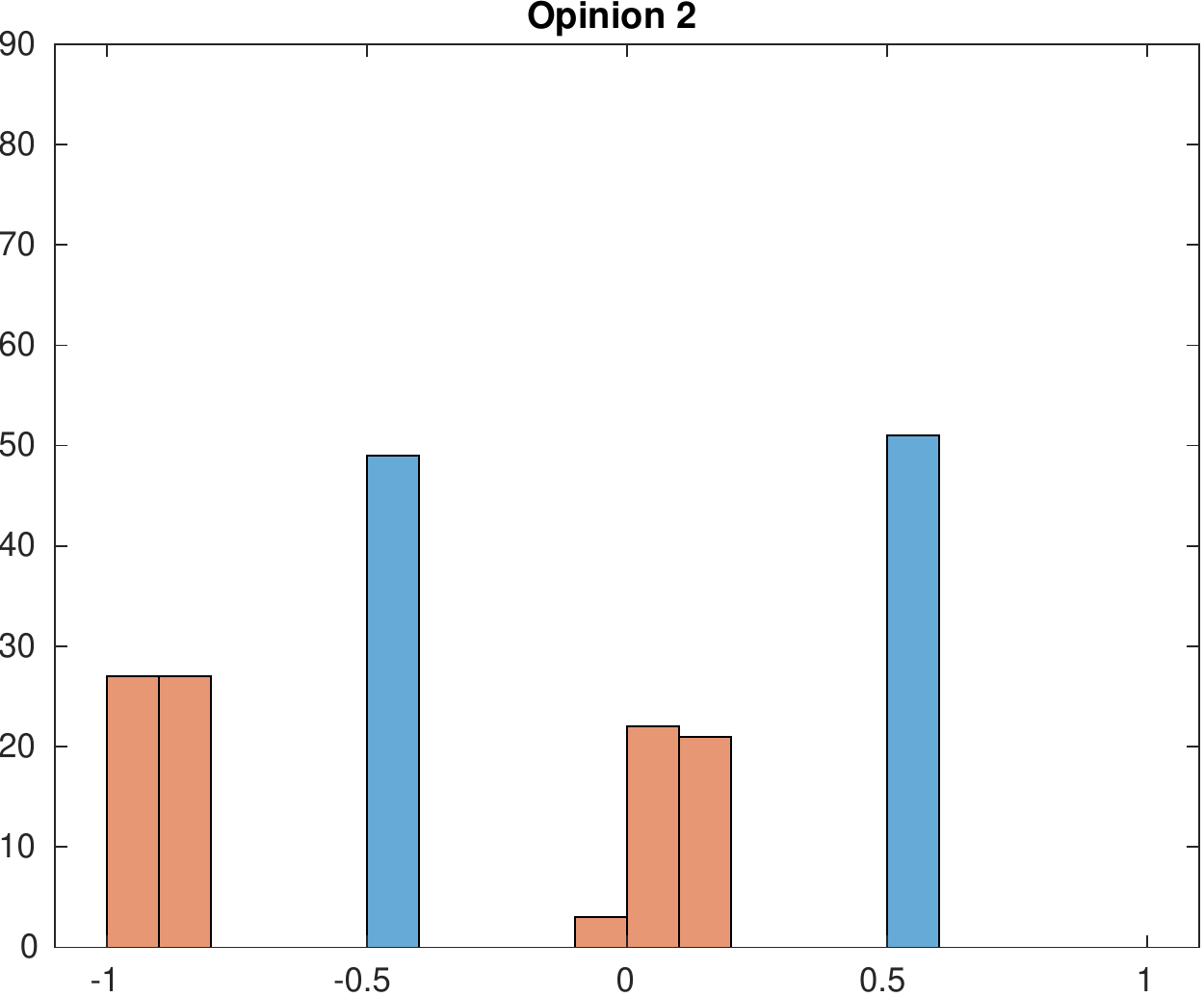}
\includegraphics[width=0.32\textwidth]{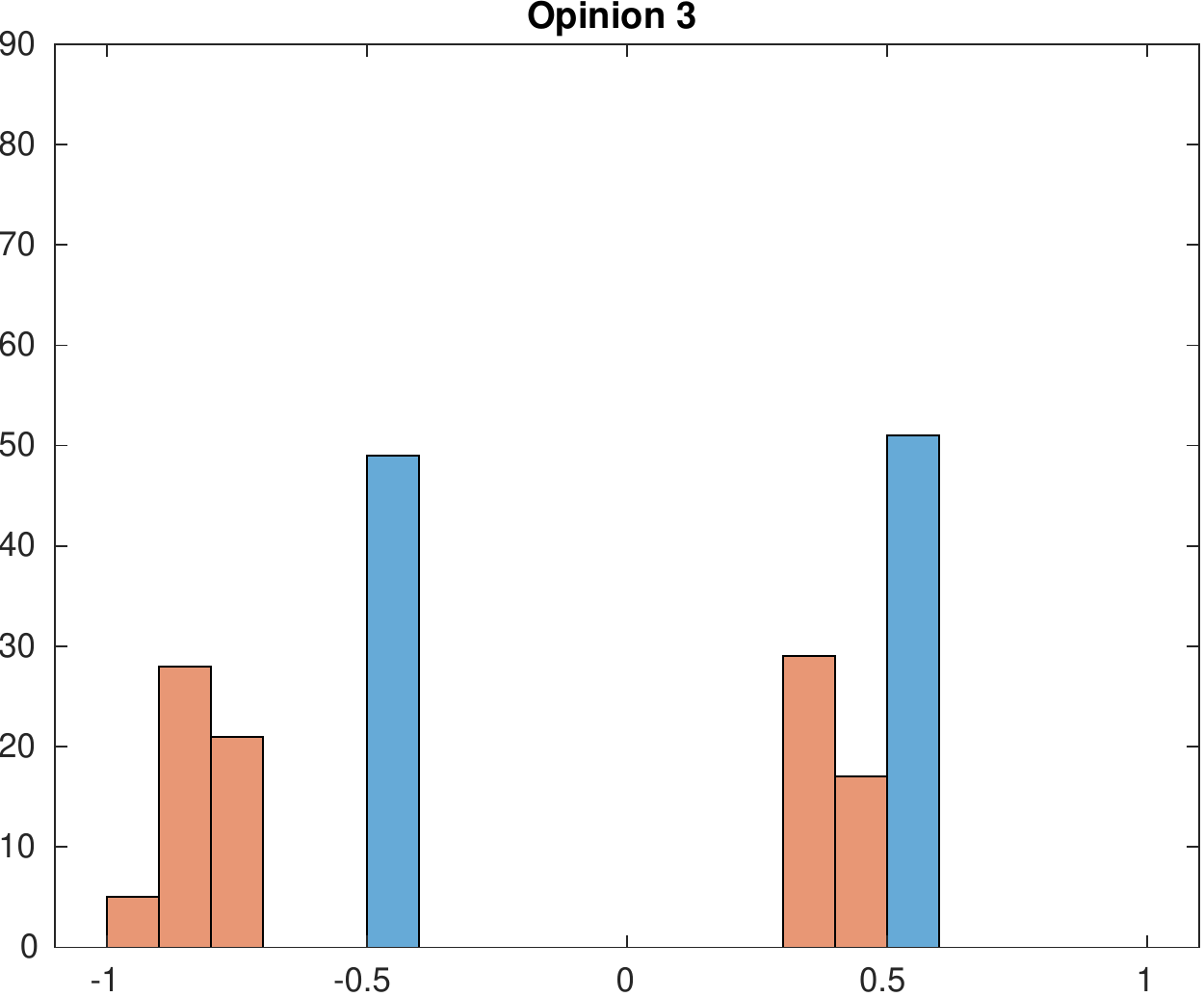}
\caption{Test 3 --- Contrarians vs Majority. Contrarians percentage $5\%$. The blue bars represent the initial data, while the orange ones the solution at time $T$.}
\label{fig:T3_c5}
\end{figure*}

\begin{figure*}[!htb]
\centering
\includegraphics[width=0.32\textwidth]{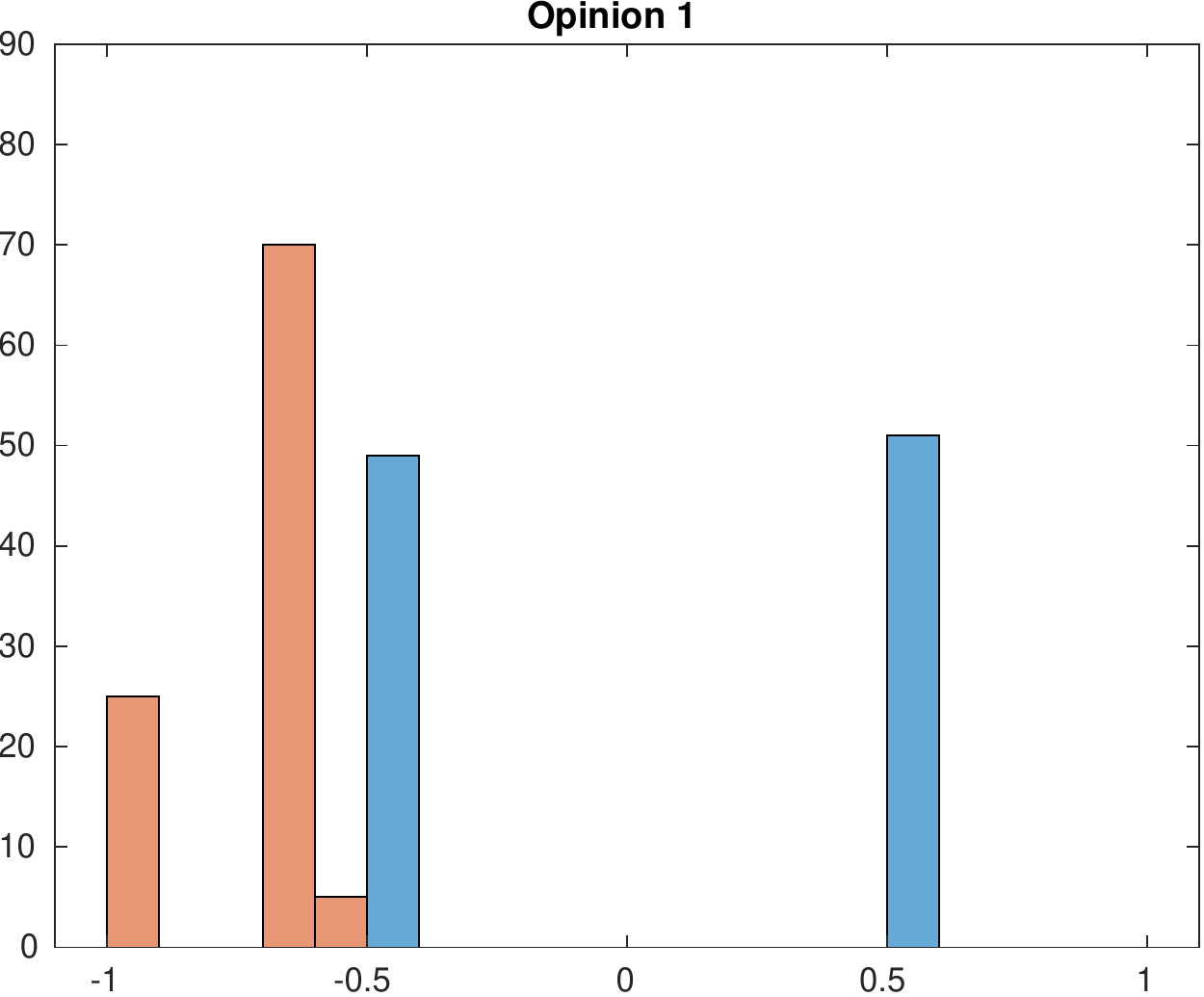}
\includegraphics[width=0.32\textwidth]{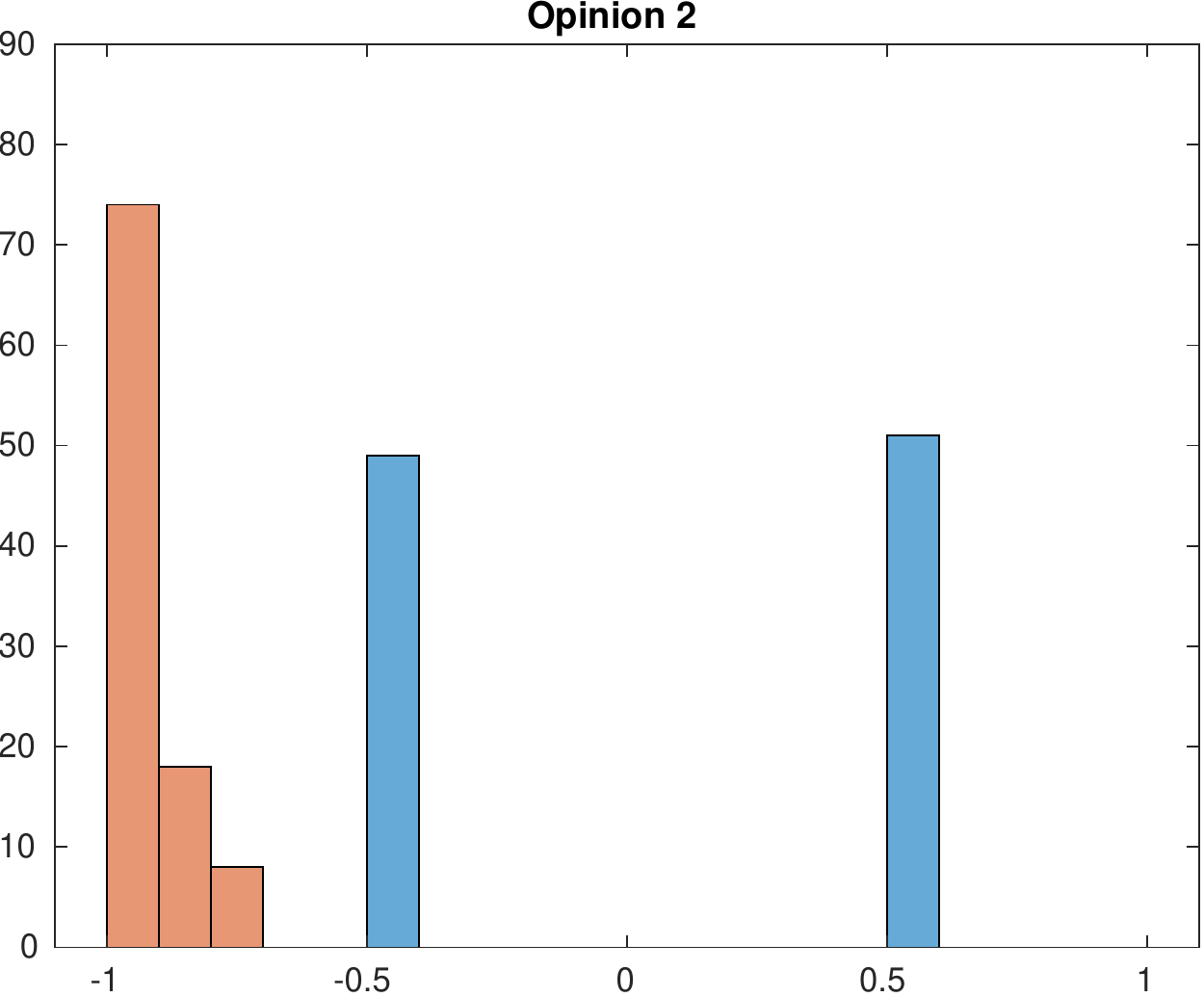}
\includegraphics[width=0.32\textwidth]{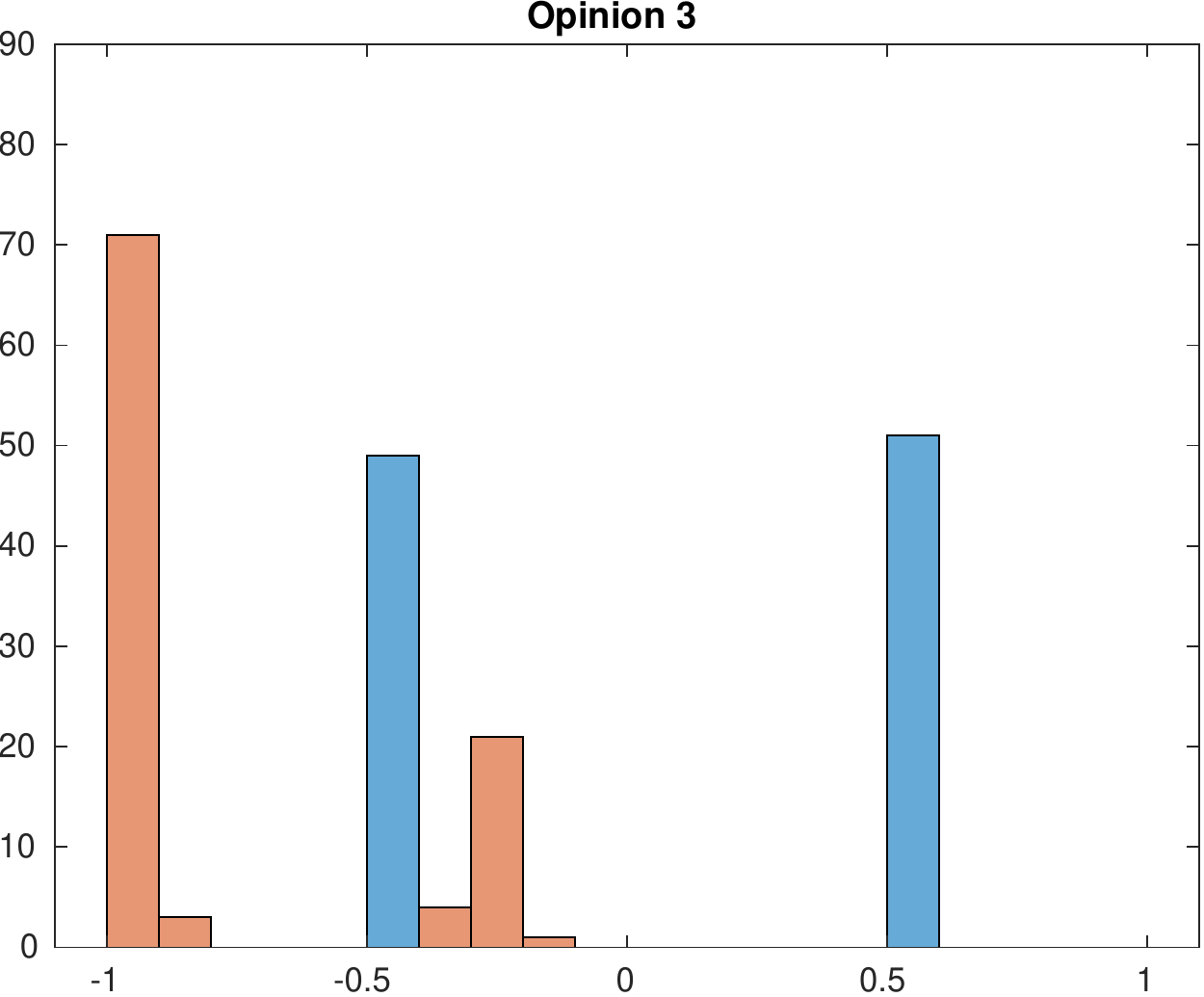}
\caption{Test 3 --- Contrarians vs Majority. Contrarians percentage $25\%$.The blue bars represent the initial data, while the orange ones the solution at time $T$.}
\label{fig:T3_c25}
\end{figure*}

Looking at Fig.\ \ref{fig:T3_c5}, we note that opinion 1 undergoes a shift towards negative values, which is better emphasised in Fig.\ \ref{fig:T3_c25}, where the percentage of contrarians is higher. In other words, notwithstanding the majority of agents expresses a positive opinion on the environmental standpoints at initial time step, the role of contrarians is still to drive this opinion towards critical stances.

\begin{figure*}[!htb]
\centering
\includegraphics[width=0.42\textwidth]{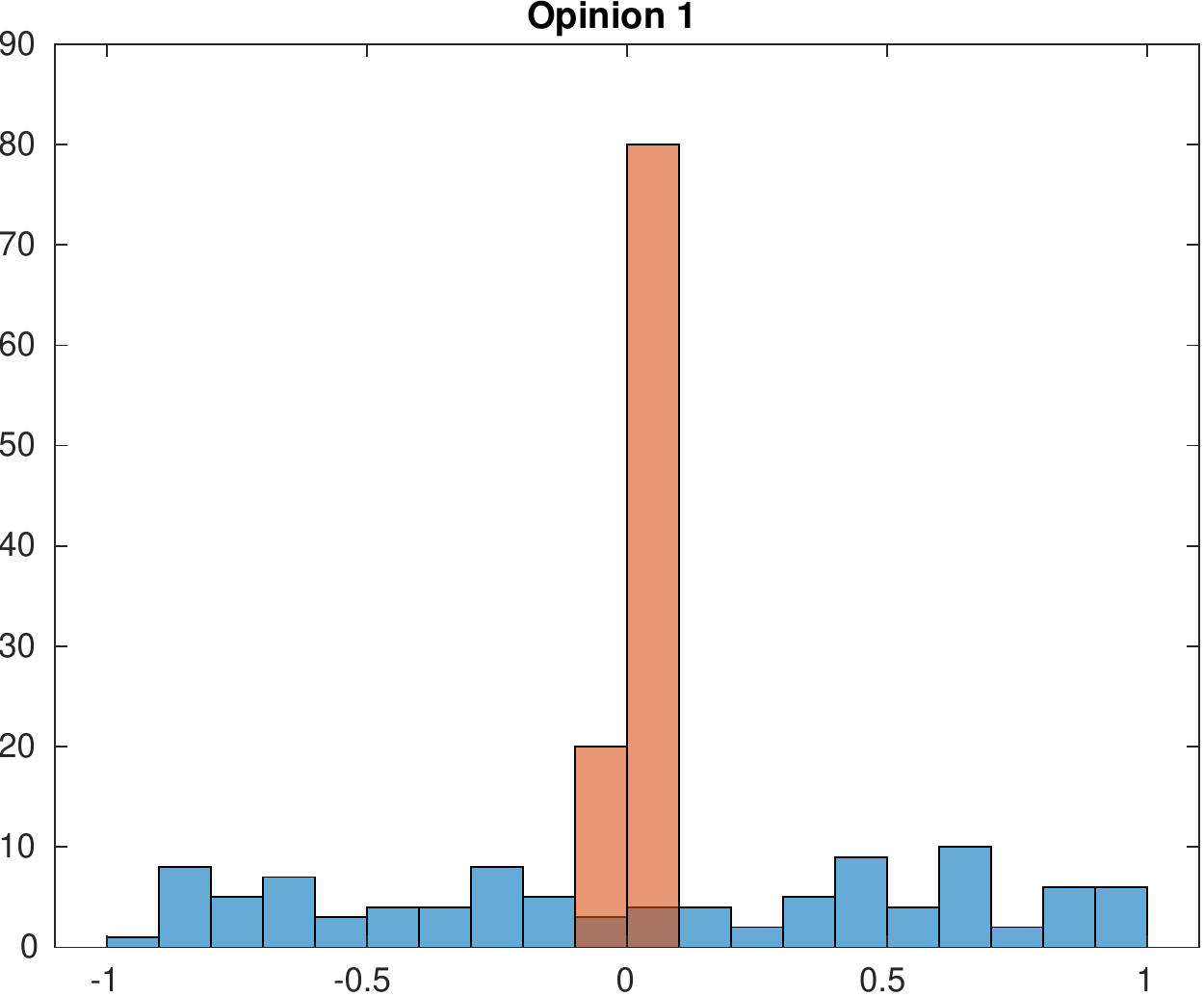}
\includegraphics[width=0.42\textwidth]{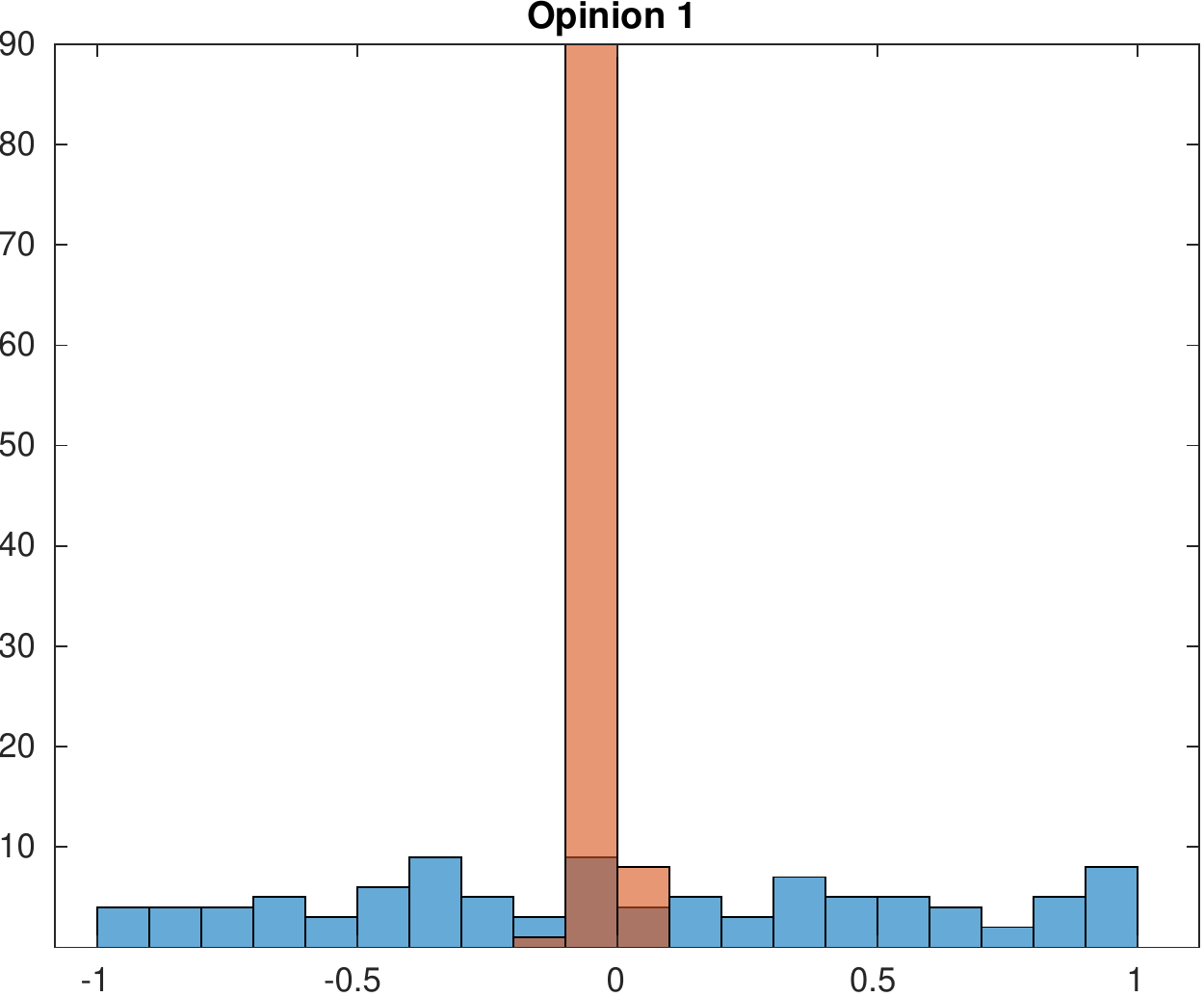}
\caption{Test 3 --- Contrarians vs Majority. Contrarians percentage $5\%$(left) and $25\%$(right) for opinion 1. The blue bars represent the initial data, while the orange ones the solution at time $T$.}
\label{fig:T3_rand}
\end{figure*}

Let us now change the initial data, choosing random values between $-1$ and $1$. In this case the opinion of the majority changes, almost, at each time step and the contrarians too. This leads to a consensus towards the central opinion in the presence of $25\%$ of contrarians, see Fig.\ \ref{fig:T3_rand}. Paradoxically, situations such as the ones already described show, instead, that the role of contrarians scaled down if they do not have an emerging opinion to oppose.

\subsection{The role of conviction}\label{sec:conv}
A crucial role in the opinion dynamics field is played by the conviction term, i.e. the strength of confidence on the opinion 1. In the previous tests we put $b_i=1$, meaning that each agent can change opinion due to the interactions with others. Let us recall that, given the agent $i$, $b_i$ expresses his/her conviction for opinion $x_{1,i}$: if $b_i=0$, agent $i$ is not willing to change his/her opinion; otherwise, if $b_i=1$, $i$ will show the maximum willingness in exchanging opinions. In the following, we assume that contrarians are the $15\%$ of the population and the coherence weights are $\alpha_{1,2}=0.75$ and $\alpha_{1,3}=0.25$.

First of all, we consider contrarians against everyone as in Section \ref{subsec:vsall}, assuming $b_i$ as a random value in $[0,0.3]$ for agents which are confident on their opinion and $b_i \in [0.7,1]$ for agents which are available in changing their opinion after interactions, for $i\in \mathcal N$. 

\begin{figure*}[!htb]
\centering
\includegraphics[width=0.32\textwidth]{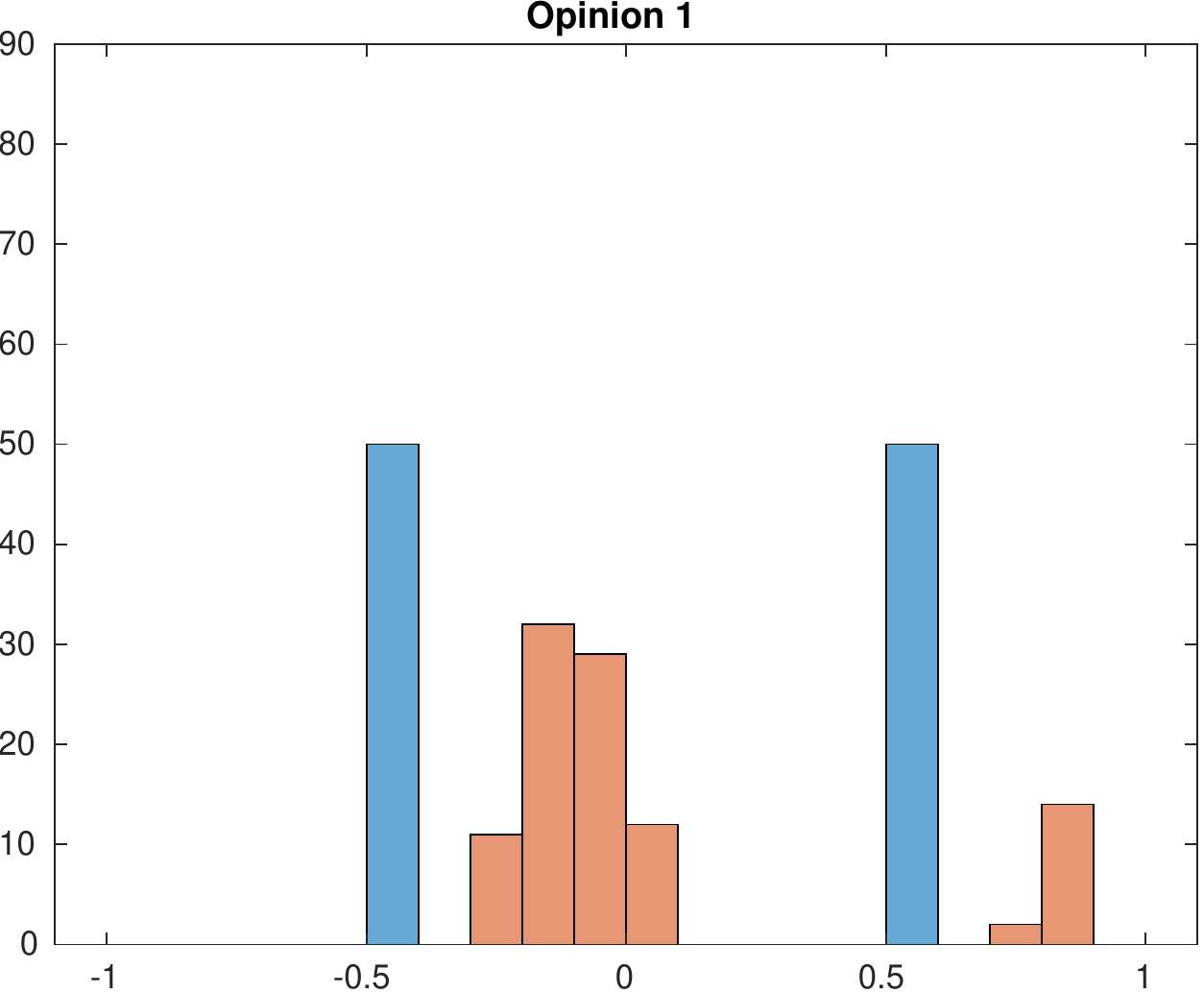}
\includegraphics[width=0.32\textwidth]{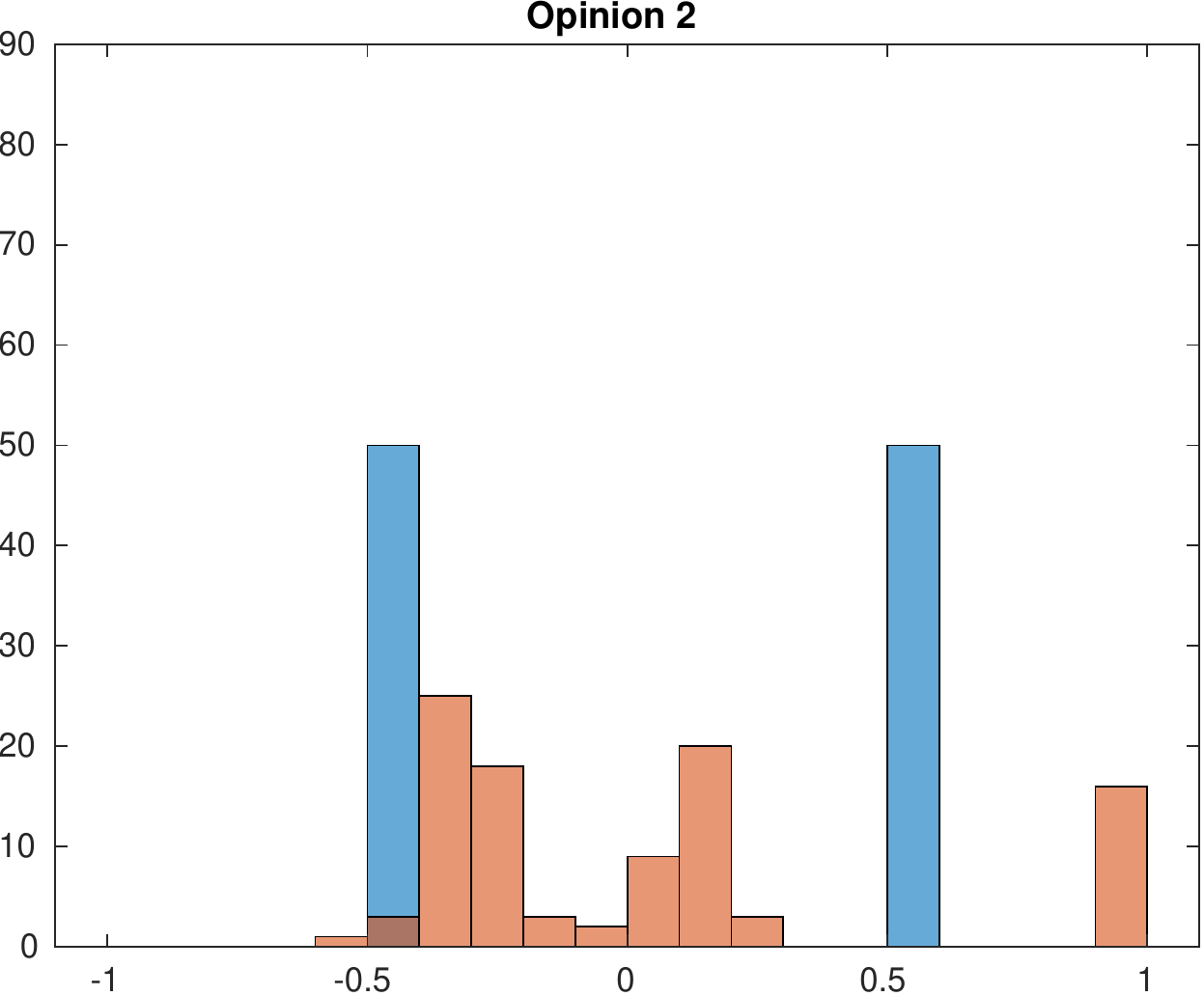}
\includegraphics[width=0.32\textwidth]{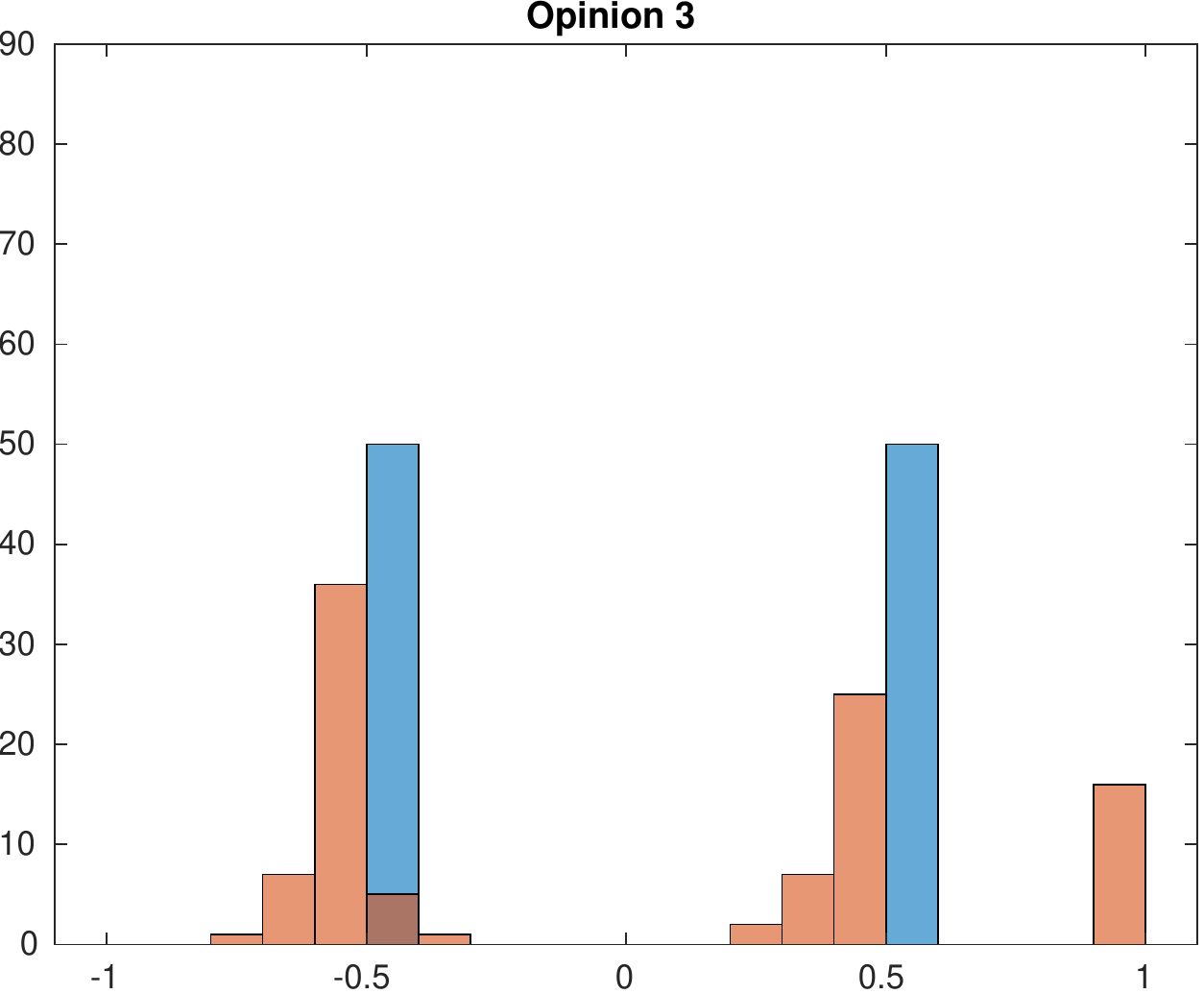}
\caption{Test 4 --- Contrarians vs All with convinced contrarians. Contrarians percentage $15\%$.The blue bars represent the initial data, while the orange ones the solution at time $T$.}
\label{fig:T4_c15_all_cont}
\end{figure*}

\begin{figure*}[!htb]
\centering
\includegraphics[width=0.32\textwidth]{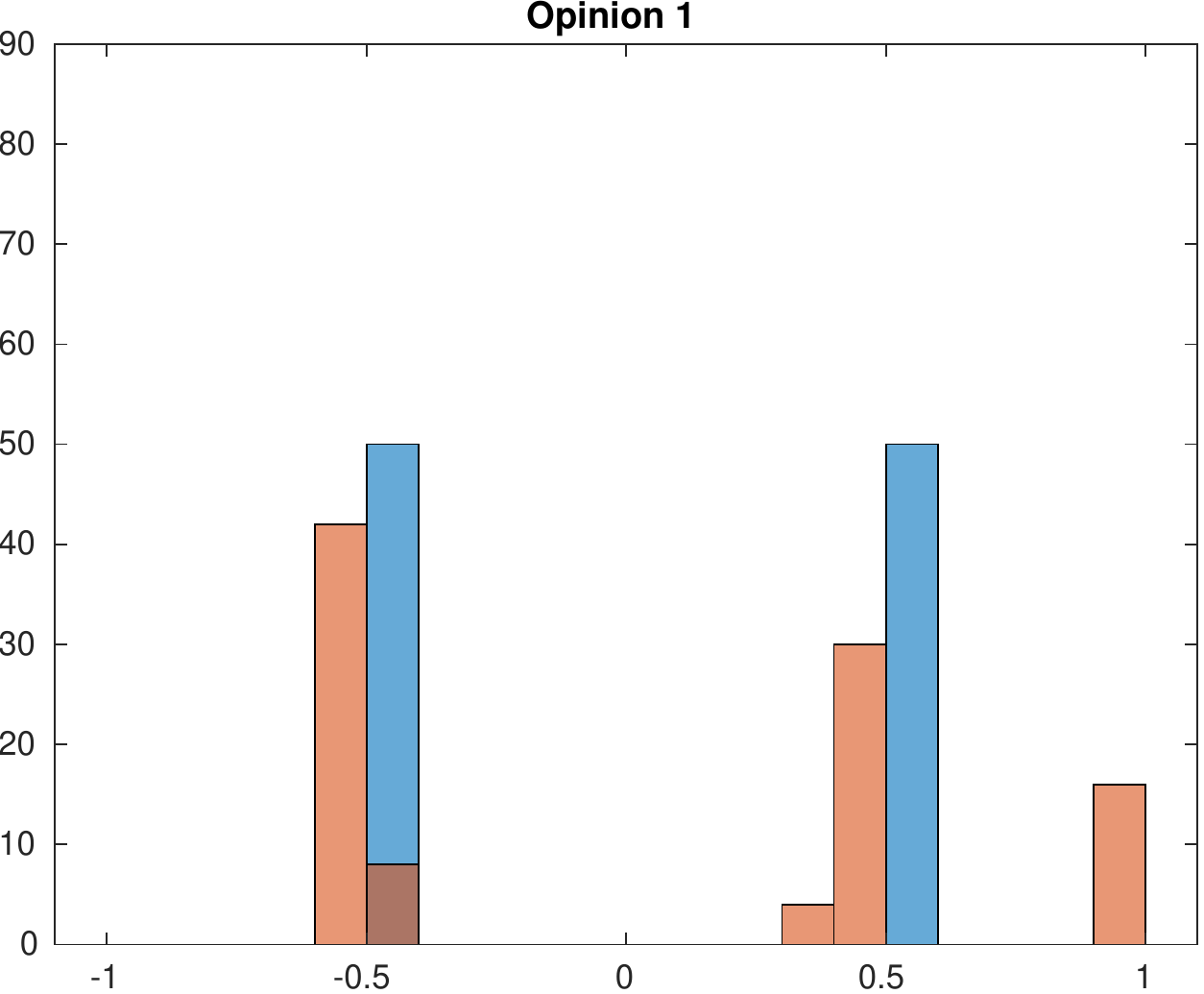}
\includegraphics[width=0.32\textwidth]{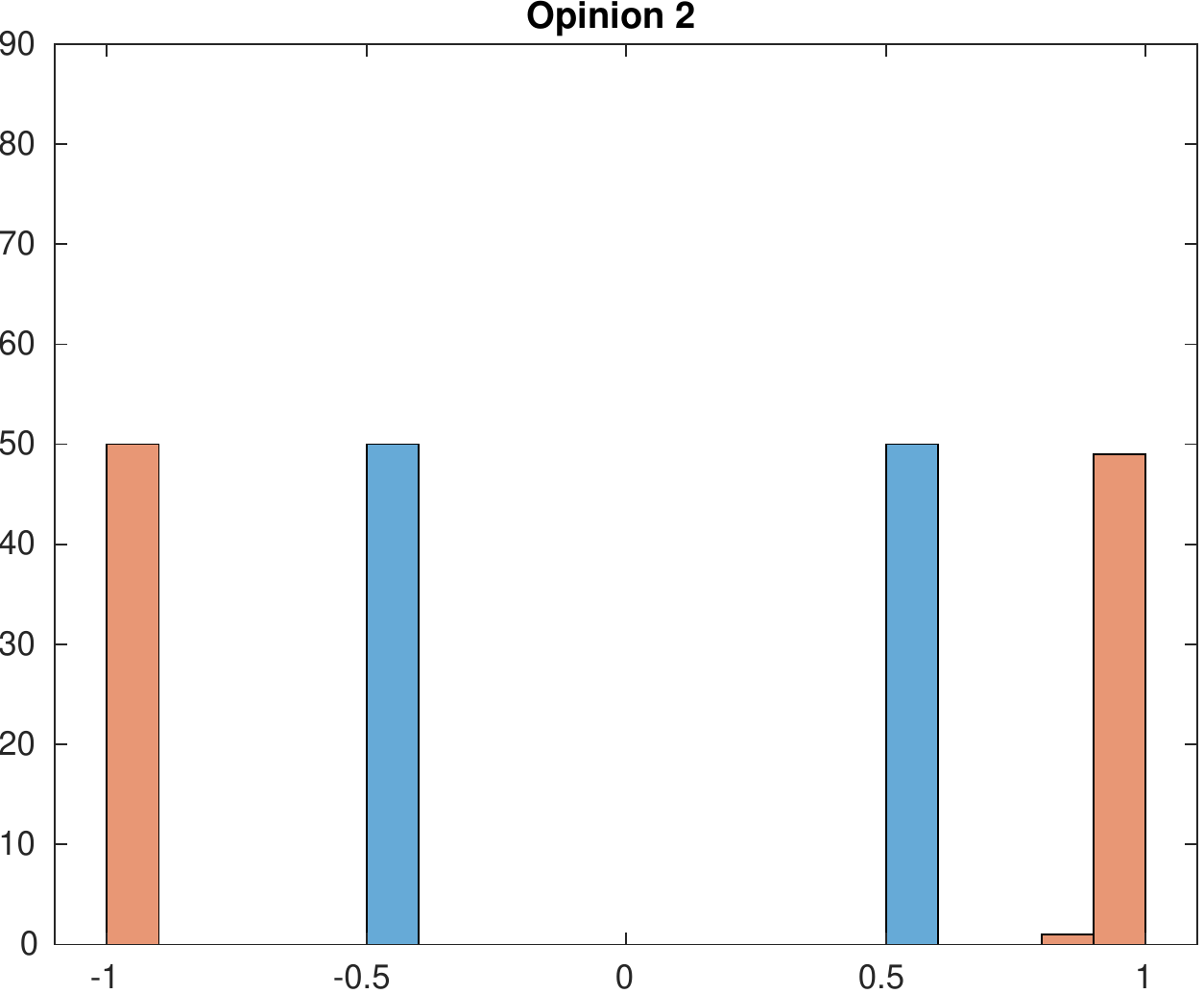}
\includegraphics[width=0.32\textwidth]{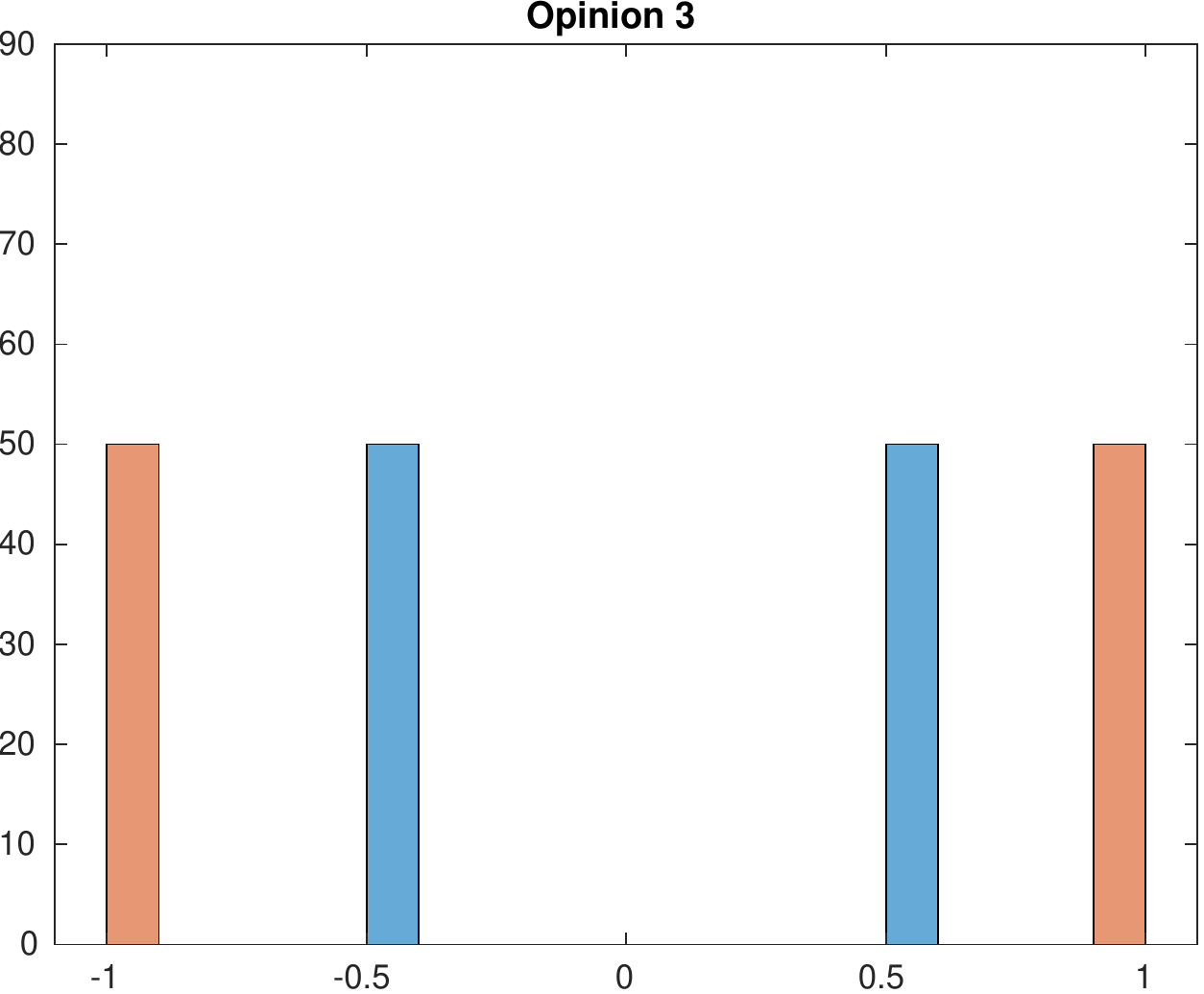}
\caption{Test 4 --- Contrarians vs All with convinced conformists. Contrarians percentage $15\%$.The blue bars represent the initial data, while the orange ones the solution at time $T$.}
\label{fig:T4_c15_all_conf}
\end{figure*}

\begin{figure*}[!htb]
\centering
\includegraphics[width=0.32\textwidth]{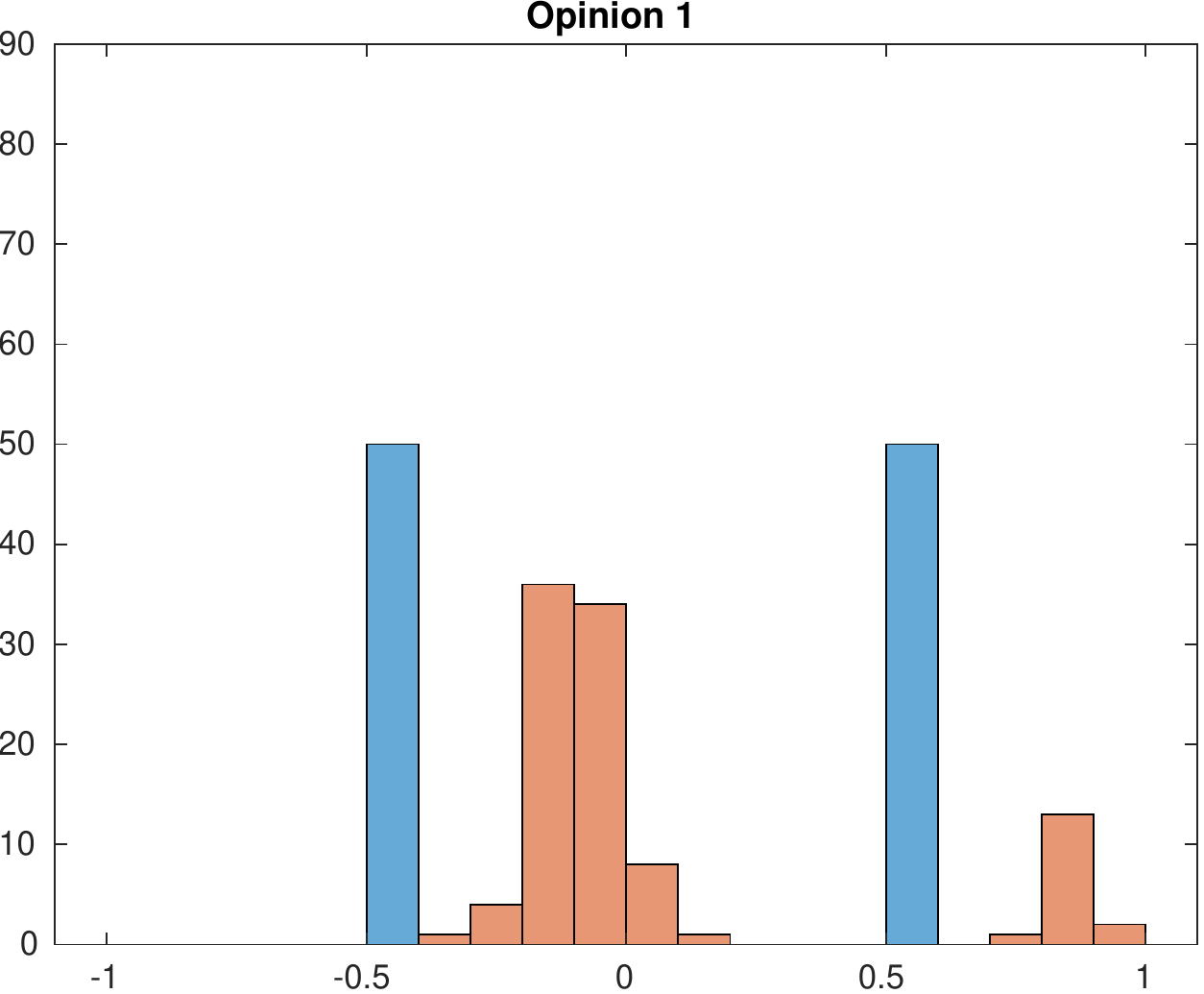}
\includegraphics[width=0.32\textwidth]{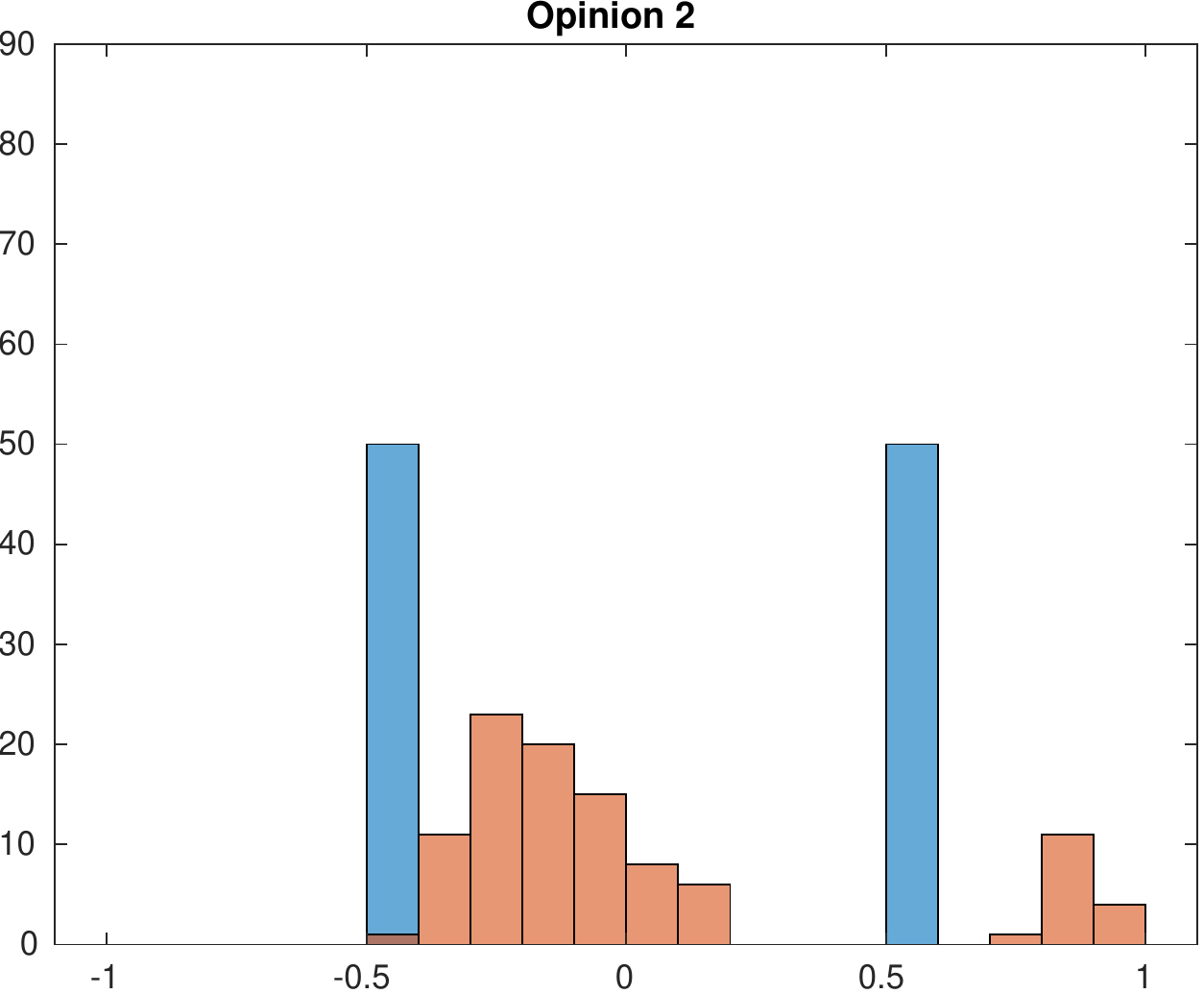}
\includegraphics[width=0.32\textwidth]{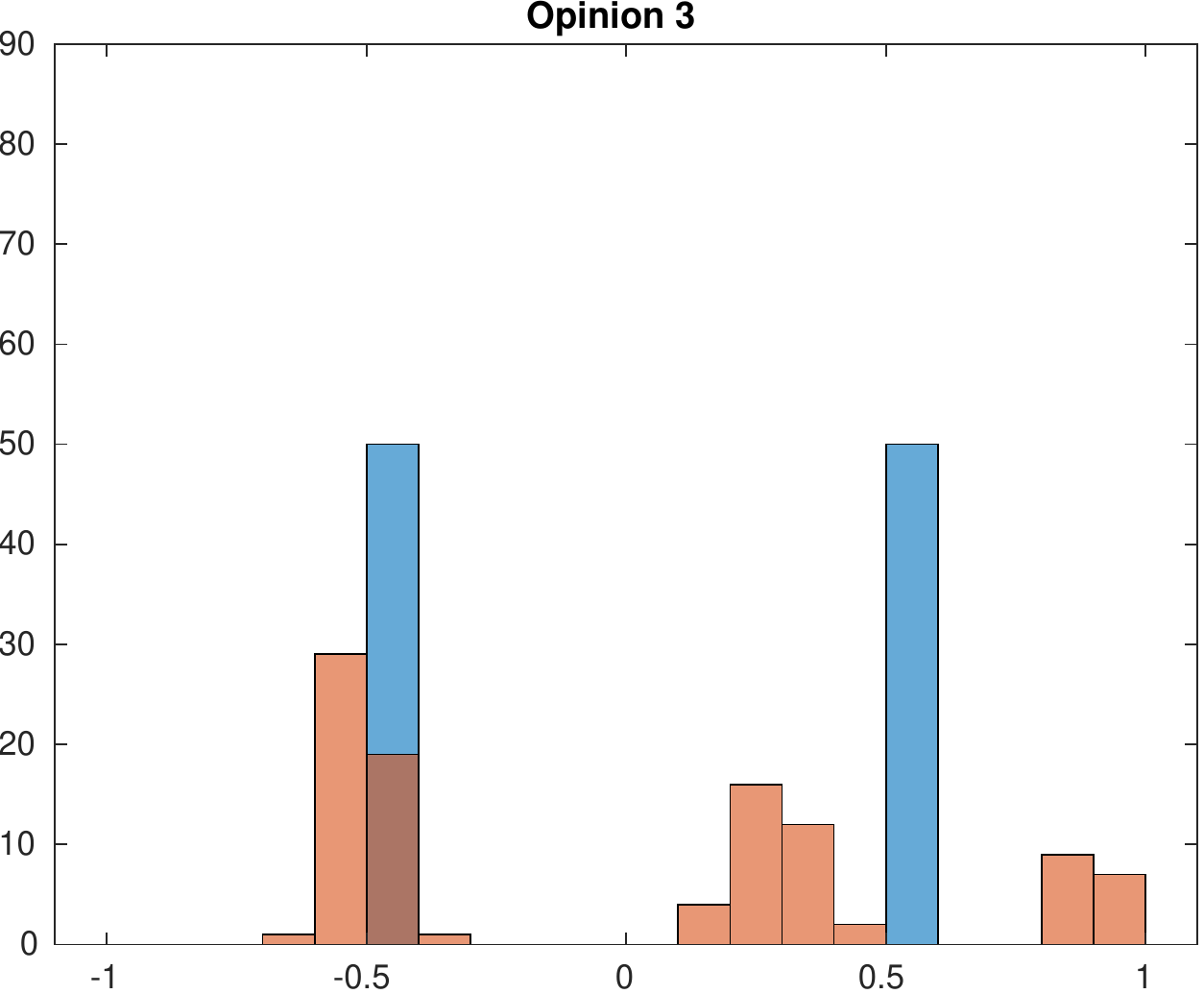}
\caption{Test 4 --- Contrarians vs All without conviction. Contrarians percentage $15\%$.The blue bars represent the initial data, while the orange ones the solution at time $T$.}
\label{fig:T4_c15_all_noconv}
\end{figure*}

If contrarians manifest high strength of confidence (and then $b$ is close to $0$), the opinion 1 shows a behaviour similar to the case of $b_i=1$ for all the population, as shown in Fig.\ \ref{fig:T4_c15_all_cont}-\ref{fig:T4_c15_all_noconv}.

On the other hand, looking at Fig.\ \ref{fig:T4_c15_all_conf}, where the conformist agents assume small values of $b_i$, all the components of the mind vector are contrasting from what we have seen so far. Indeed opinion 2 and opinion 3 are segregated towards the extreme values.


\begin{figure*}[!htb]
\centering
\includegraphics[width=0.32\textwidth]{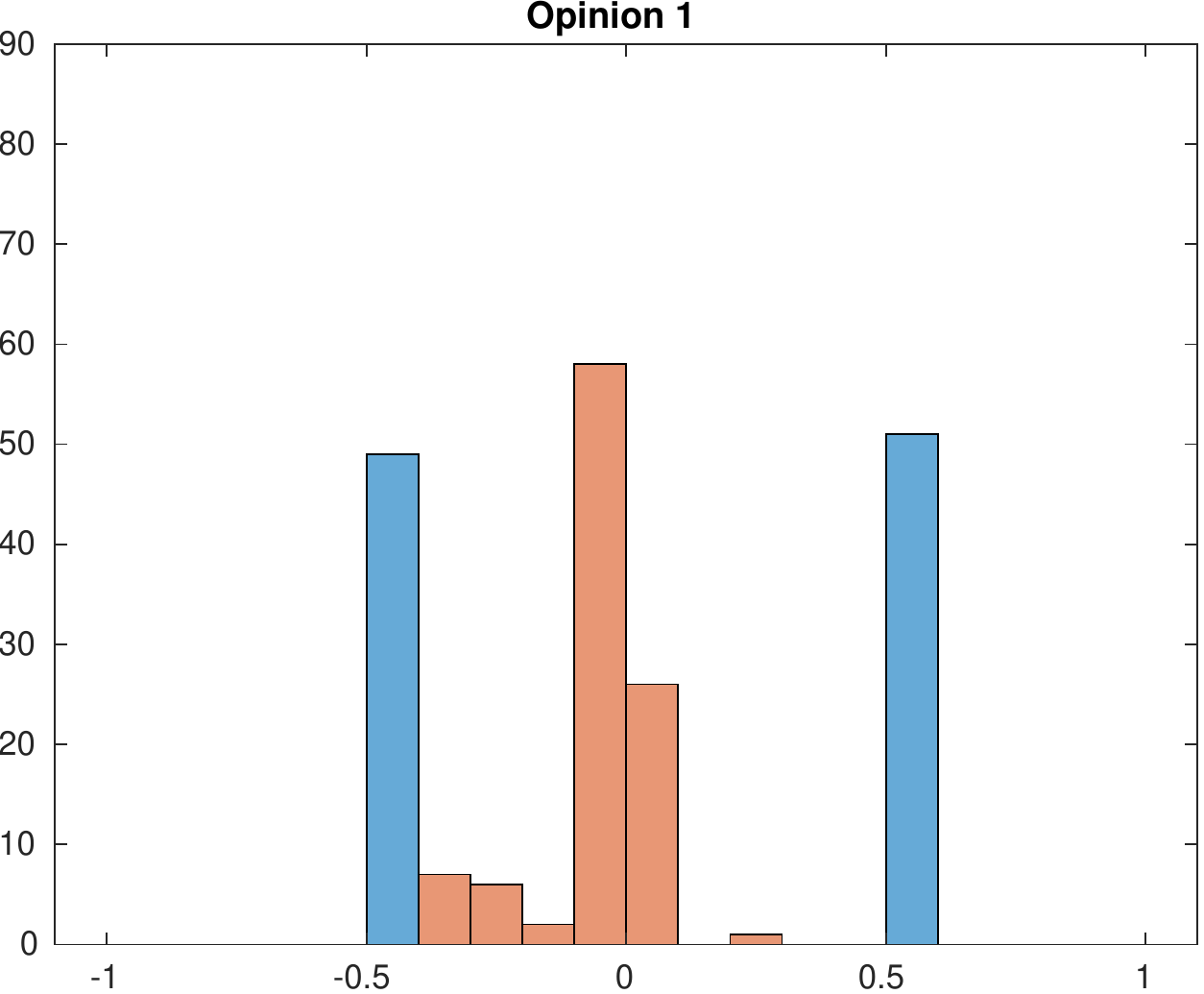}
\includegraphics[width=0.32\textwidth]{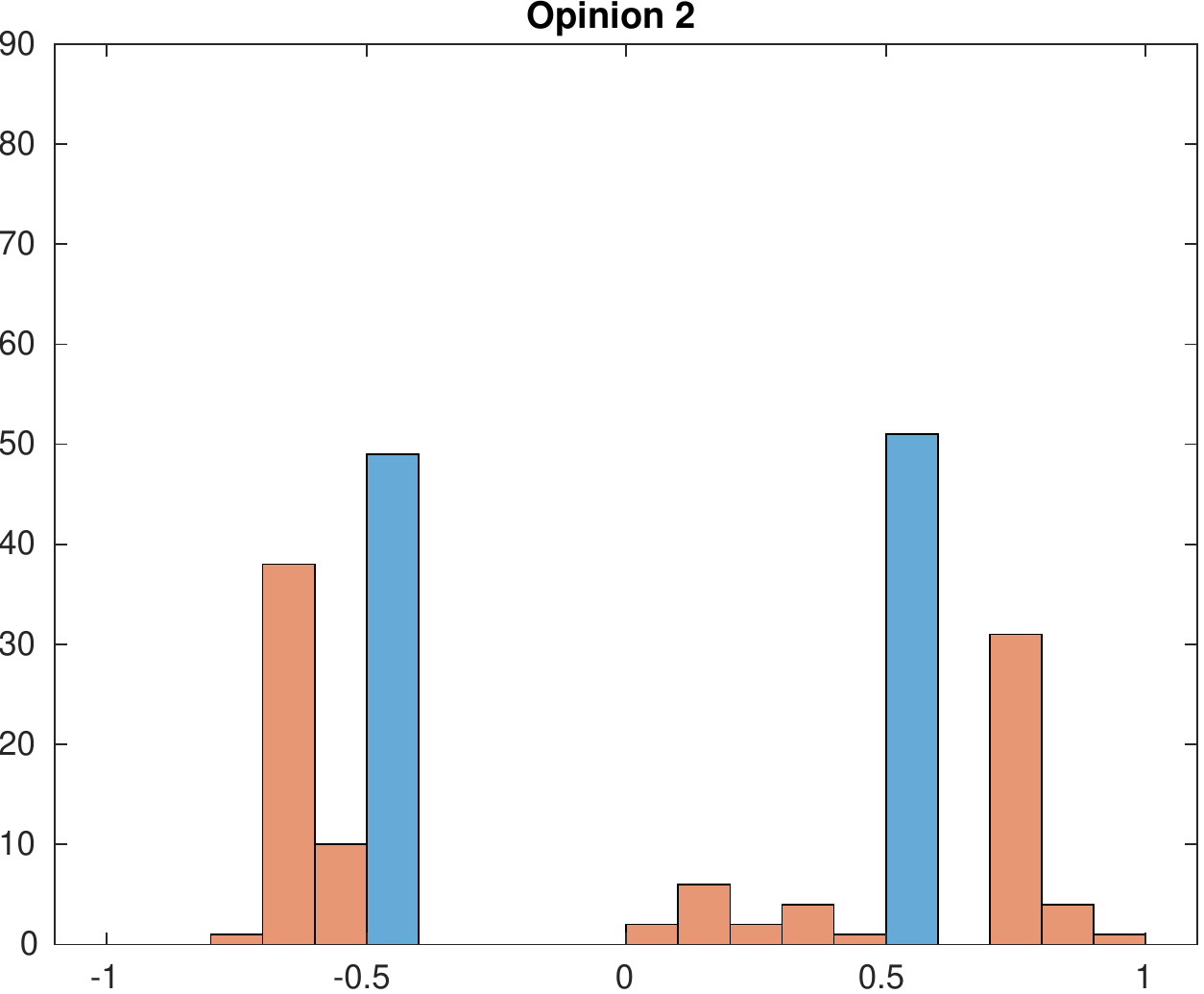}
\includegraphics[width=0.32\textwidth]{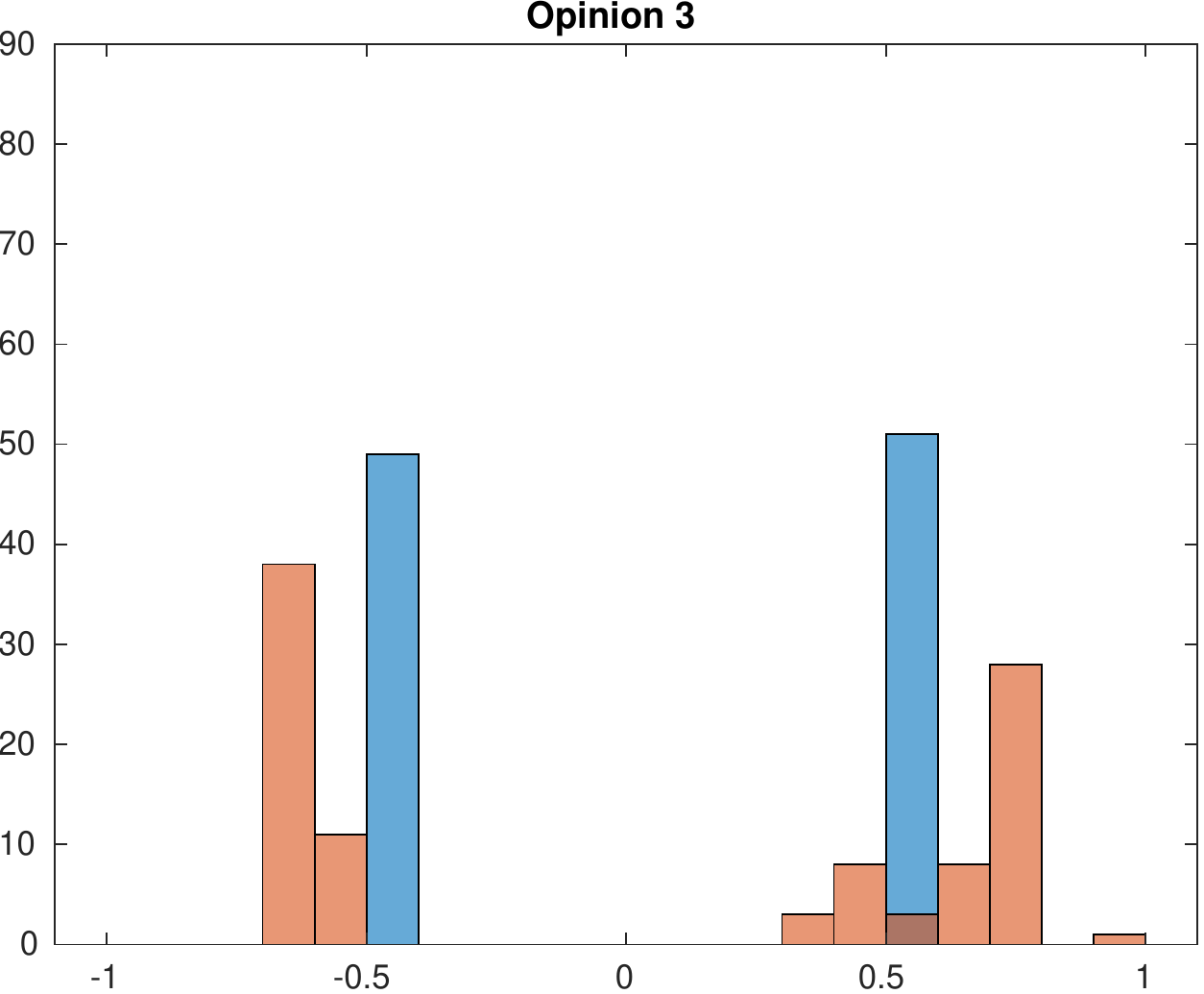}
\caption{Test 4 --- Contrarians vs Majority with convinced contrarians. Contrarians percentage $15\%$.The blue bars represent the initial data, while the orange ones the solution at time $T$.}
\label{fig:T4_c15_maj_cont}
\end{figure*}

\begin{figure*}[!htb]
\centering
\includegraphics[width=0.32\textwidth]{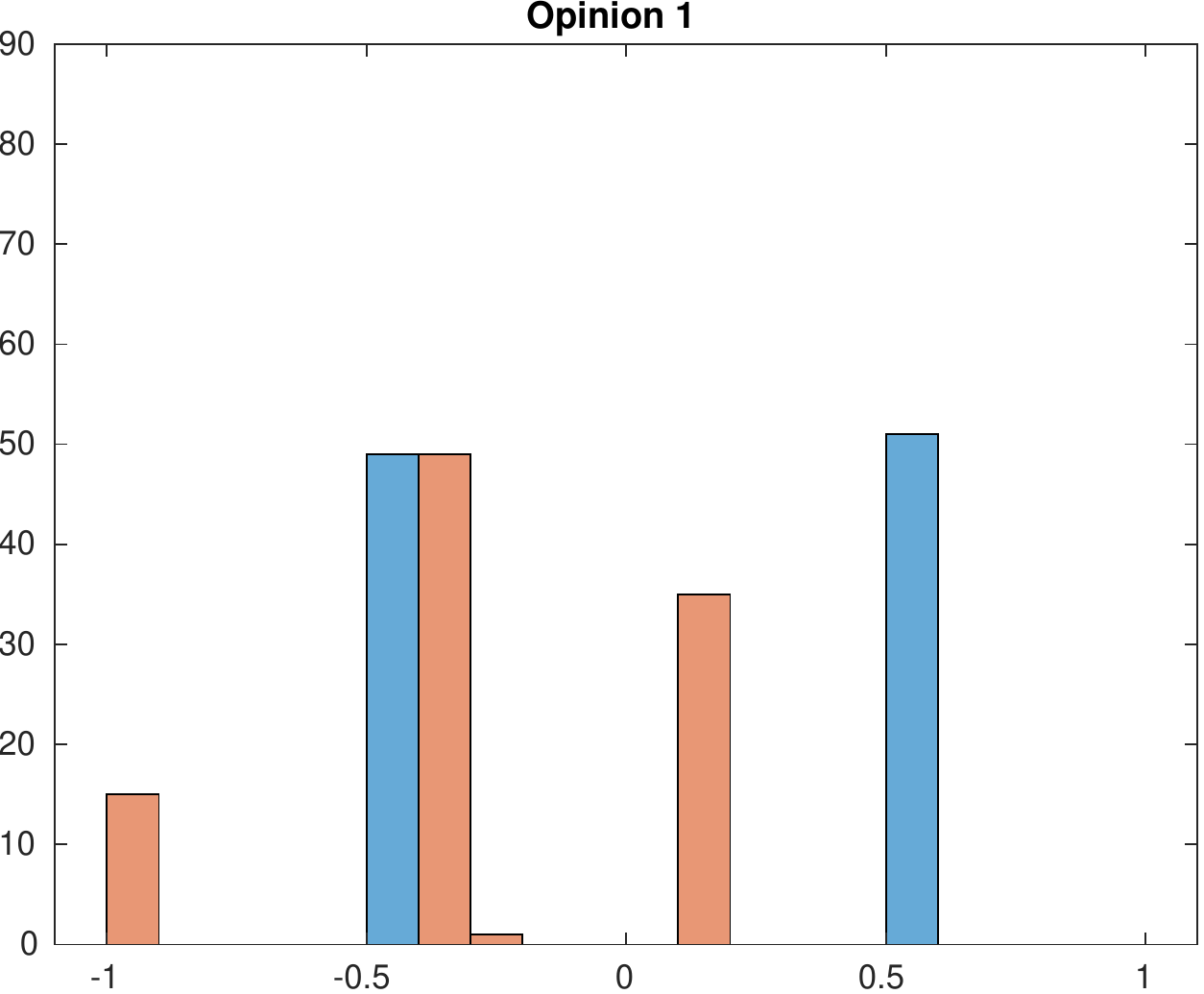}
\includegraphics[width=0.32\textwidth]{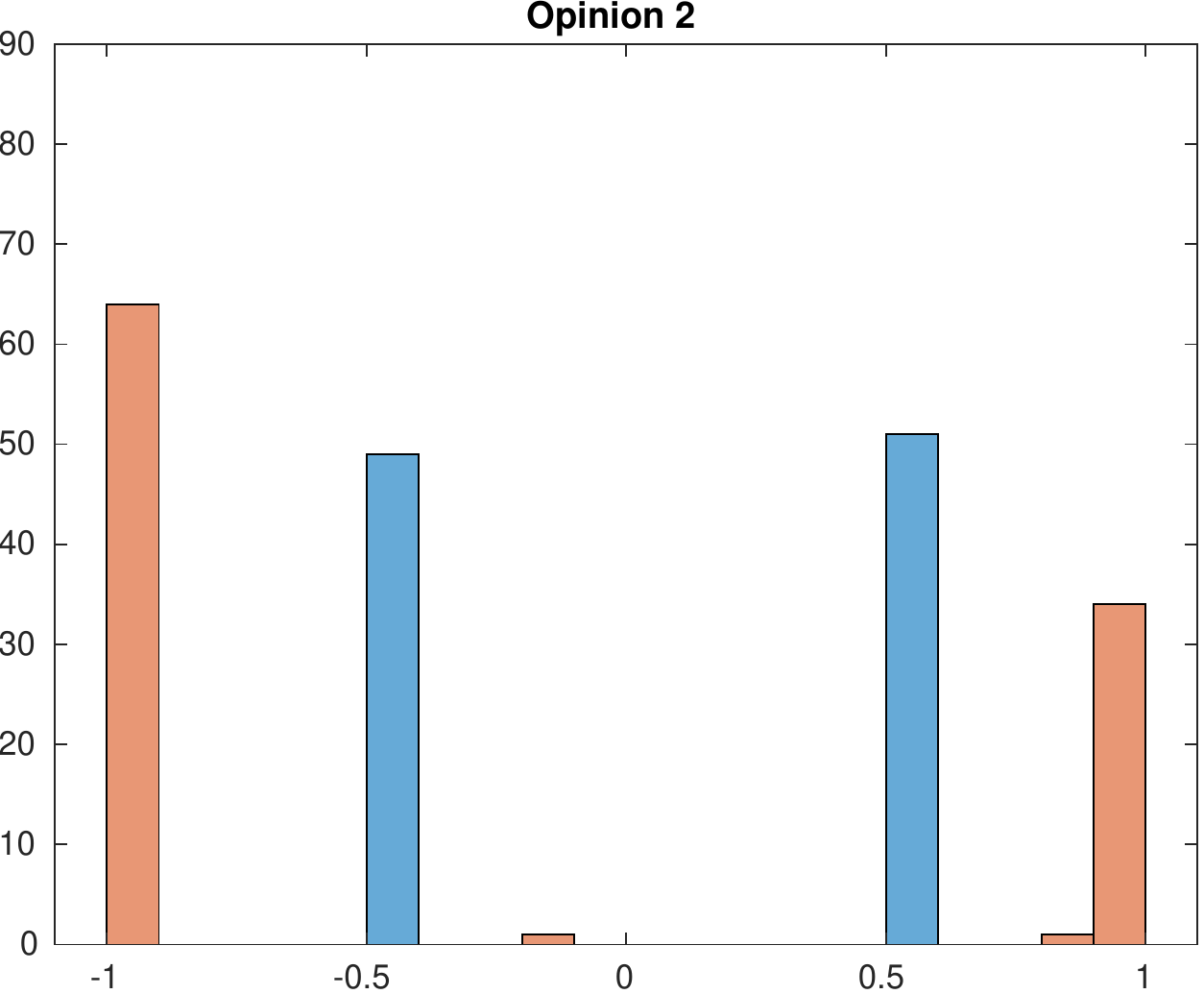}
\includegraphics[width=0.32\textwidth]{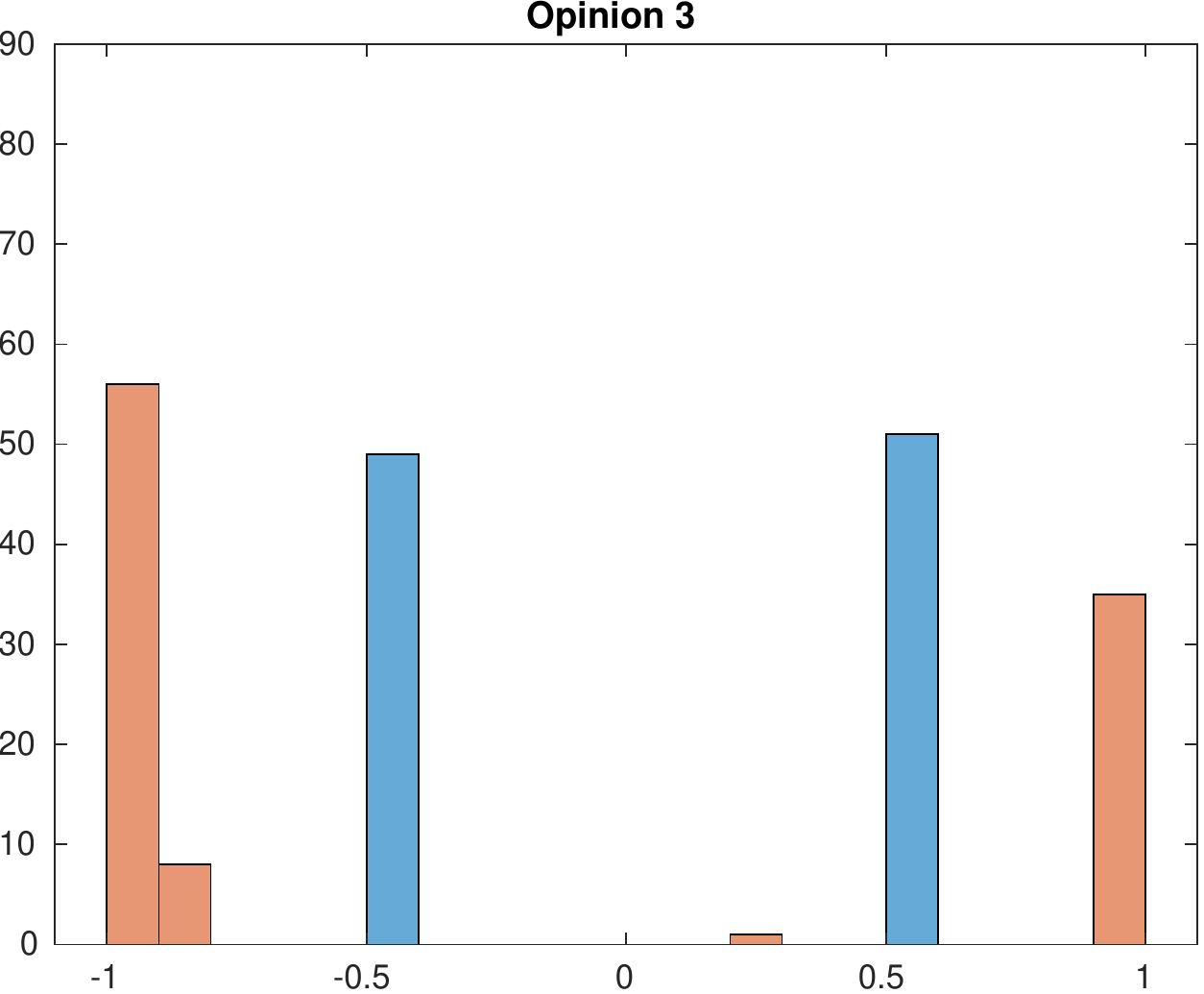}
\caption{Test 4 --- Contrarians vs Majority with convinced conformists. Contrarians percentage $15\%$.The blue bars represent the initial data, while the orange ones the solution at time $T$.}
\label{fig:T4_c15_maj_conf}
\end{figure*}

As far as concern the contrarians against the opinion of the majority, we see again a segregation on opinions 2 and 3 for convinced conformists (i.e. conformists that manifest high strength of confidence and then near-zero values for $b$) in Fig.\ \ref{fig:T4_c15_maj_conf}. Moreover looking at Fig.\ \ref{fig:T4_c15_maj_cont}, we observe that for convinced contrarians the opinion 1 assumes values close to 0 while opinions 2 and 3 have a different behaviour.


\begin{figure*}[!htb]
\centering
\includegraphics[width=0.32\textwidth]{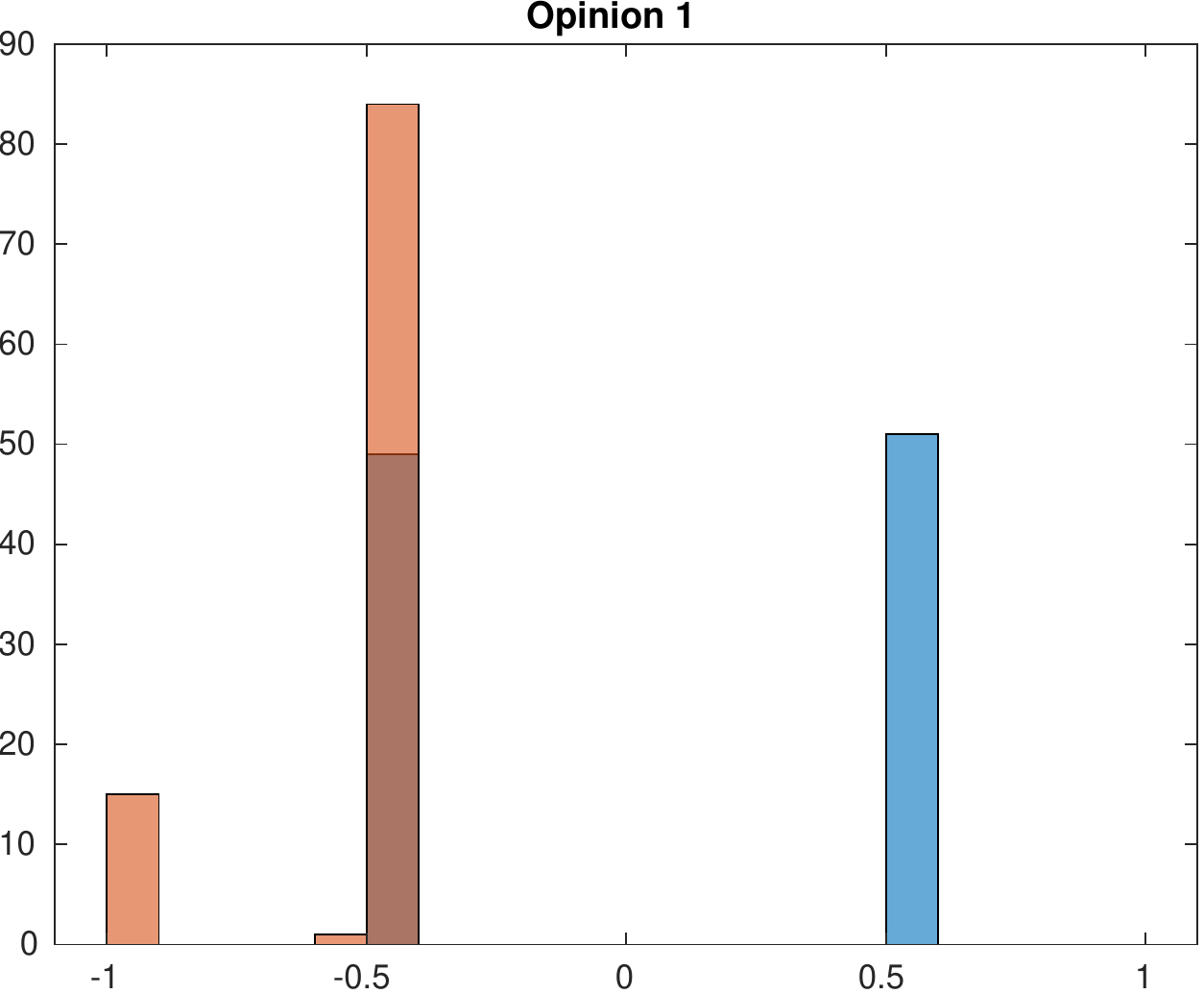}
\includegraphics[width=0.32\textwidth]{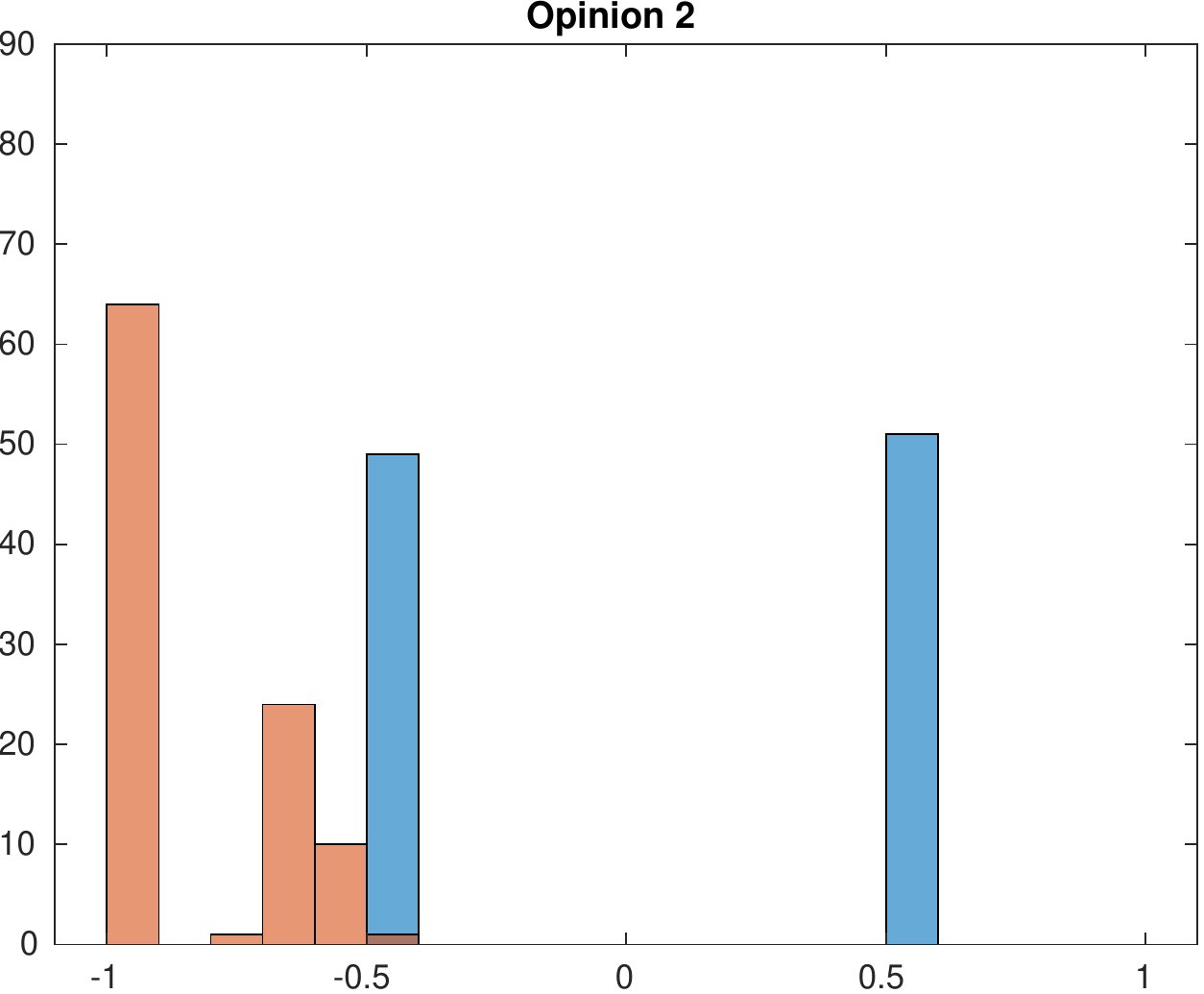}
\includegraphics[width=0.32\textwidth]{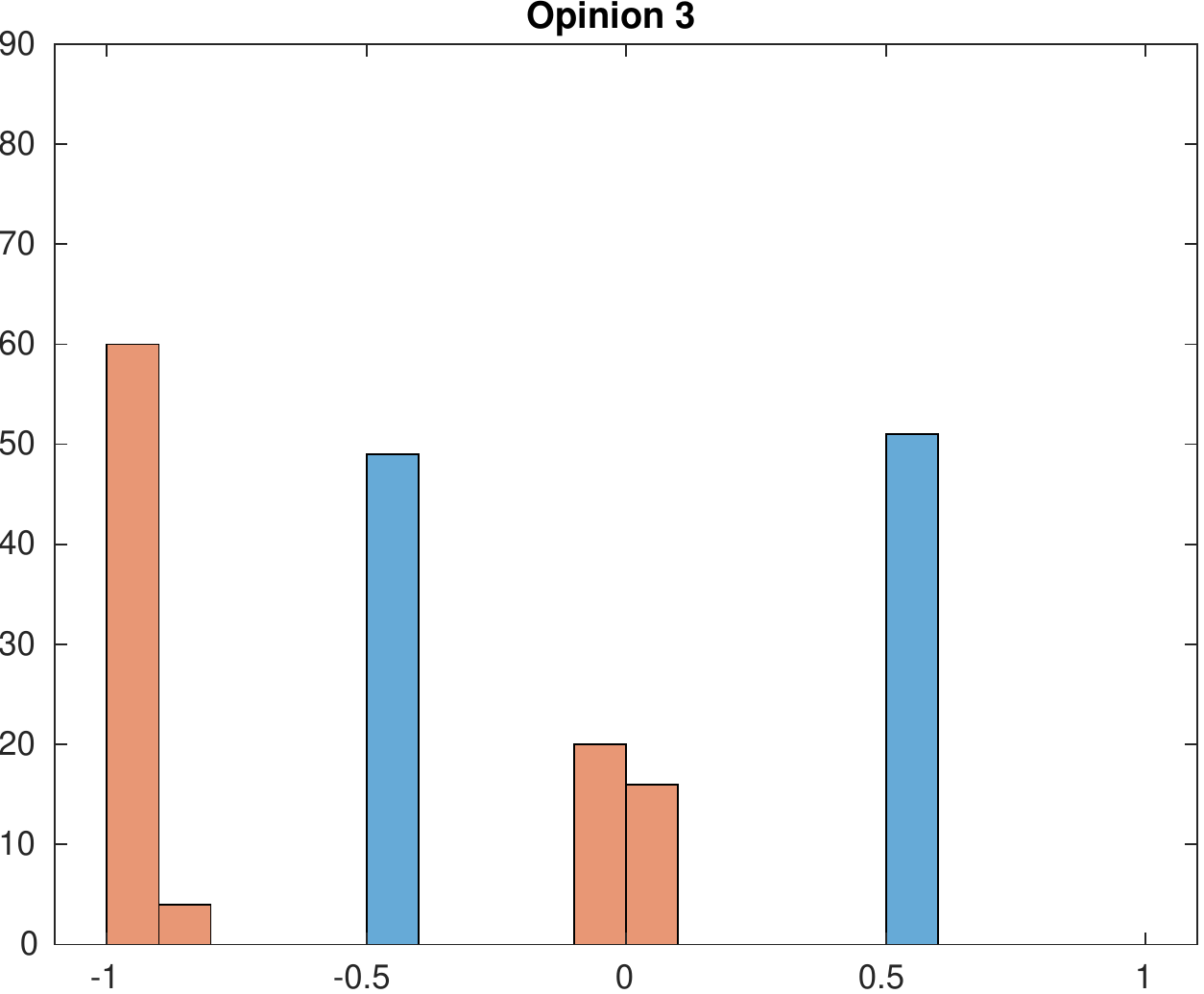}
\caption{Test 4 --- Contrarians vs Majority without conviction. Contrarians percentage $15\%$.The blue bars represent the initial data, while the orange ones the solution at time $T$.}
\label{fig:T4_c15_maj_noconv}
\end{figure*}

\begin{figure*}[!htb]
\centering
\includegraphics[width=0.32\textwidth]{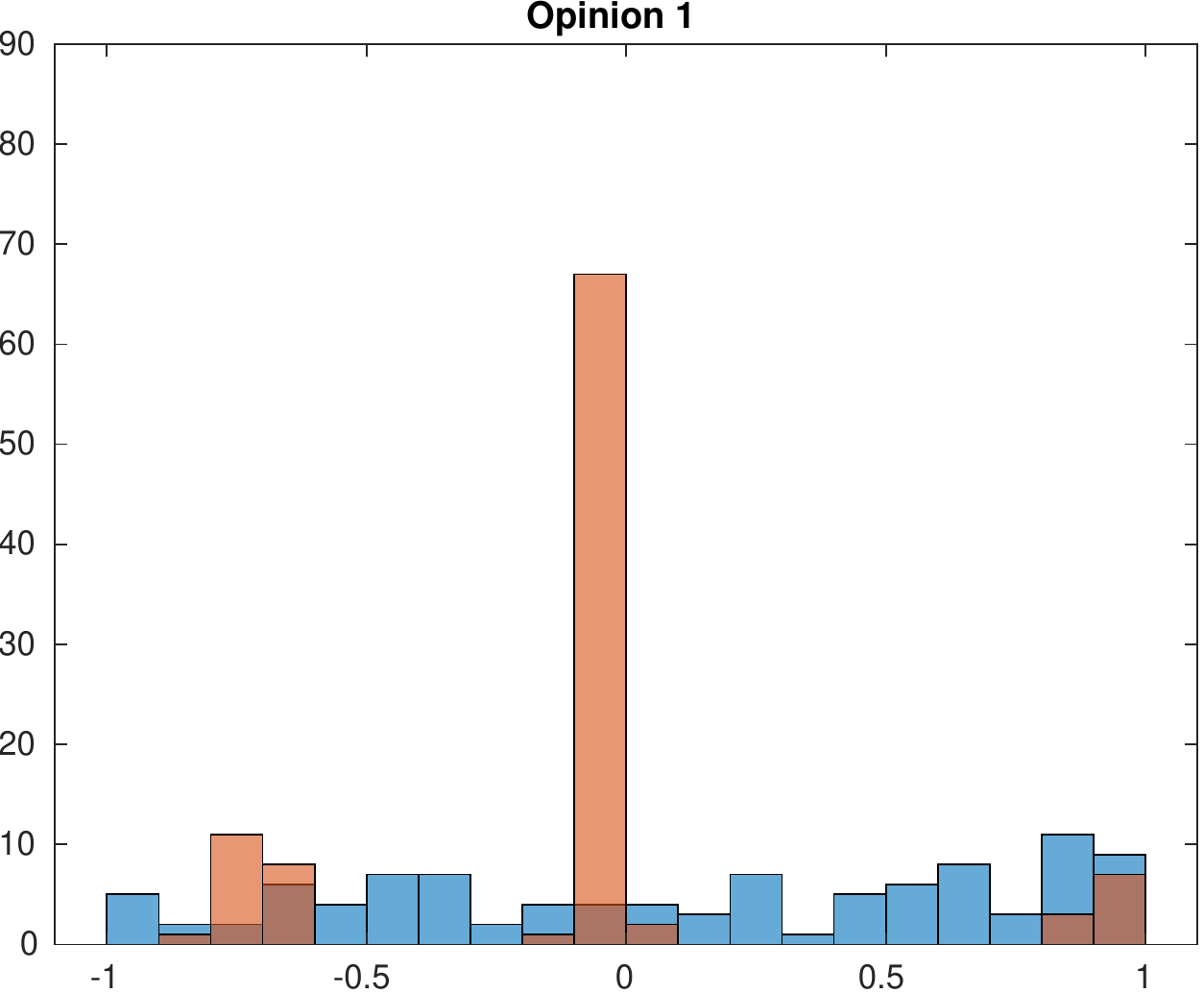}
\includegraphics[width=0.32\textwidth]{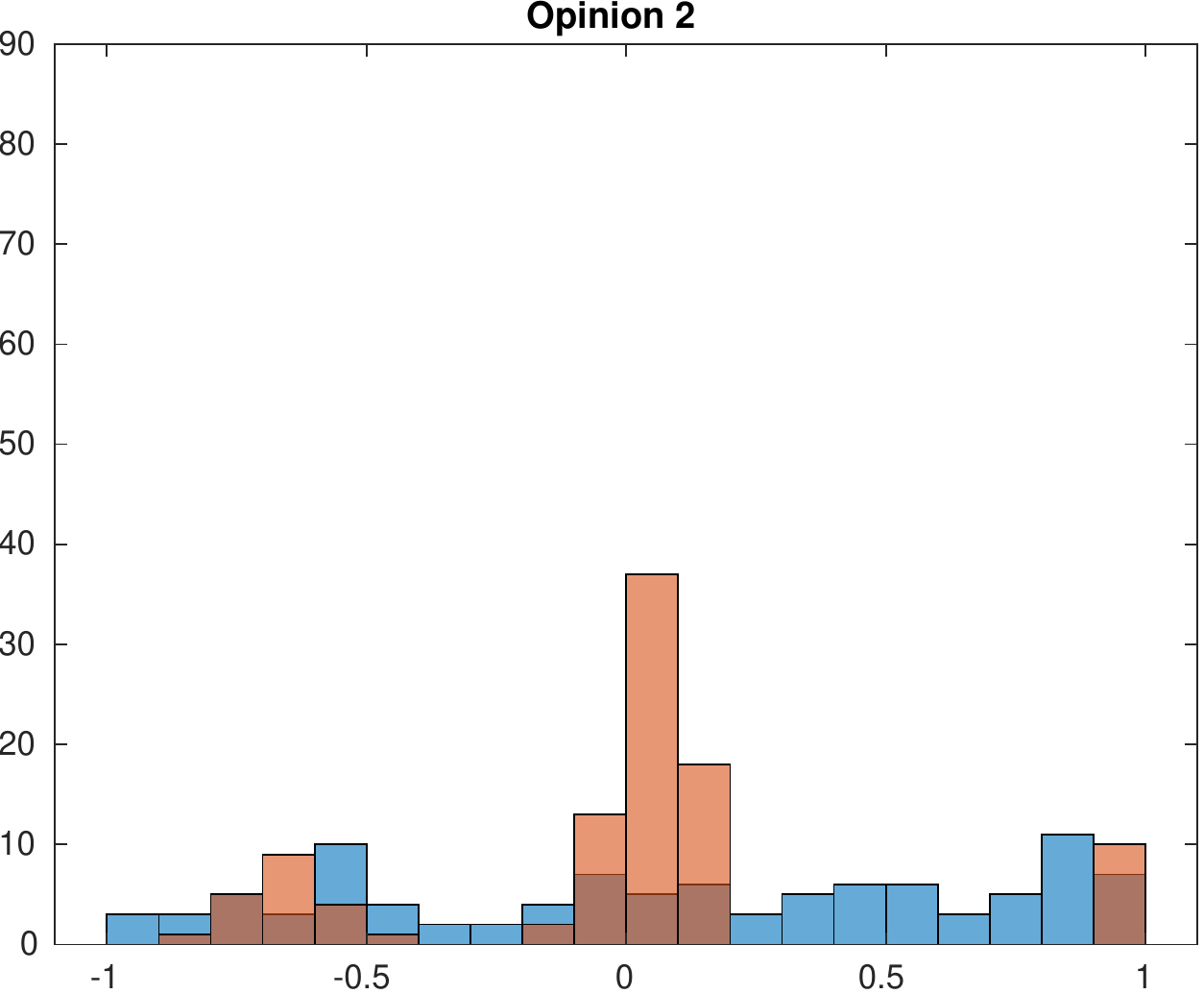}
\includegraphics[width=0.32\textwidth]{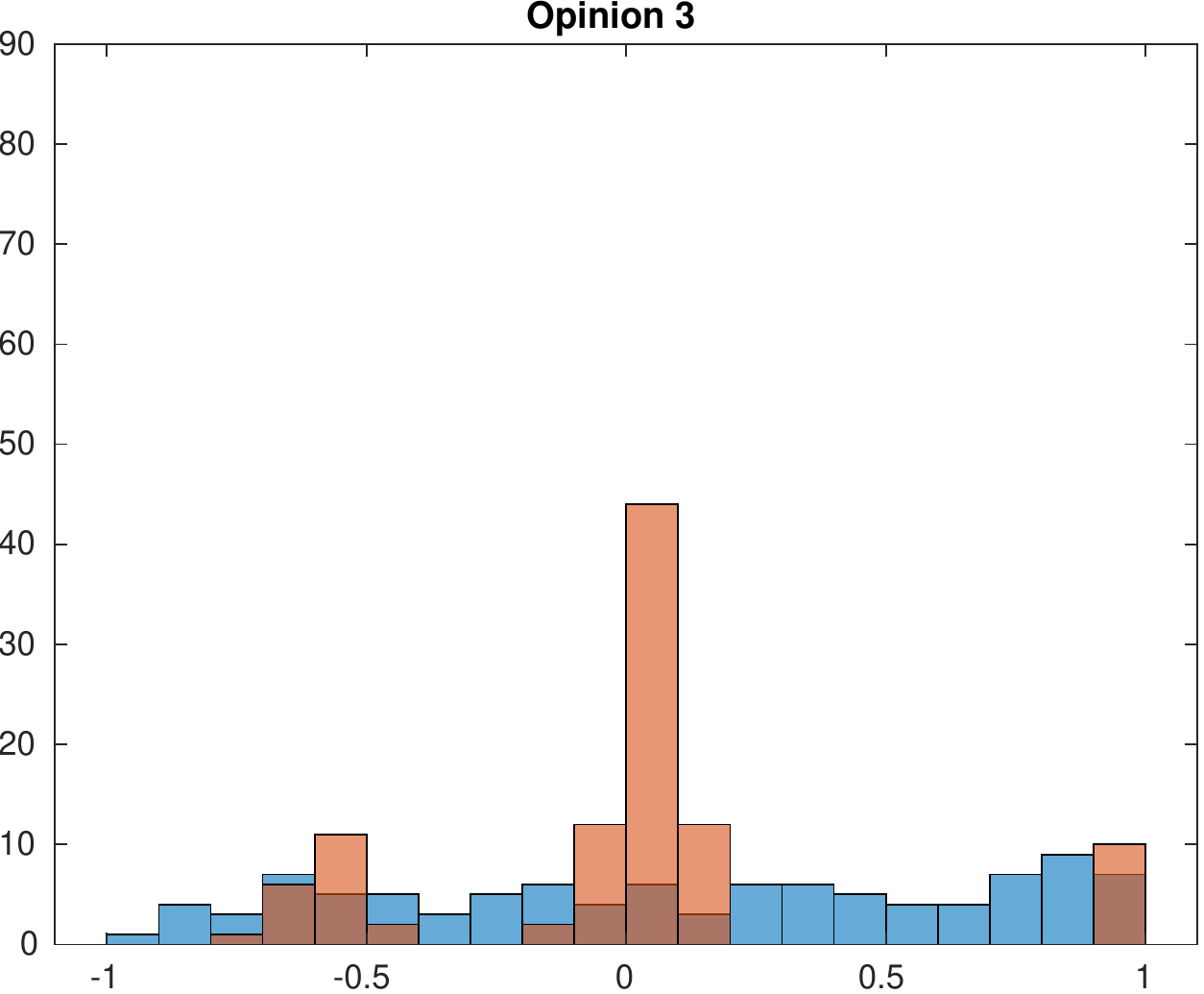}
\caption{Test 4 --- Contrarians vs Convicted agents with convinced contrarians. Contrarians percentage $15\%$.The blue bars represent the initial data, while the orange ones the solution at time $T$.}
\label{fig:T4_c15_lead_cont}
\end{figure*}
\begin{figure*}[!htb]
\centering
\includegraphics[width=0.32\textwidth]{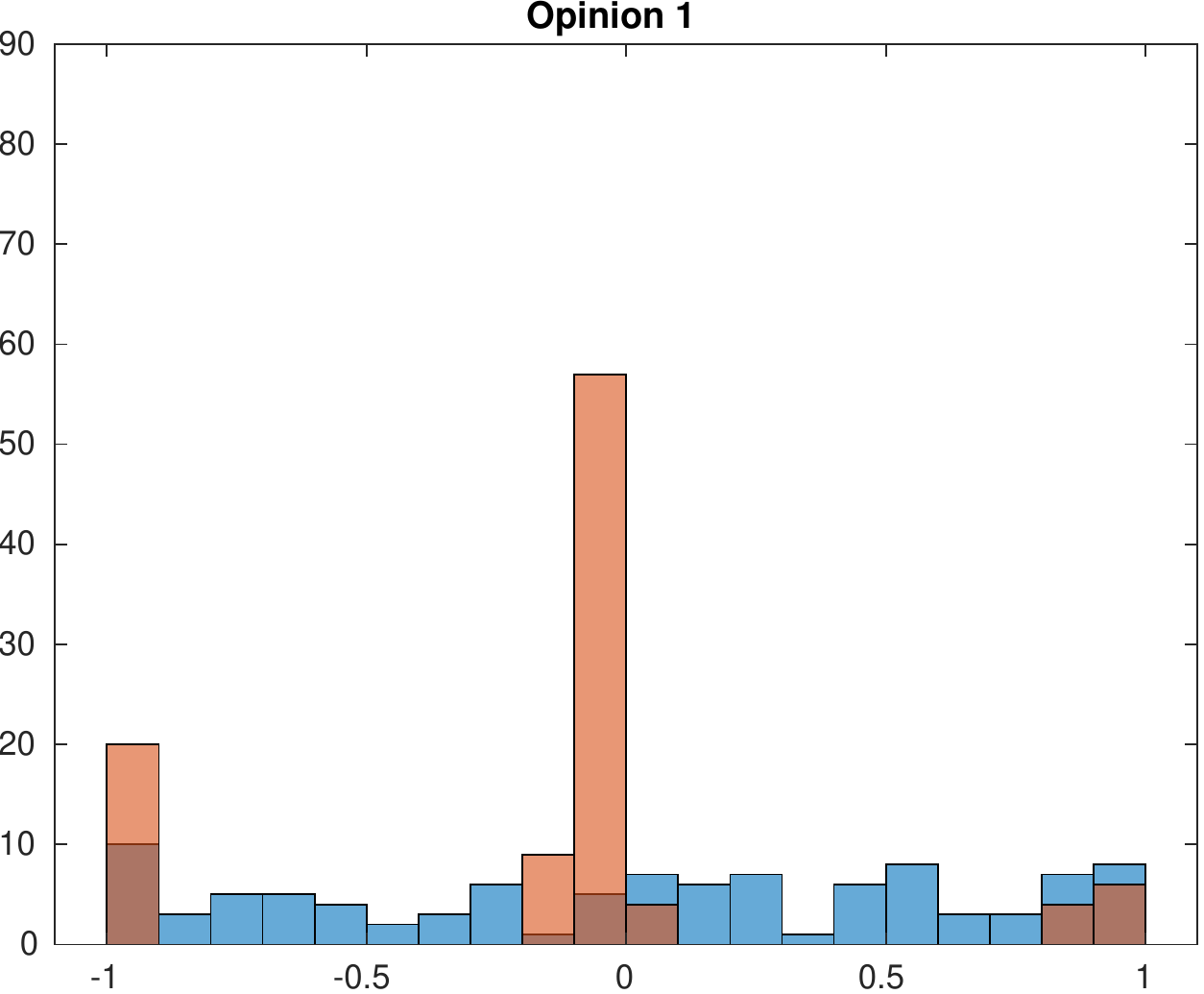}
\includegraphics[width=0.32\textwidth]{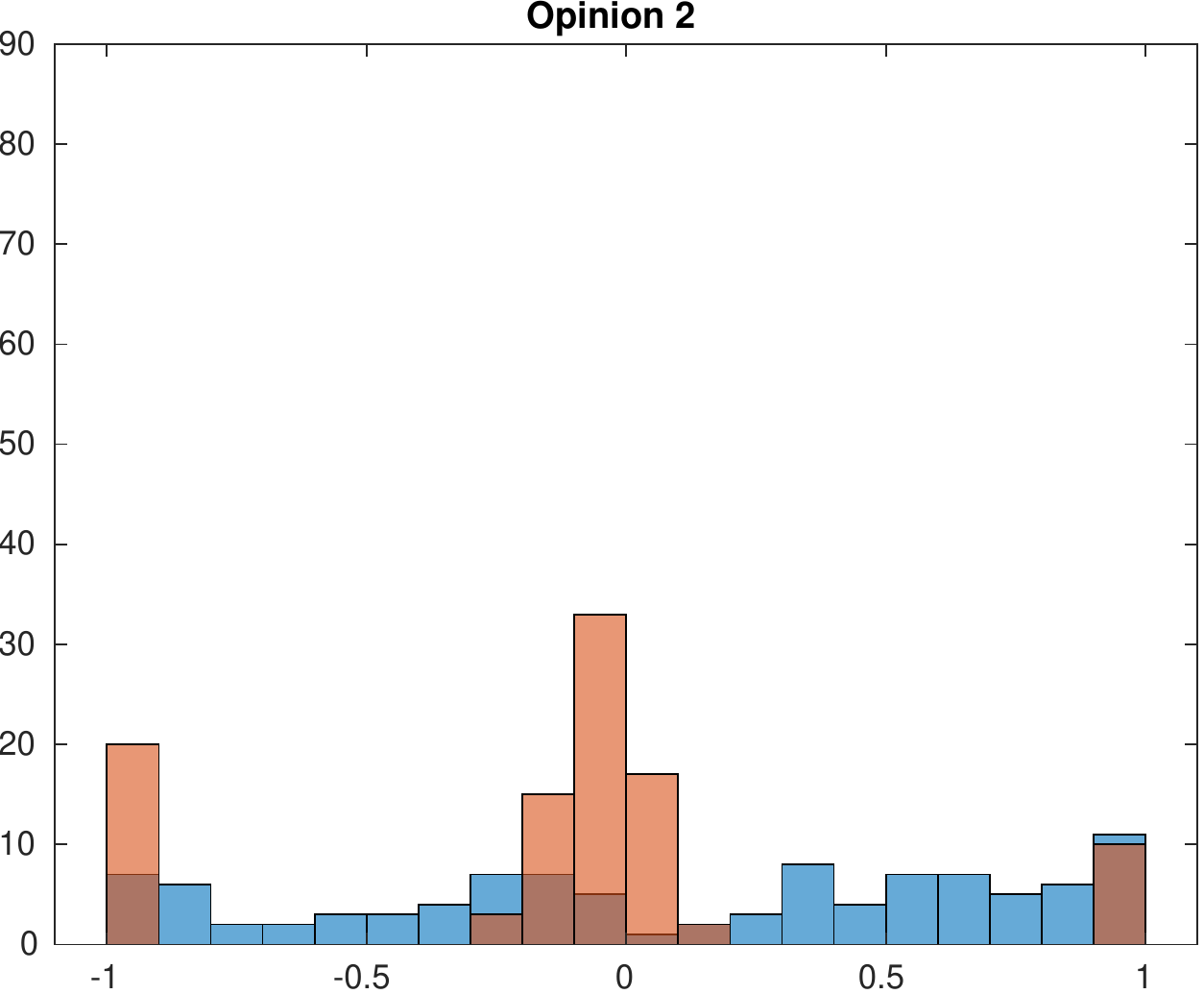}
\includegraphics[width=0.32\textwidth]{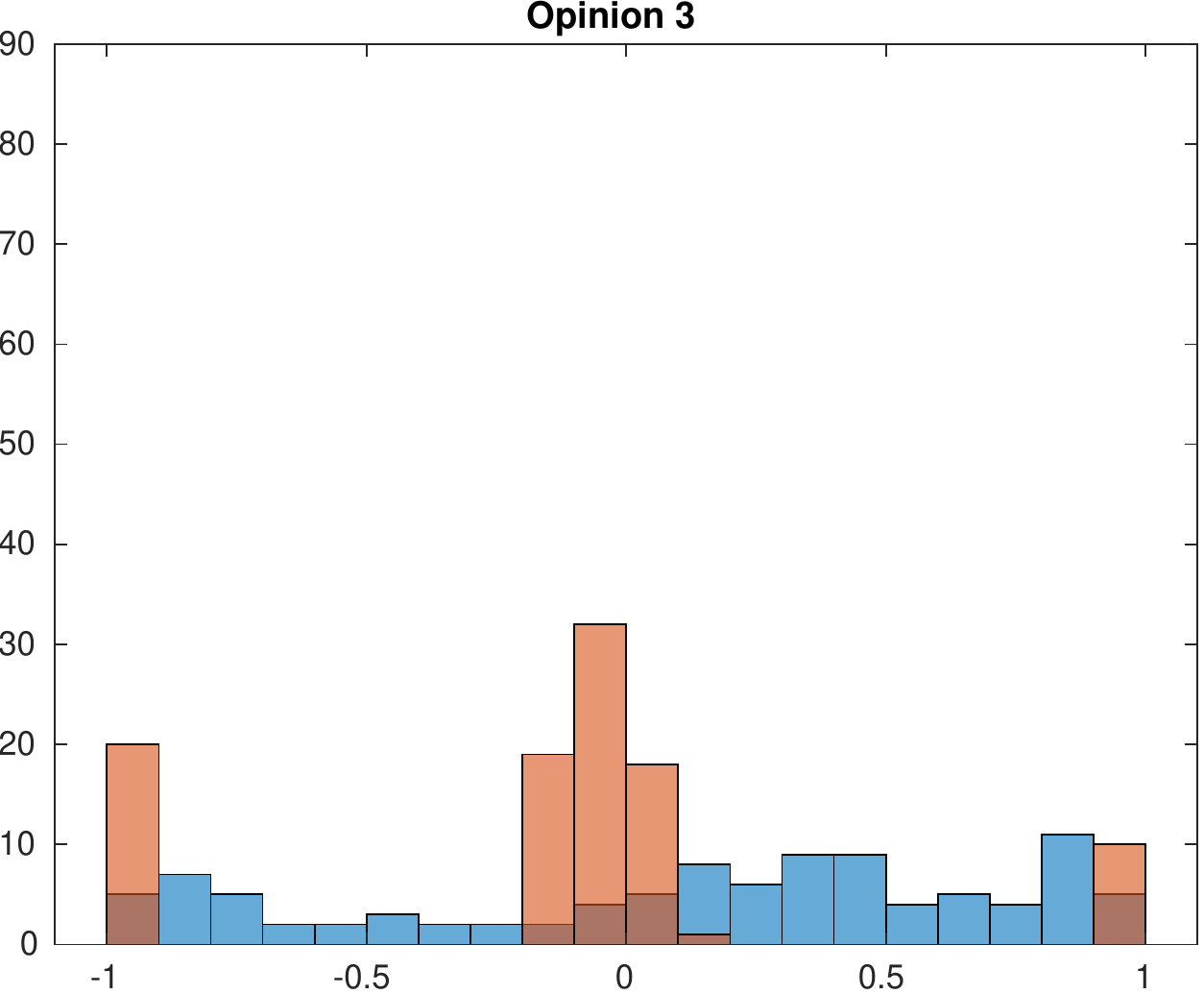}
\caption{Test 4 --- Contrarians vs Convicted agents with convinced conformists. Contrarians percentage $15\%$.The blue bars represent the initial data, while the orange ones the solution at time $T$.}
\label{fig:T4_c15_lead_conf}
\end{figure*}
In the case of fans --- see Section \ref{sec:vsconag} --- the impact of the coefficients $b_i\neq 1$ is less evident, since the contrarians agents have to assume opinions opposite to the members of fan club, which in turn do not change their opinions. Hence, effect of conviction term on opinion forming is less marked than in the other cases when the fan club is present, because in our model: (i) the fans do not change their idea (they have exactly $b=0$); (ii) the contrarians assume opinions opposite to fans, so opinion values do not change even for different values of $b$; (iii) the conformists adapt to others, and besides they can interact as much with fans as with contrarians.

\subsection{Test on feedback effect}
\label{test_hype}
In this test we focus on the influence that other topics may have on the evolution of the main one, indeed we discretize the Eq. \eqref{eq:feedback} in the case of $q=3$:
\begin{align}
    x^{n+1}_{1,i}&= x^{n}_{1,i}+ I(x^{n}_{1,j},x^{n}_{1,i})+\notag \\
    &+\alpha_{2,1}\operatorname{sgn}(x^n_{2,j})\frac{|x^n_{2,i}-\operatorname{sgn}(x^n_{2,i}x^n_{2,j})x^n_{2,j}|}{2}+ \notag \\
    &+\alpha_{3,1}\operatorname{sgn}(x^n_{3,j})\frac{|x^n_{3,i}-\operatorname{sgn}(x^n_{3,i}x^n_{3,j})x^n_{3,j}|}{2}+\notag \\
    &+\sqrt{2\mu}B^n_i
\end{align}
In this way we are able to generalize all the models seen so far. In particular we focus now on the fan club contrarians' type, see Section \ref{cont_vs_lead}.

As initial data we consider a uniformly distributed random values between $-1$ and $1$ and the coefficient $\alpha_{2,1}=0.08, \alpha_{3,1}=0.05$. For the other parameters we assume the same as in Section \ref{cont_vs_lead}.
Moreover, let us consider the percentage of contrarians equals to $10\%$ of the population.

\begin{figure*}[!htb]
\centering
\includegraphics[width=0.32\textwidth]{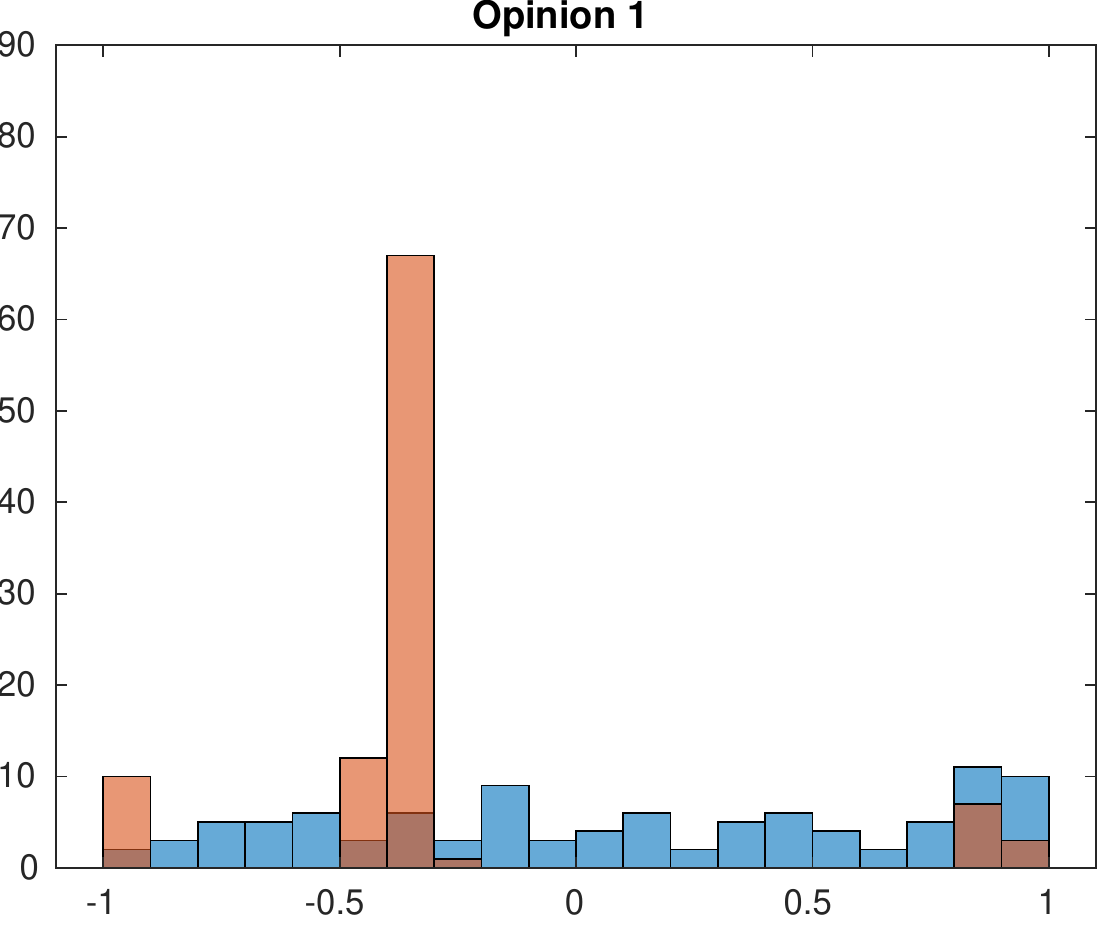}
\includegraphics[width=0.32\textwidth]{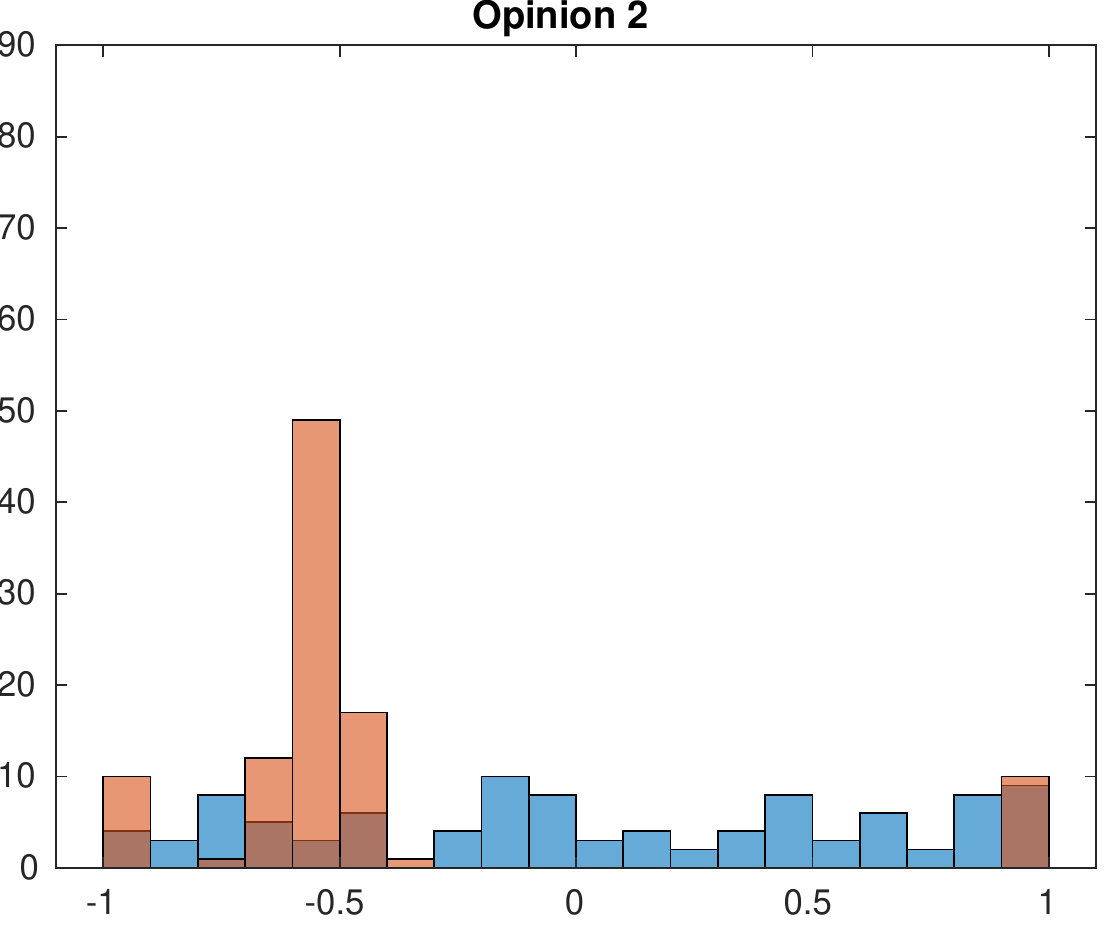}
\includegraphics[width=0.32\textwidth]{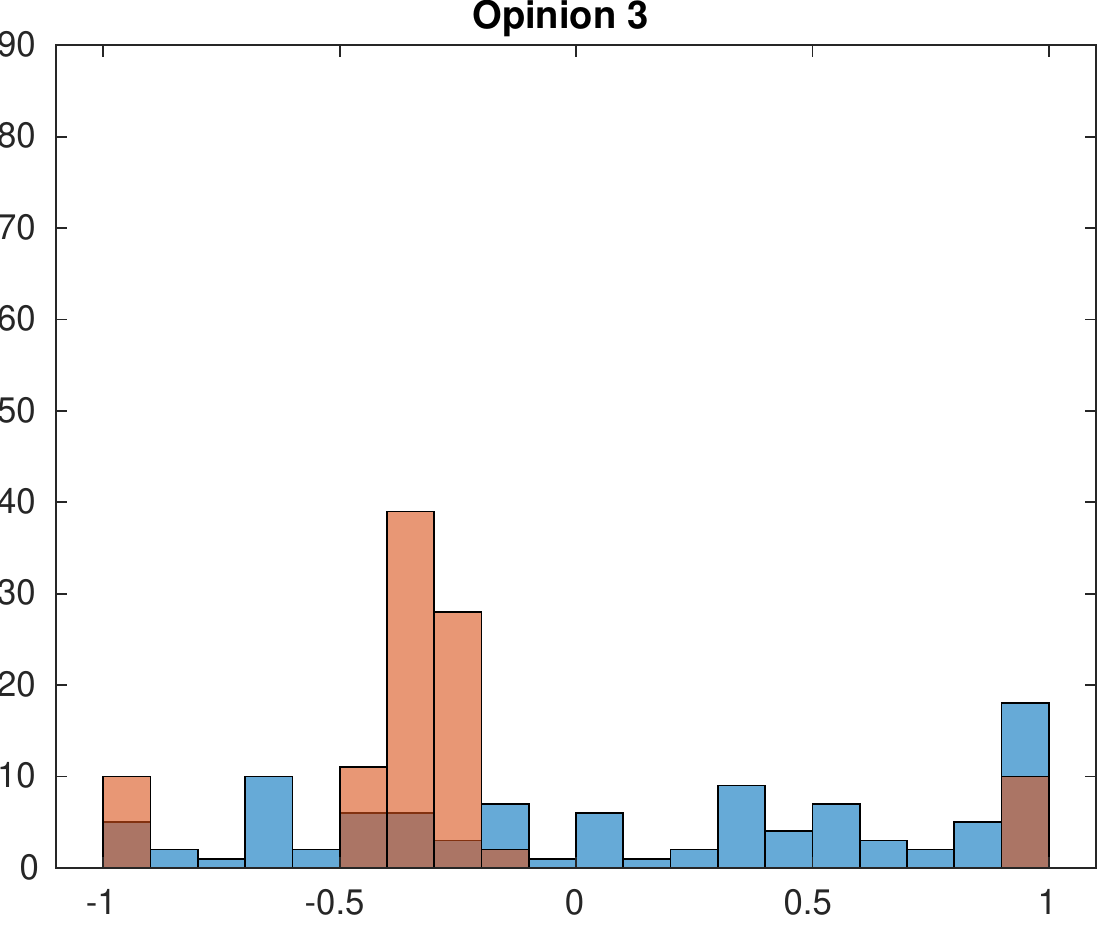}
\caption{Test H ---  Contrarians percentage $10\%$.The blue bars represent the initial data, while the orange ones the solution at time $T$.}
\label{fig:TH_c10}
\end{figure*}

By comparing Fig.\ \ref{fig:T2_c10}-\ref{fig:TH_c10}, it is evident the influence of the opinions concerning other topics (i.e. $x_{k,i}$ for $k>1$) on the whole mind vector evolution. Opinions 1, 2 and 3 of the entire society are dragged leftward by contrarians agents, on the other side respect to the fans' positions.

Let us note that the behaviour shown in Fig. \ref{fig:TH_c10} is very similar to that of Fig. \ref{fig:T2_c20}, even if the number of contrarians agents in the latter is the half of the first. These numerical experiments are consistent with the results of web scraping and polarization analysis evidenced in Section \ref{intro}. Twitter users can not only ``upload'' their opinions on environmental policies (in our case the opinion 2) according to a pre-determined political bias (in our case the opinion 1) but this could be influenced by opinions 2 and 3, thus forming a feedback mechanism. For this reason, even a small group of contrarians can have a great effect: when such a feedback mechanism is present, the social function of contrarians in the opinion formation of a group of agents seems to be reinforced. Compared to Fig. \ref{fig:T2_c20}, this means that the opinions concerning climate issues (i.e. opinions 2) can be driven toward more critical stances (i.e. negative values) even with a smaller number of contrarians


\section{Conclusions}

In this paper, we have moved away from the theme of the perception of climate risk and we focused instead on the opinions exchanges within the public debate on climate change, by suggesting that the public's hesitation towards it could be the results of a complex process in opinion dynamics. We have taken inspiration from the appearance of Greta Thunberg on the public scene.

From the analysis of tweets, we observed that the discussions on Twitter regarding Greta were based on topics related to each other and this gave us the idea for the model formalized in our work: an individual expresses opinions on different issues and these opinions are connected and influenced by each other beside, of course, the interaction with other individuals. Hence, we developed an agent based model able to reproduce the coherence effect between different topics. 

Moreover we postulated the existence of some mechanisms inside the society. We first observed the phenomenon induced by Greta Thunberg appearance, then we tried to explain it by conjecturing the existence of contrarians in the society.
Indeed different contrarians' dynamics are modeled and investigated in details.

We also presented several numerical experiments which show that the model is capable to describe correctly the insights coming from our study-case. 
The following points summarized the main findings of numerical tests:
\begin{itemize}
    \item Contrarians vs ``fan club''. By increasing the number of contrarians, it is possible to drive the opinions concerning climate issues toward more critical stances.
    This happens despite the presence of a group of Greta's fans acting as leaders. This result is more pronounced if the evolution of the ``main" for agent $i$ is affected by the other topics too, see  Section \ref{test_hype}.
    \item Contrarians vs Majority. If the opinions at initial time step is chosen randomly between $-1$ and $1$, a consensus towards the central opinion emerged also for high percentage of contrarians agents. Paradoxically this shows that the role of contrarians scaled down if they do not have an emerging opinion to oppose. 
    \item Conviction term. In the presence of fans, the impact of conviction coefficients is less evident, since the contrarians agents have to assume opinions opposite to the members of fan club, which in turn do not change their opinions.
\end{itemize}

However, the microscopic model adopted in this work does not let us to handle the population of a whole country, since the number of agents would be huge. Moreover we assume that all the people has a Twitter account that, in general, it is not true. Also, some users may change their opinions without posting them on Twitter. Agents that are not active enough cannot be tracked in their real-time opinion evolution. Therefore, our results should be just treated as a large-scale sample of the real social network. As a future work, we could project a captivating Twitter application able to attract users to express their feelings towards the topic of this paper should be developed (in this case, the popularity of the application would measure the reasonableness of the data collected by it). This motivates us to further investigations on the scale limit for obtaining a macroscopic model, following the idea of mean field limit, or to build a realistic model capable of predicting the evolutionary trend of public opinion (e.g. by using actual data to train parameters). Moreover, we collected tweets written in Italian; in future works, tweets in other languages (e.g. English) could be considered, and comparisons between different countries could be done.

\section*{Acknowledgements}
The authors sincerely thank the anonymous reviewer for critical feedback that helped us to improve the clarity and precision of our results and presentation.

The authors would like to acknowledge Emiliano Cristiani, Fabio Camilli and Luigi Teodonio for their stimulating discussions and suggestions about the topic of the paper.

The first author is member of the INdAM Research group GNCS. Moreover the author thanks the Deutsche Forschungsgemeinschaft (DFG, German Research Foundation) for the financial support through 20021702/GRK2326, 333849990/IRTG-2379, HE5386/18-1,19-1,22-1 and under Germany's Excellence Strategy EXC-2023 Internet of Production 390621612. The funding through HIDSS-004 is acknowledged.

The second author is member of the Association for Mathematics Applied to Economics and Social Sciences (AMASES).

\bibliographystyle{tfcad}
\bibliography{interactcadsample}

\end{document}